\documentclass[12pt,preprint]{aastex}

\begin{document}

\title{Light Curve Templates and Galactic Distribution of RR Lyrae Stars from Sloan Digital Sky Survey Stripe 82}

\author{
Branimir Sesar\altaffilmark{\ref{Washington}},
\v{Z}eljko Ivezi\'{c}\altaffilmark{\ref{Washington}},
Skyler H.~Grammer\altaffilmark{\ref{Washington}},
Dylan P.~Morgan\altaffilmark{\ref{Washington}},
Andrew C.~Becker\altaffilmark{\ref{Washington}},
Mario Juri\'{c}\altaffilmark{\ref{IAS}},
Nathan De Lee\altaffilmark{\ref{UF}},
James Annis\altaffilmark{\ref{FNAL}},
Timothy C.~Beers\altaffilmark{\ref{MSU}},
Xiaohui Fan\altaffilmark{\ref{UA}},
Robert H.~Lupton\altaffilmark{\ref{Princeton}},
James E.~Gunn\altaffilmark{\ref{Princeton}},
Gillian R.~Knapp\altaffilmark{\ref{Princeton}},
Linhua Jiang\altaffilmark{\ref{UA}},
Sebastian Jester\altaffilmark{\ref{MPIA}},
David E.~Johnston\altaffilmark{\ref{Princeton}},
Hubert Lampeitl\altaffilmark{\ref{FNAL}}
}

\altaffiltext{1}{University of Washington, Dept.~of Astronomy, Box
                           351580, Seattle, WA 98195-1580\label{Washington}}
\altaffiltext{2}{Institute for Advanced Study, 1 Einstein Drive,
                           Princeton, NJ 08540\label{IAS}}
\altaffiltext{3}{University of Florida, Dept.~of Astronomy, PO Box 112055,
                           Gainesville, FL 32611-2055\label{UF}}
\altaffiltext{4}{Fermi National Accelerator Laboratory, Box 500,
                           Batavia, IL 60510\label{FNAL}}
\altaffiltext{5}{Michigan State University, Dept.~of Physics \& Astronomy, and
                           JINA: Joint Institute for Nuclear Astrophysics, 
                           East Lansing, MI 48824-2320\label{MSU}}
\altaffiltext{6}{Steward Observatory, University of Arizona, Tucson, AZ
                           85721-0065\label{UA}}
\altaffiltext{7}{Princeton University Observatory, Princeton,
                           NJ 08544-1001\label{Princeton}}
\altaffiltext{8}{Max-Planck-Institut f\"{u}r Astronomie, K\"onigstuhl 17,
                           D-69117 Heidelberg, Germany\label{MPIA}}

\begin{abstract}
We present an improved analysis of halo substructure traced by RR Lyrae stars in
the SDSS stripe 82 region. With the addition of SDSS-II data, a revised
selection method based on new $ugriz$ light curve templates results in a sample
of 483 RR Lyrae stars that is essentially free of contamination. The main result
from our first study persists: the spatial distribution of halo stars at
galactocentric distances 5--100 kpc is highly inhomogeneous. At least 20\% of
halo stars within 30 kpc from the Galactic center can be statistically
associated with substructure. We present strong direct evidence, based on both
RR Lyrae stars and main sequence stars, that the halo stellar number density
profile significantly steepens beyond a Galactocentric distance of $\sim$30 kpc,
and a larger fraction of the stars are associated with substructure. By using a
novel method that simultaneously combines data for RR Lyrae and main sequence
stars, and using photometric metallicity estimates for main sequence stars
derived from deep co-added $u$-band data, we measure the metallicity of the
Sagittarius dSph tidal stream (trailing arm) towards R.A.$\sim2^h-3^h$ and
$Dec\sim0^\circ$ to be 0.3 dex higher ($[Fe/H]=-1.2$) than that of surrounding
halo field stars. Together with a similar result for another major halo
substructure, the Monoceros stream, these results support theoretical
predictions that an early forming, smooth inner halo, is metal poor compared to
high surface brightness material that have been accreted onto a later-forming
outer halo. The mean metallicity of stars in the outer halo that are not
associated with detectable clumps may still be more metal-poor than the bulk of
inner-halo stars, as has been argued from other data sets.  

\end{abstract}


\section{Introduction\label{introduction}}

Studies of the Galactic halo can help constrain the formation history of the
Milky Way and the galaxy formation process in general. For example, within the
framework of hierarchical galaxy formation \citep{fbh02}, the spheroidal
component of the luminous matter should reveal substructures such as tidal tails
and streams \citep{jhb96,hw99,bkw01,har01}. The number of these substructures,
due to mergers and accretion over the Galaxy's lifetime, may provide a crucial
test for proposed solutions to the ``missing satellite'' problem \citep{bkw01}.
Substructures are expected to be ubiquitous in the outer halo (galactocentric
distance $>15-20$ kpc), where the dynamical timescales are sufficiently long for
them to remain spatially coherent \citep{jhb96,may02}, and indeed many have been
discovered
\citep[e.g.,][]{iba97,yan00,ive00,bel06,gri06,vz06,bel07a,new07,jur08}. Various
luminous tracers, such as main sequence turn-off stars, RR Lyrae variables, or
red giants are used to detect halo substructures, and of them, RR Lyrae stars
have proven to be especially useful.

RR Lyrae stars represent a fair sample of the old halo population \citep{smi95}.
They are nearly standard candles ($\langle M_V \rangle = 0.59 \pm 0.03$ at
$[Fe/H]=-1.5$, \citealt{cc03}), are sufficiently bright to be detected at large
distances ($5-120$ kpc for $14 < V < 21$), and are sufficiently numerous to
trace the halo substructure with good spatial resolution ($\sim1.5$ kpc at 30
kpc). Fairly complete and relatively clean samples of RR Lyrae stars can be
selected using single-epoch colors \citep{ive05}, and if multi-epoch data exist,
using variability (\citealt{ive00}; QUEST, \citealt{viv01}; \citealt{ses07},
\citealt{dle08}; SEKBO, \citealt{kel08}; LONEOS-I, \citealt{mic08}). A useful
comparison of recent RR Lyrae surveys in terms of their sky coverage, distance
limits, and sample size is presented by \citet{kel08} (see their Table 1).
Compared to other surveys, the so-called stripe 82 survey (an equatorial strip
defined by declination limits of $\pm$1.27$^\circ$ and extending from
R.A.$\sim$20$^h$ to R.A.$\sim$4$^h$) based on Sloan Digital Sky Survey data
(SDSS; \citealt{yor00}) is the deepest one yet obtained (probing distances to
$\sim110$ kpc), and the only survey with available 5-band photometry.

In an initial study based on SDSS stripe 82 data, we discovered a complex
spatial distribution of halo RR Lyrae stars at distances ranging from 5 kpc to
$\sim$100 kpc \citep[hereafter S07]{ses07}. The candidate RR Lyrae sample was
selected using colors and low-order light curve statistics. In this paper, we
present a full light curve analysis enabled by the addition of new stripe 82
observations. The resulting sample of RR Lyrae stars is highly complete and
essentially free of contamination. The extended observations now allow the
determination of flux-averaged magnitudes, thus providing better luminosity and
distance estimates. This reduces the uncertainty in the spatial distribution of
RR Lyrae stars, and increases the overall accuracy of computed density maps.

While this paper was in preparation, another study of RR Lyrae stars in stripe
82, based on essentially the same data set, was announced \citep{wat09}. In
addition to analysis that overlaps with their work, here we also discuss
\begin{itemize}
\item
An RR Lyrae template construction for the SDSS bandpasses, and analysis of the
template behavior
\item
Template fitting and visual inspection of best-fits for single-band and color
light curves, which results in an exceedingly clean final sample
\item
Extended color range for the initial selection, which results in about 20\% more
stars than in the sample analyzed by Watkins et al.
\item
A step-by-step analysis of the sample incompleteness at the faint end
\item
A new method for constraining the metallicity of spatially coherent structures
that are detected using main sequence and RR Lyrae stars
\item
A comparison of the observed RR Lyrae spatial distribution with predictions
based on an oblate halo model constrained by SDSS observations of main sequence
stars.
\end{itemize}
We emphasize, however, that both studies obtain consistent main results: strong
substructure in the spatial distribution of halo RR Lyrae stars and evidence for
a steepening of the best-fit power-law description beyond a galactocentric
radius of $\sim$30 kpc.

Our data set and the initial selection of candidate RR Lyrae stars are described
in \S~\ref{data}, while the construction of light curve templates is presented
in \S~\ref{ugriz_templates}. Various properties of the resulting RR Lyrae
sample, with emphasis on the spatial distribution, are analyzed in
\S~\ref{sec:spatial}. In \S~\ref{sec:MS} we compare the spatial distribution of
RR Lyrae stars and main sequence stars and constrain the metallicity
distribution of the Sgr tidal stream. Our main results are summarized and
discussed in \S~\ref{discussion}.

\section{The Data\label{data}}

\subsection{An Overview of the SDSS Imaging and Stripe 82 Data\label{sdss_data}}

The Sloan Digital Sky Survey (SDSS; \citealt{amc08}) provides homogeneous and
deep ($r<22.5$) photometry in five bands ($u$, $g$, $r$, $i$, and $z$) of more
than 7,500 deg$^2$ of the North Galactic Cap, and three stripes in the South
Galactic Cap totaling 740 deg$^2$. The central stripe in the South Galactic Cap,
stripe 82 ($20^h 32^m <\alpha_{J2000.0} < 04^h 00^m$,
$-1.26\arcdeg<\delta_{J2000.0} < +1.26\arcdeg$, $\sim280$ deg$^2$) was observed
repeatedly (not all imaging runs cover the entire region) to enable a deep
co-addition of the data and to enable the discovery of variable objects. It was
the largest source of multi-epoch data during the first phase of the SDSS
(SDSS-I), with a median of 10-20 observations per source (S07, \citealt{bra08})
and it remains so thanks to additional imaging runs obtained during the SDSS-II
Supernova (SN) survey \citep{fri08}. Some of the SDSS-II SN imaging runs were
taken in bad photometric conditions (e.g., with fast-changing cloud coverage),
rendering the photometric data unreliable if not recalibrated. We have followed
the \citet{ive07} prescription (see their Section 3) and recalibrated 180
SDSS-II imaging runs to 1\% photometry (magnitude zero-point uncertainty
$\la0.01$ mag). The repeated photometric measurements demonstrate that the
photometric precision is about 0.02 mag at the bright end and reaches 0.05 mag
at magnitudes of 20, 21.5, 21.5, 21.5, and 19 for the $ugriz$ bands,
respectively (S07). The morphological information from the images allows
reliable star-galaxy separation to $r\sim21.5$ \citep{lup02}.

\begin{figure}
\epsscale{1.0}
\plotone{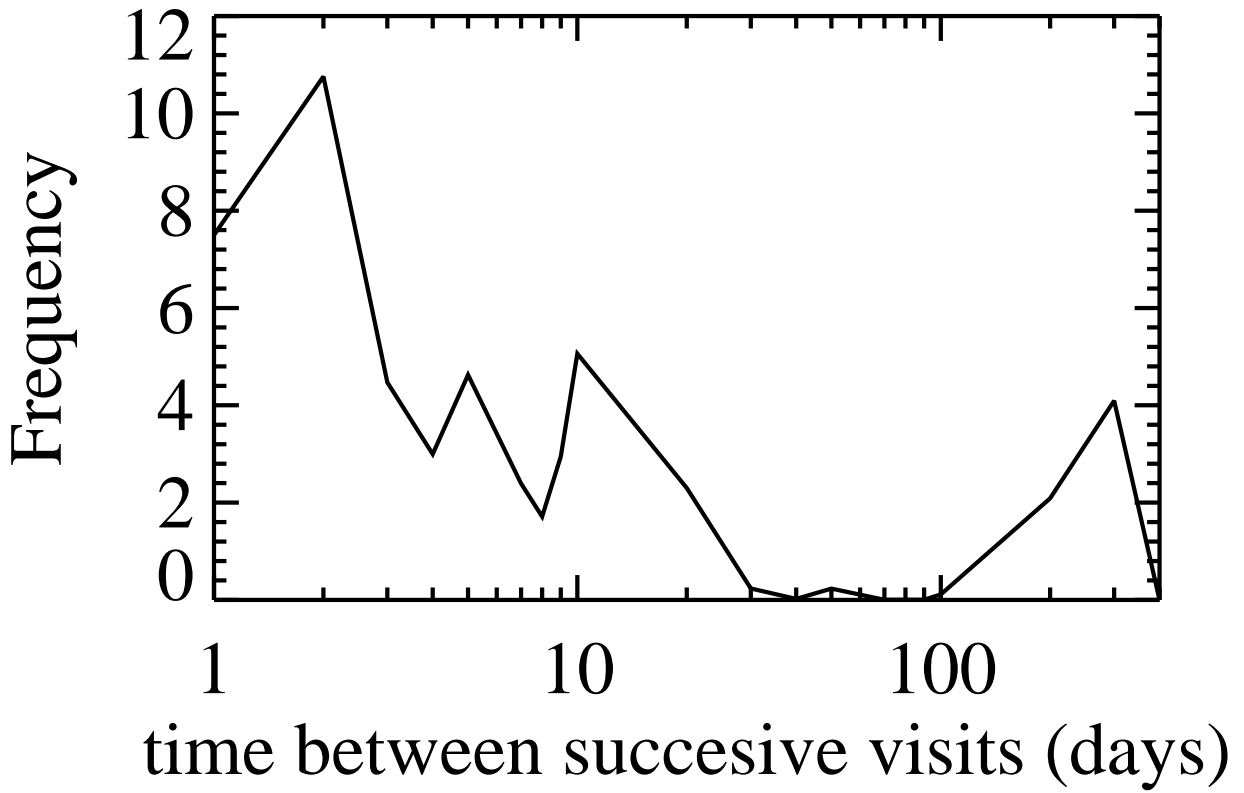}
\caption{
The cadence of SDSS stripe 82 observations. On average, the objects were most
often re-observed every two days (the SDSS-II SN Survey cadence), followed by
5-day, 10-day and yearly re-observations.
\label{s82_cadence}}
\end{figure}

We have combined SDSS-I imaging runs used in S07 with recalibrated SDSS-II SN
imaging runs, and created a catalog with $\sim4.4$ million unique unresolved
(point source) objects. Each object in the catalog was observed at least 4 times
in the $ugr$ bands, and the cadence of re-observations is shown in
Figure~\ref{s82_cadence}. On average, the SDSS stripe 82 objects were most
often re-observed every two days (the SDSS-II SN Survey cadence), followed by
5-day, 10-day and yearly re-observations. We time-average repeated $ugriz$
observations (median PSF magnitudes) and correct them for interstellar
extinction using the map from \citet{SFD98}. We also compute various low-order
statistics, such as the root-mean-square (rms) scatter ($\sigma$), $\chi^2$ per
degree of freedom ($\chi^2$), and light-curve skewness ($\gamma$) for each
$ugriz$ band and each object (see S07 for details). For objects brighter than
$g=21$, the median number of observations per band is 30.

The constructed catalog is a superb dataset for the selection and identification
of RR Lyrae stars. The three-fold increase in the median number of observations
per object over the S07 catalog allows a reliable measurement of the period of
pulsation and enables the construction of period-folded $ugriz$ light curves. As
we describe in the next Section, RR Lyrae stars can be unambiguously identified
using their period-folded $ugriz$ light curves, and an essentially complete
sample of stripe 82 RR Lyrae stars can then be constructed.

\subsection{The Selection of RR Lyrae Stars\label{RRLyrae_catalog}}

In this section we describe the selection of 483 RR Lyrae stars in SDSS stripe
82. The positions and photometry (MJD, $ugriz$ magnitudes and errors) of
selected RR Lyrae stars are provided in the electronic edition of the Journal;
the data format is provided in Table~\ref{example_photometry}.

In S07 we defined color and variability cuts specifically designed for the
selection of candidate RR Lyrae stars. We adopt those selection criteria, and
expand the probed $u-g$ color range from $u-g=0.98$ (used in S07) to $u-g=0.7$.
This expansion is motivated by theoretical predictions obtained from recent
RR Lyrae pulsation models by \citet{mar06}. They find that, for a given
effective temperature, more metal-poor stars have bluer $u-g$ colors. A similar
behavior is also detected in main sequence stars (\citealt{ive08} and references
therein). Thus, the expanded $u-g$ color range might increase completeness for
more metal-poor RR Lyrae stars; at the same time the sample contamination is
not an issue because light curves are now well sampled.

We start the RR Lyrae candidate selection by making the following color
\begin{eqnarray}
0.7 < u-g < 1.35\label{color_cuts1}\\
-0.15 < g-r < 0.40\label{color_cuts2} \\
-0.15 < r-i < 0.22\label{color_cuts3} \\
-0.21 < i-z < 0.25\label{color_cuts4},
\end{eqnarray}
and variability cuts
\begin{itemize}
\item
The number of $g$ band observations, $N_{obs,g} > 15$
\item
The chi-square per degree of freedom in the $g$ and $r$ bands, $\chi^2(g) > 3$
and $\chi^2(r) > 3$
\item
The rms scatter in the $g$ and $r$ bands, $\sigma(g) > 0.05$ mag and
$\sigma(r) > 0.05$ mag
\item
The ratio of the rms scatter in the $g$ and $r$ bands,
$0.7<\sigma(g)/\sigma(r)\leqslant2.5$, and
\item
The $g$ band light curve skewness, $\gamma(g) \leqslant 1$, to avoid eclipsing
binaries (see S07 for details).
\end{itemize}
Note that the colors in Equations~\ref{color_cuts1}-\ref{color_cuts4} are
obtained from time-averaged (median) $ugriz$ magnitudes, corrected for
extinction using the maps from \citet{SFD98}. These criteria are designed to
reject stars that are definitely not RR Lyrae, and all but one
($N_{obs,g} > 15$) should be satisfied by every true RR Lyrae star. The second
variability criterion can introduce incompleteness at the faint end, because
photometric errors increase with magnitude. We address this incompleteness
effect in detail in \S~\ref{magnitude_completeness}. Using these criteria we
select the initial sample of 3449 candidate RR Lyrae stars (about 1 in 7 are
true RR Lyrae stars; the fraction of RR Lyrae stars in the full sample is about
1 in 10$^4$), and proceed to measure their light curve periods in the next step.

To estimate the period of pulsation for candidate RR Lyrae stars, we use an
implementation of the {\em Supersmoother} routine \citep{rei94}. The routine
uses a running mean or running linear regression on the data to fit the
observations as a function of phase to a range of periods, constrained to be
between 0.2 and 1.0 days (see Figure 9 in \citealt{em08}). The main advantage of
this routine is that the light curve shape need not be known in advance. We
supply the Modified Julian date (MJD) of observations, $g$-band magnitudes and
their errors, and the routine outputs the first 5 periods with the lowest
chi-square values. The $g$-band magnitudes are used instead of $u$- or $r$-band
magnitudes because they have higher signal-to-noise ratios than the $u$-band
magnitudes, and because RR Lyrae stars have a larger amplitude of variability in 
the $g$ band than in the $r$ band (\citealt{ive00}, S07). The precision of
periods is about a few seconds, and was determined by comparing periods
estimated from $g$- and $r$-band data.

Using each of the five periods supplied by the {\em Supersmoother} routine, we
period-fold a $g$-band light curve and then fit to it one of eight $V$-band
template RR Lyrae light curves. The $V$-band template light curves were obtained
from \citet[6 RR type $ab$ templates]{lay98} and K.~Vivas (priv.~comm., 2 RR
type $c$ templates), and are used due to the lack of $g$-band templates in the
literature. The fitting is performed in the least square sense, with the epoch
of the maximum brightness $\phi^g_0$, peak-to-peak amplitude $A_{g}$, and mean
magnitude $g_0$ as free parameters. The precision of the epoch of maximum
brightness is on the order of a few minutes, and was determined by comparing
epochs of maximum brightness estimated from $g$- and $r$-band light curves
fitted with $V$-band templates.

To robustly quantify the quality of a template fit we define a $\chi^2$-like
parameter 
\begin{equation}
\zeta = median(|m_{observed}^i - m_{template}|/\sigma_{observed}^i)\label{abs_sig_dev},
\end{equation}
where $m_{observed}$ and $\sigma_{observed}$ are the observed magnitude and its
uncertainty, $m_{template}$ is the magnitude predicted by the template, and
$i=1, N_{obs}$, where $N_{obs}$ is the number of observations. Here we use the
median to minimize the bias in $\zeta$ due to poor observations (outliers). The
template with the lowest $\zeta$ value is selected as the best fit, and the
best-fit parameters are stored.

Initially, we wanted to use $\zeta$ to separate RR Lyrae stars from non-RR Lyrae
stars. Low $\zeta$ values, indicating good agreement between the observed light
curves (data) and template light curves (models), would indicate that a
candidate is an RR Lyrae star. Likewise, high $\zeta$ values would indicate that
a candidate is not an RR Lyrae star. However, visual inspection of light curves
initially classified as non-RR Lyrae (high $\zeta$ values) showed that some of
the light curves are actually quite RR Lyrae-like, but with a lot of scatter
around the template-predicted values. We suspect that these stars are RR Lyrae
stars that undergo changes in their period, light curve shape, or amplitude as a
function of time (e.g., the Blazhko effect; \citealt{bla07,kol08}). By putting a
hard limit on $\zeta$ such RR Lyrae stars would be rejected, and the
completeness of our sample (the fraction of recovered RR Lyrae stars) would
decrease. To avoid this, we look for additional criteria beyond the light curve
shape that can be used to distinguish true RR Lyrae stars from non-RR Lyrae
stars. For this purpose we have created a ``training'' set of 479 RR Lyrae stars
that were selected by visual inspection of their light curves, and use it to
explore the variability phase space of RR Lyrae stars. Hereafter we refer to the
training set as the template-accepted RR Lyrae stars.

\begin{figure}
\epsscale{0.75}
\plotone{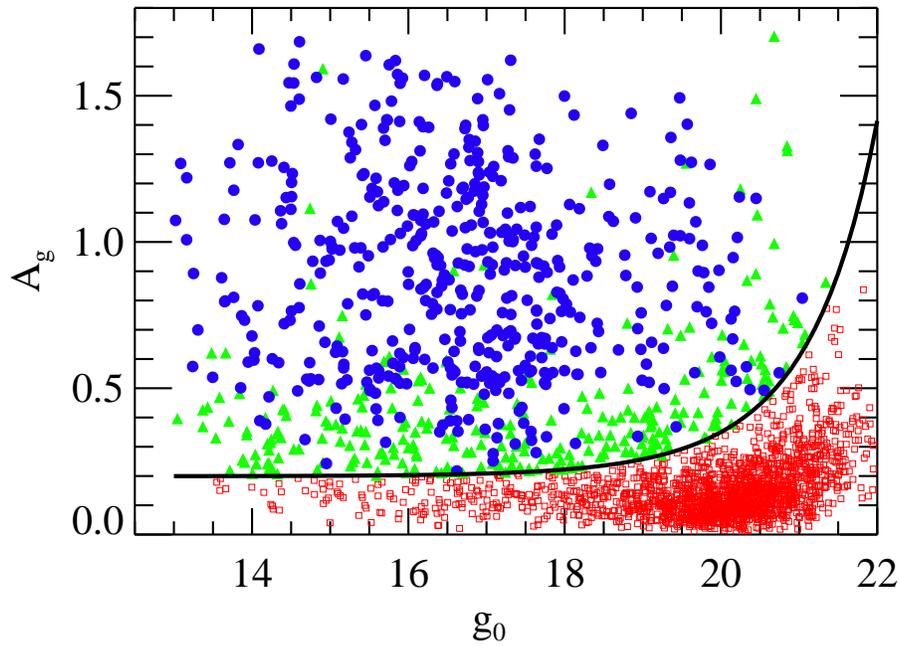}
\caption{
The distribution of $g$-band amplitudes ($A_{g}$) and mean magnitudes ($g_0$)
determined from best-fit templates for 3449 RR Lyrae candidates. The solid line
separates candidates into template-accepted ({\em blue dots}) and ambiguous
candidates ({\em green triangles}), which lie above the line, and
template-rejected ({\em red open squares}) which lie below the line.
\label{amp_vs_mag0_all}}
\end{figure}

Based on the distribution of template-accepted RR Lyrae stars shown in
Figure~\ref{amp_vs_mag0_all}, we define a line that separates candidate RR Lyrae
into template-accepted and ambiguous (above the line), and template-rejected
(2608 candidates below the line). The ambiguous candidates (362 objects) did not
have at first visually convincing (RR Lyrae-like) light curves, but based on
their variability properties (rms scatter in the $g$, $r$ and $g-r$) some of
them might be true RR Lyrae stars.

The question of whether some of the ambiguous candidates are true RR Lyrae stars
is especially important at the faint end. As shown in
Figure~\ref{amp_vs_mag0_all}, the distribution of template-accepted RR Lyrae
stars has a sharp cut-off around $g_0\sim20.5$, suggesting a limit on the
spatial distribution of halo stars. A similar cut-off in the distribution of RR
Lyrae stars was also discussed by \citet{ive00}. If it can be shown that the
ambiguous candidates with $g_0\sim20.5$ have properties (beside their light
curves) that are inconsistent with the properties of template-accepted RR Lyrae
stars, they could be excluded from the sample, and the claim that the halo
cut-off is real would become much stronger.

\begin{figure}
\epsscale{0.4}
\plotone{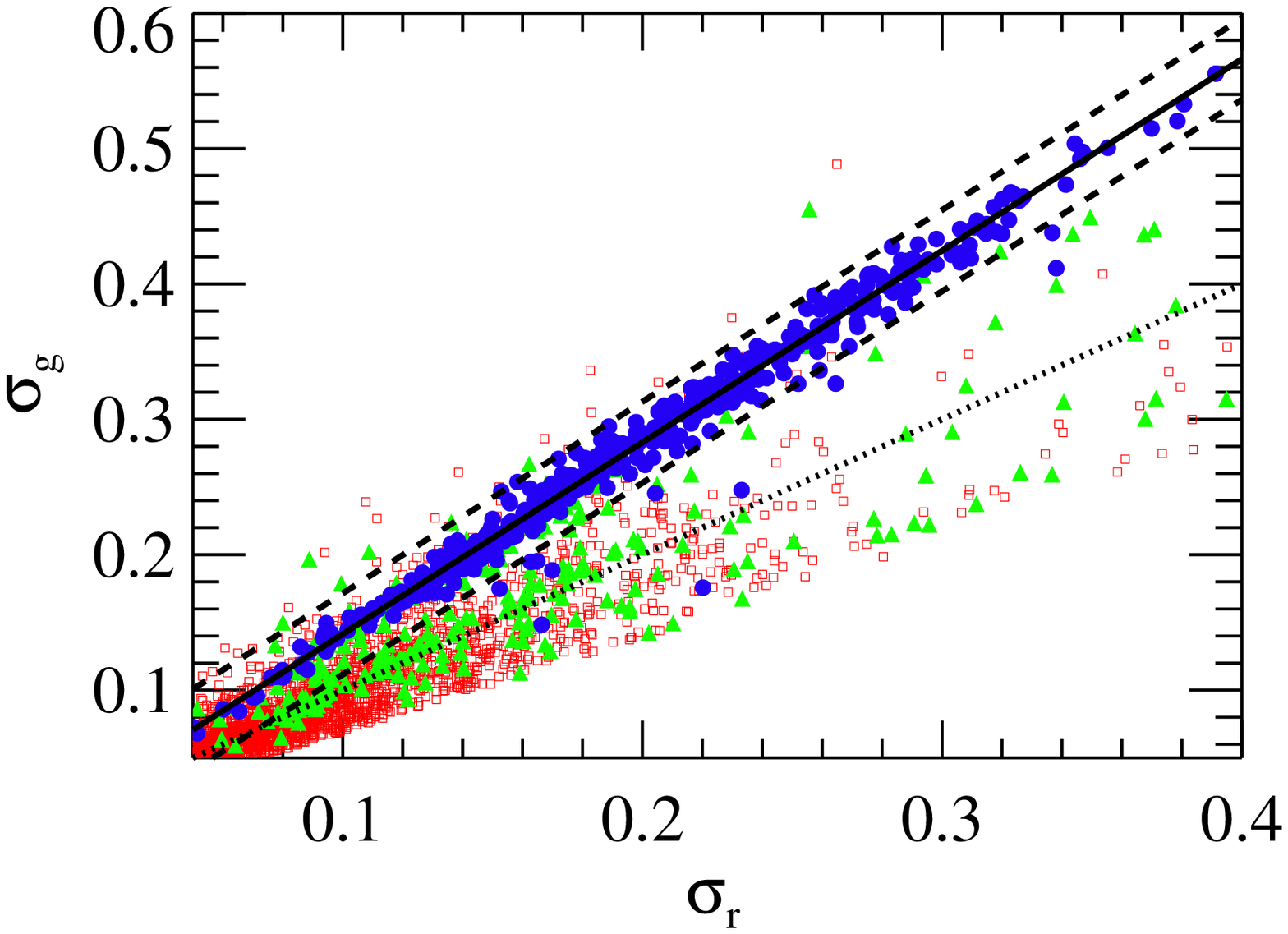}

\plotone{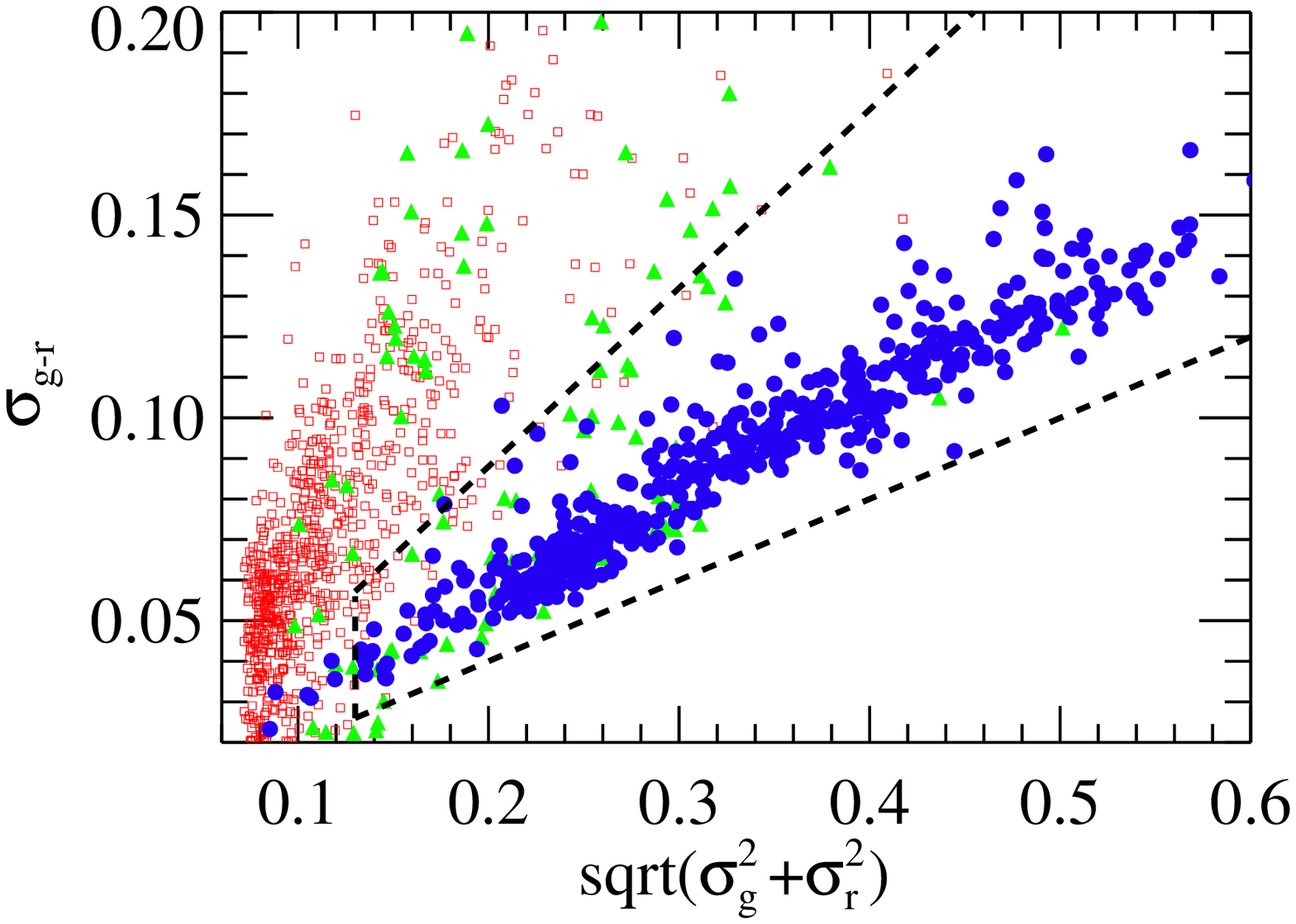}

\plotone{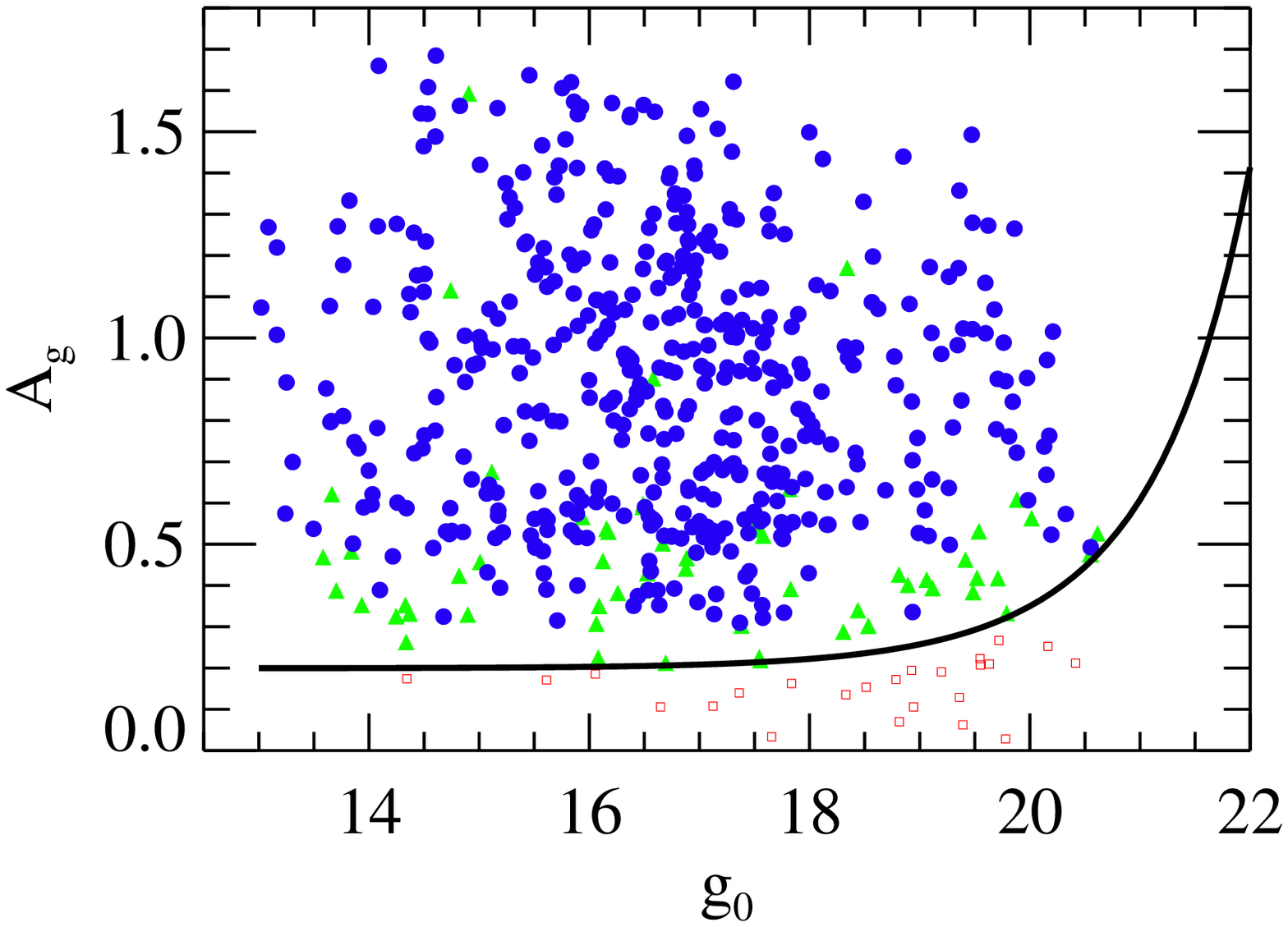}
\caption{
{\em Top}: The $\sigma_g$ vs.~$\sigma_r$ distribution of template-accepted
({\em blue dots}), template-rejected ({\em red open squares}), and ambiguous
candidates ({\em green triangles}) before the $|\sigma_g - 1.42\sigma_r|<0.03$
selection (region between dashed lines). Note that the majority of
template-rejected and ambiguous candidates do not follow the
$\sigma_g = 1.42\sigma_r$ relation typical of RR Lyrae stars ({\em solid line}),
but rather the $\sigma_g = \sigma_r$ relation ({\em dotted line}) that indicates
non-pulsational variability. {\em Middle}: The $\sigma_{g-r}$
vs.~$sqrt(\sigma_g^2+\sigma_r^2)$ distribution of candidates from the top panel
after the $|\sigma_g - 1.42\sigma_r|<0.03$ selection is applied. The dashed
lines show the selection based on the $\sigma_{g-r}$ and
$sqrt(\sigma_g^2+\sigma_r^2)$ values, where $\sigma_{g-r}$ is the rms scatter in
the $g-r$ color. {\em Bottom}: The $A_{g}$ vs.~$\langle g \rangle$ distribution
of candidates selected from the middle panel's dashed box region. Note that
almost all of the ambiguous candidates with $g_0\sim20.5$ shown in
Fig.~\ref{amp_vs_mag0_all} are removed after top and middle panel cuts are made,
while the template-accepted candidates remain.
\label{auto_class}}
\end{figure}

\begin{figure}
\epsscale{1.0}
\plotone{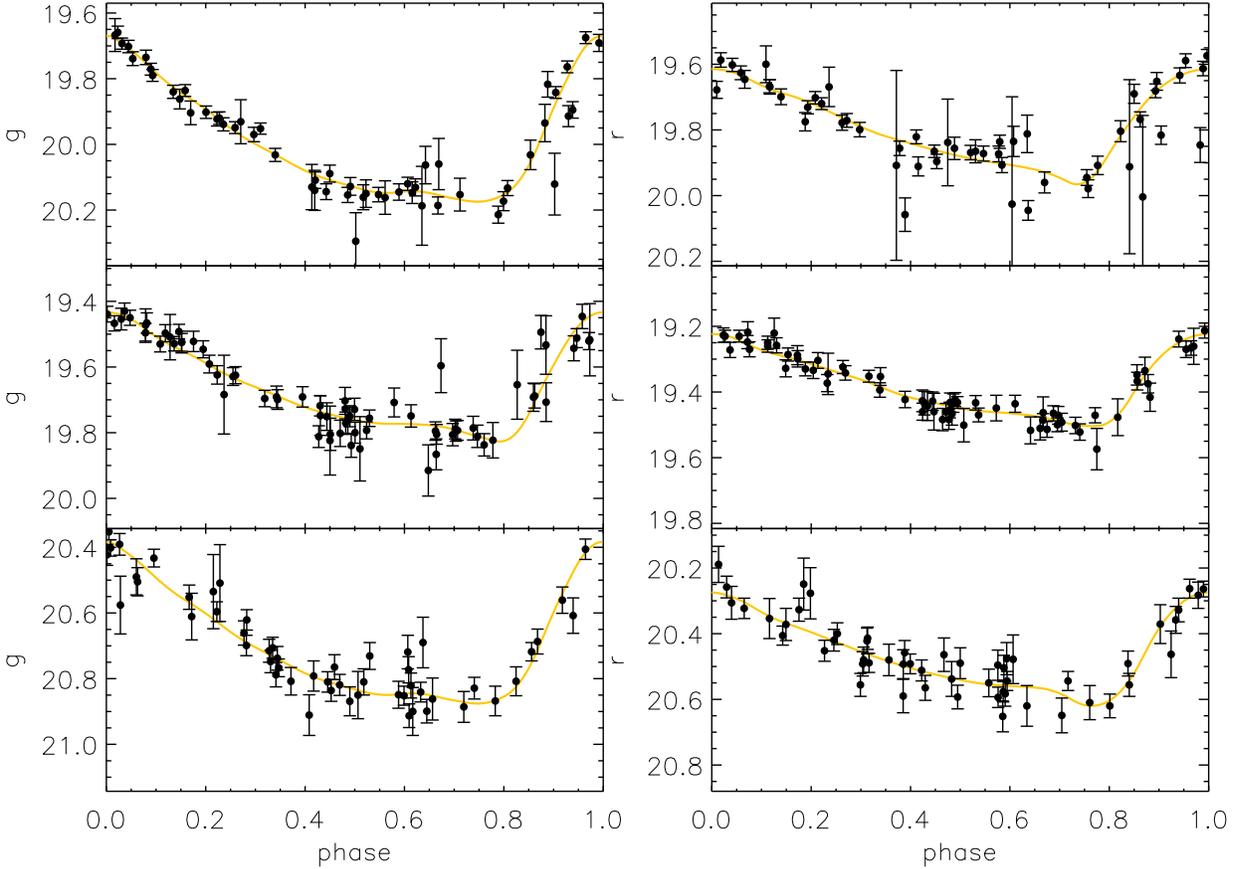}
\caption{
The top and middle panel show the $g$- ({\em left}) and $r$-band ({\em right})
light curves of two RR Lyrae stars with the observed $\sigma_g$ slightly smaller
than the predicted $\sigma_g=1.42\sigma_r$ value (by $\sim0.1$ mag). The bottom
panel shows the $g$- and $r$-band light curves of a RR Lyrae star with
$\sigma_{g-r}$ greater by $\sim0.05$ mag than the cut-off dashed line in
Figure~\ref{auto_class} ({\em bottom}). The solid lines show the best-fit $g$-
and $r$-band templates (see Section~\ref{ugriz_templates}). All three stars have
slightly peculiar $\sigma_g$, $\sigma_r$, and $\sigma_{g-r}$ values, but are
still included in the final RR Lyrae sample as their light curves are very much
RR Lyrae-like.
\label{outliers}}
\end{figure}

Aiming to define the RR Lyrae variability space, we now consider how the three
groups of candidates distribute in the $\sigma_g$ vs.~$\sigma_r$ diagram, where
the $\sigma_g$ and $\sigma_r$ are the root-mean-square (rms) scatter in the $g$
and $r$ bands, respectively. This projection of phase space is useful as it
separates pulsationally variable stars from other variable sources (S07). As
shown in Figure~\ref{auto_class} ({\em top}), the majority of template-accepted
RR Lyrae stars exist in a narrow region centered on the
$\sigma_g = 1.42\sigma_r$ (the rms scatter around $\sigma_g = 1.42\sigma_r$ is
$0.006$ mag) line typical of RR Lyrae stars (\citealt{ive00}; S07), while the
majority of template-rejected and ambiguous candidates follow the
$\sigma_g = \sigma_r$ relation that suggests non-pulsational variability. It is
interesting to note that 13 template-accepted RR Lyrae stars have lower observed
$\sigma_g$ values (by $>0.03$ mag) than what is predicted by the
$\sigma_g=1.42\sigma_r$ ($5\sigma$) relation. The $g$- and $r$-band light curves
of two such stars are shown in Figure~\ref{outliers} ({\em top} and
{\em middle}). Based on the Figure~\ref{auto_class} ({\em top}) plot, we
conclude that the majority of RR Lyrae stars ($97\%$) have
$|\sigma_g - 1.42\sigma_r|<0.03$, and select all candidates that pass this
condition for further analysis. The 13 template-accepted RR Lyrae stars found
outside the $|\sigma_g - 1.42\sigma_r|<0.03$ region are also included as we
believe they are RR Lyrae stars (based on their light curves), but with peculiar
variability properties (slightly lower than expected rms scatter in the $g$
band).

The candidates that pass the $|\sigma_g - 1.42\sigma_r|<0.03$ condition also
separate well in the $\sigma_{g-r}$ vs.~$\sqrt{\sigma_g^2+\sigma_r^2}$ plot
(Figure~\ref{auto_class} [{\em middle}]), which is sensitive to the covariance
between the $g$ and $r$ bands. Using the distribution of template-accepted
candidates as an indicator of where RR Lyrae stars exist in this plot, we make
another selection and again plot the selected candidates in the $g$ band
amplitude vs.~mean magnitude plot (Figure~\ref{auto_class} [{\em bottom}]).
There are eight template-accepted RR Lyrae stars that do not pass the
$\sigma_{g-r}$ vs.~$\sqrt{\sigma_g^2+\sigma_r^2}$ selection, one of which is
shown in Figure~\ref{outliers} ({\em bottom}). As in the previous paragraph, we
choose to keep these candidates as they might be RR Lyrae stars with peculiar
variability properties (higher than expected rms scatter in the $g-r$ color).

\begin{figure}
\epsscale{1.0}
\plotone{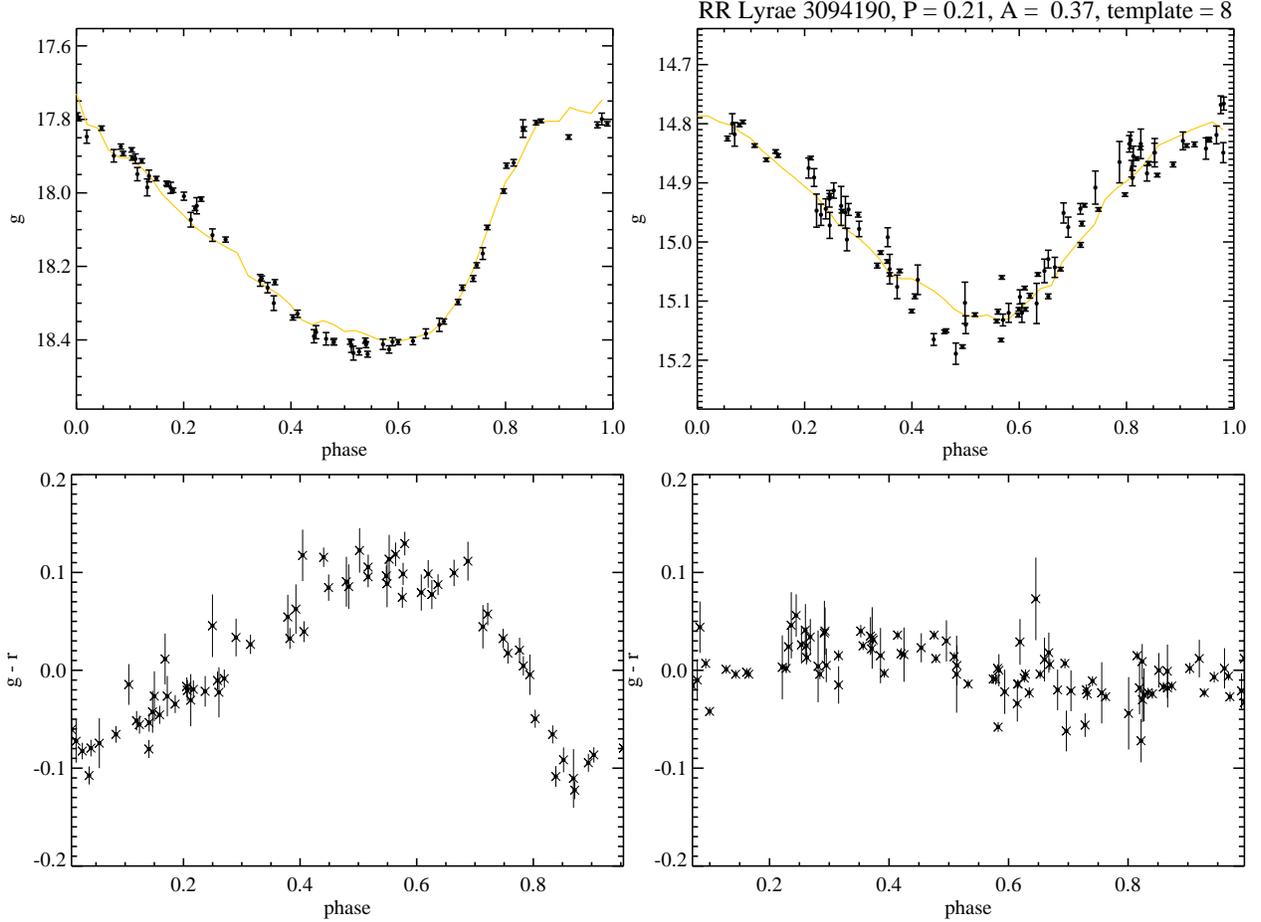}
\caption{
The $g$ ({\em top}) and $g-r$ ({\em bottom}) light curves of a RRc star ({\em
left}) and a candidate eclipsing binary ({\em right}). The best fit $V$-band
templates are shown as solid lines. Even though their $g$-band light curves are
quite similar, and their periods are in the $0.2<P<0.43$ days range typical of
RRc stars, their $g-r$ light curves are quite different. The eclipsing binary is
outside the Figure~\ref{auto_class} ({\em middle}) selection box as it has a
much smaller rms scatter in the $g-r$ color ($\sigma_{g-r}\sim0.02$ mag) than
the RRc star ($\sigma_{g-r}\sim0.07$ mag).
\label{eclipsing}}
\end{figure}

The utility of the $\sigma_{g-r}$ selection is demonstrated in
Figure~\ref{eclipsing}, where we compare $g$ and $g-r$ light curves of a type
$c$ RR Lyrae (hereafter RRc) and a candidate eclipsing binary star. Even though
their $g$-band light curves are quite similar and their periods match those of
RRc stars, their $g-r$ light curves are quite different. The $g-r$ light curve
of the eclipsing binary varies much less that that of the RRc star, as the
binary is most probably comprised of two similar effective temperature stars. On
the other hand, the effective temperature of the RRc star changes due to
pulsations, and the variations in the $g-r$ light curve are much greater.

A comparison of Figures~\ref{amp_vs_mag0_all} and~\ref{auto_class}
({\em bottom}) shows that the majority of ambiguous and template-rejected
candidates do not have variability properties ($\sigma_g$, $\sigma_r$,
$\sigma_{g-r}$) that are consistent with variability properties of
template-accepted RR Lyrae stars. Only 18\% of ambiguous and 1\% of
template-rejected candidates pass our $\sigma_g-\sigma_r-\sigma_{g-r}$
variability cuts, while only a small fraction of template-accepted candidates
(4\%) is rejected. Most importantly, almost all of the ambiguous candidates
with $g_0\ga20.5$ were rejected based on their variability properties.

These results suggest that the candidates initially classified as
template-accepted are indeed true RR Lyrae stars, but that some ambiguous and
template-rejected candidates might also be true RR Lyrae stars. We have
re-examined light curves of ambiguous and template-rejected candidates that
pass the $\sigma_g-\sigma_r-\sigma_{g-r}$ variability cuts, and re-classified
four as template-accepted. The 21 template-accepted candidates that were
slightly outside of our selection criteria were also kept, as the visual
inspection of their $g$ and $g-r$ light curves strongly suggests they are indeed
RR Lyrae stars. Finally, we report 483 stars in our SDSS stripe 82 RR Lyrae
sample, and estimate it to be essentially complete and free of non-RR Lyrae
stars.

Our high degree of completeness is achieved by relaxing color and variability
cuts from S07, and by accepting stars that are slightly outside the selection
region if their light curves are strongly RR Lyrae-like. We find that the median
$g-r$, $r-i$, and $i-z$ colors of stars in the final sample are well within the
ranges defined by Equations~\ref{color_cuts2}-\ref{color_cuts4}, and 20 of the
selected stars ($\sim4\%$) have median $u-g$ color in the $0.85<u-g<0.98$ range.
This finding validates the expansion of the $u-g$ range blueward to 0.7 mag
(from $u-g=0.98$ used in S07). The visual inspection of $g$ and $g-r$ light
curves ensures that non-RR Lyrae stars are not included in the final sample.

When time-averaged colors are not available (as is the case for the majority of
SDSS stars), a range of single-epoch colors might be more useful for the
selection of RR Lyrae stars. We find that RR Lyrae stars have single-epoch
colors in these ranges:
\begin{eqnarray}
0.75 < u-g < 1.45 \\
-0.25 < g-r < 0.4 \\
-0.2 < r-i < 0.2 \\
-0.3 < i-z < 0.3\label{single_epoch_colors}.
\end{eqnarray}

\section{The $ugriz$ RR Lyrae Templates\label{ugriz_templates}}

The availability of a large sample of RR Lyrae stars with densely sampled light
curves presents us with a unique opportunity to derive template RR Lyrae light
curves for the SDSS photometric system. While template RR Lyrae light curves for
the SDSS $ugriz$ bands do exist in the literature (e.g., see \citealt{mar06}),
they are based on pulsation models, and not on actual observations. Recognizing
the importance of such templates for future large sky surveys that will use the
SDSS filter set, such as the Large Synoptic Survey Telescope (LSST;
\citealt{ive08}) and Pan-STARRS \citep{kai02}, we now describe the construction
of template $ugriz$ RR Lyrae light curves.

\subsection{Construction of $ugriz$ RR Lyrae Templates\label{construction}}

Our goal is to construct the smallest set of template $ugriz$ RR Lyrae light
curves that will model the light curves present in our sample at a level of
uncertainty set by photometric errors ($\sim0.02$ mag). To achieve this goal, we
use an iterative approach to refine the initial set of prototype templates.

First, for each $ugriz$ band we find RR Lyrae stars that have at least 50
observations in a band with photometric errors less than 0.05 mag. The light
curves of such RR Lyrae are period-folded, and a B-spline is interpolated
through the data if the largest gap in the phase coverage is smaller than 0.05.
The RR Lyrae stars with phase coverage gaps greater than 0.05 are not used in
the construction of the RR Lyrae templates. The light curves with interpolated
B-splines ($\sim$100 per band) are hereafter referred to as prototype templates,
and are normalized to have values in the $[0,1]$ range and maxima at $phase=0$.

In the next step, we fit the prototype templates to each of the 483 light curves
and qualify a fit as acceptable if $\zeta < 1.3$. Here it is possible to have
more than one acceptable template fit per light curve. The RR Lyrae stars with
acceptable fits are grouped by prototype templates, and their observed light
curves are period-folded and normalized using best-fit period, amplitude, mean
magnitude, and epoch of the maximum brightness. The normalized light curves are
grouped on a per template basis, and a B-spline is interpolated through the
grouped data. With this procedure we smooth out prototype templates by averaging
over similarly shaped light curves. In other words, the main purpose of the
prototype templates is to select self-similar subsets of observed light curves.

Since the number of smoothed prototype templates is about 100 per band, it is
possible that some of them are redundant and that they may be replaced by a
single template without decreasing the generality of the template set. To
establish which templates are redundant, we fit smoothed templates to RR Lyrae
light curves and accept template fits with $\zeta<1.3$. Again, it is possible to
have more than one acceptable template fit per light curve.

We rank the smoothed templates by the number of light curves they fit, and sort
them in descending order (the highest ranking template fits the greatest number
of light curves). Starting with the second-ranked template, we count the number
of light curves that were {\em not} fitted by the highest ranking template. We
move down the template list and for each template count only light curves that
were not already fitted by all higher ranking templates. We stop moving down the
template list when we reach a template that fits only one light curve.

The templates that survive this pruning are compared with each other, and if
there are two with maximum difference less than 0.02 (the average photometric
error), we keep the one that fits a greater number of light curves. Our final,
pruned template set contains 12 $u$-band (1 is type $c$), 23 $g$-band (2 are
type $c$), 22 $r$-band (2 are type $c$), 22 $i$-band (2 are type $c$), and 19
$z$-band templates (1 is type $c$). The templates are provided in the electronic
edition of the Journal. The light curve parameters of stripe 82 RR Lyrae stars
obtained from best-fit $ugriz$ templates are listed Table~\ref{RR_LCparams}.

\begin{figure}
\epsscale{0.44}
\plotone{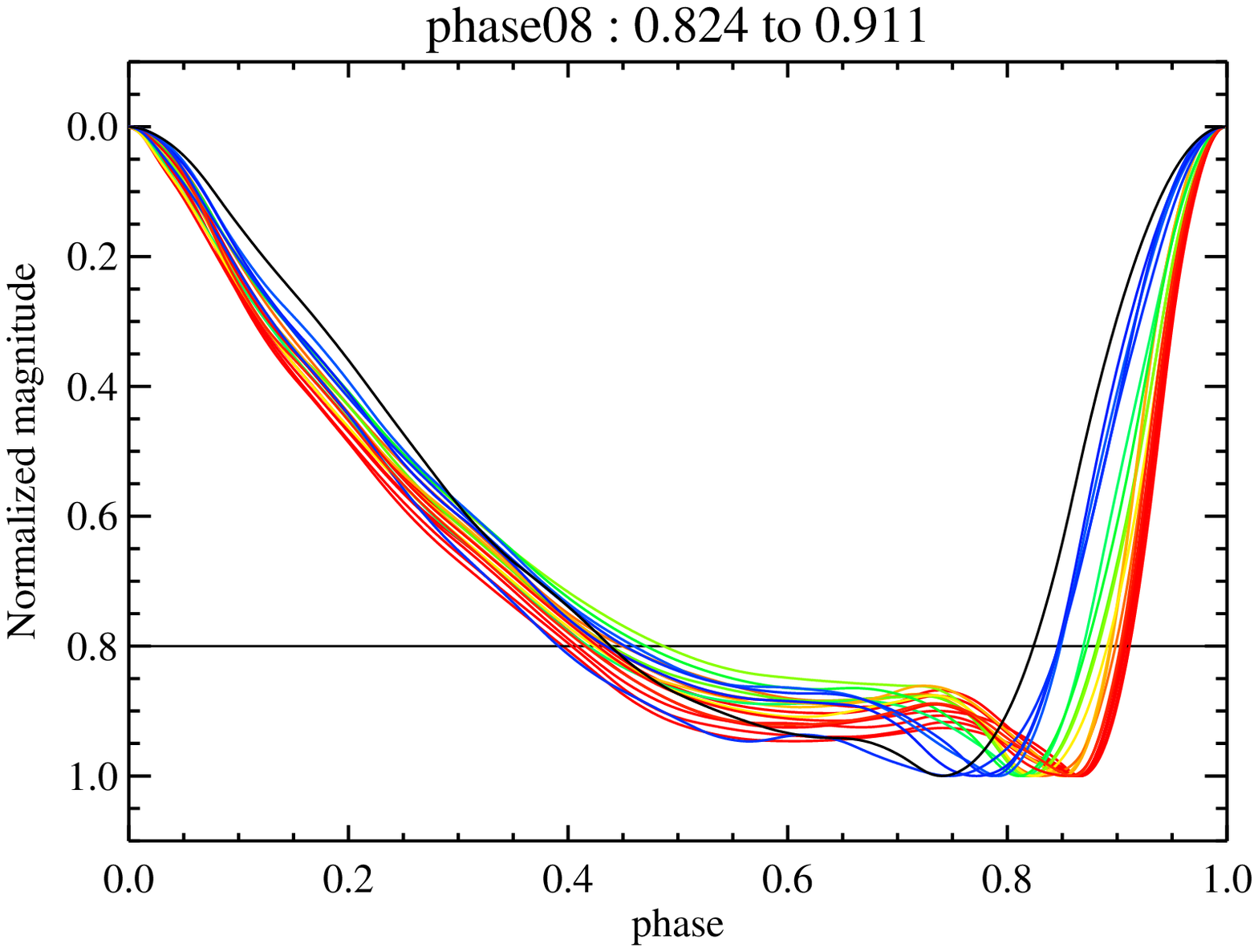}

\plotone{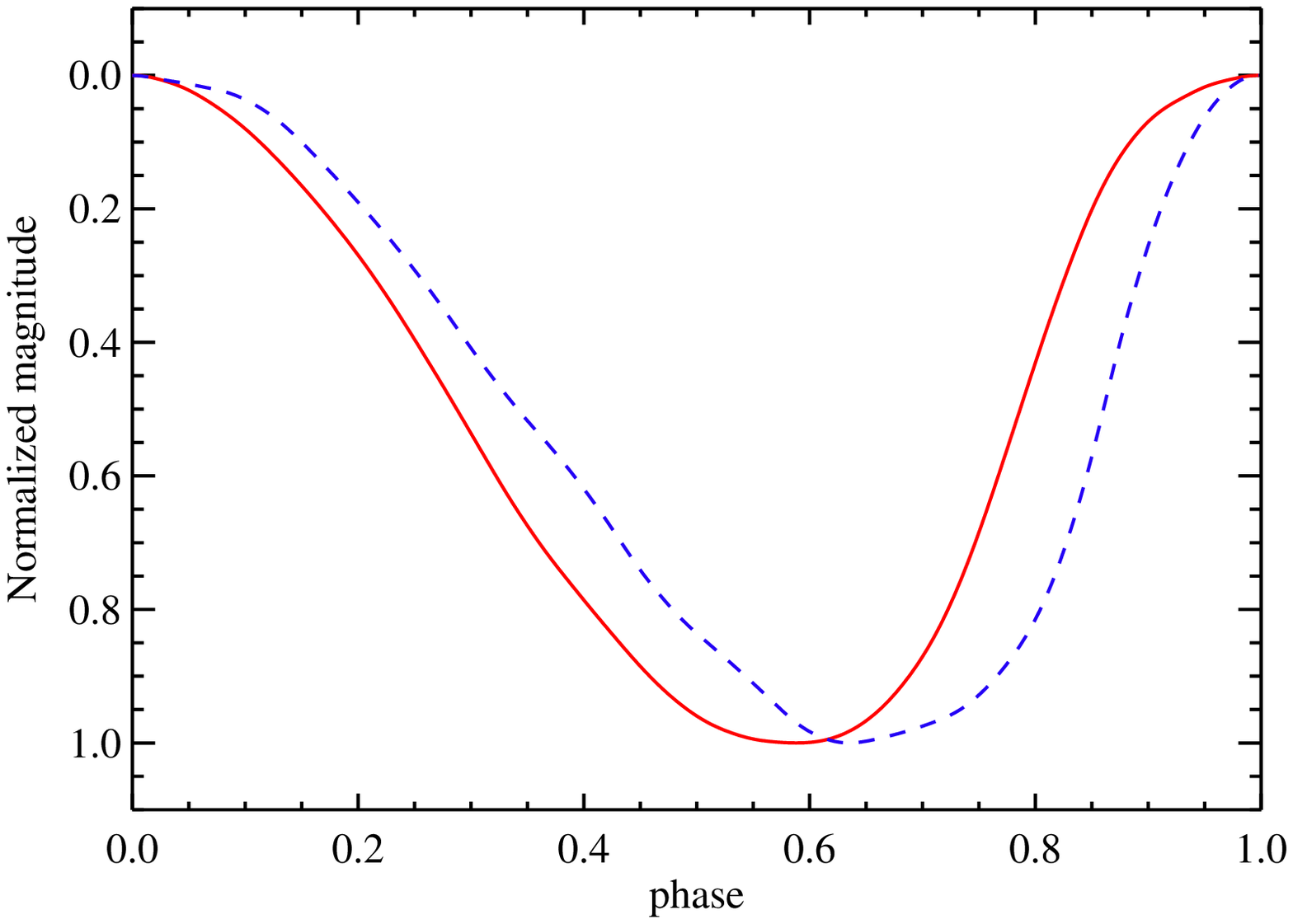}
\caption{
{\em Top}: The 22 $g$-band RR $ab$-type template light curves, color-coded from
black to red using $\phi_{0.8}^g$, defined as the phase at which a $g$-band
template value is equal to 0.8. The $\phi_{0.8}$ is similar in principle to the
rise time used by \citet{sks81} and \citet{lay98} (shorter rise time means
higher $\phi_{0.8}$).
{\em Bottom}: A comparison of two $g$ band RR $c$-type template light curves.
The template showed by the dashed line has a higher $\phi_{0.8}^g$ value than
the template showed by the solid line (0.80 vs.~0.72).
\label{template_comparisons}}
\end{figure}

As illustrated in Figure~\ref{template_comparisons}, the set of $\sim20$ $ab$
templates per band appears consistent with a continuous distribution of light
curve shapes. This essentially one-dimensional template family can be
conveniently parametrized by the phase at which the normalized template (with
unit amplitude) crosses the value of 0.8 on the rising part of light curve
(hereafter $\phi_{0.8}$). This parameter measures the rise time, as a fraction
of the period, and is related to other parametrizations of the light curve
shape, such as the coefficients of a Fourier expansion (see
\S~\ref{LCmetallicity}).

On the other hand, $c$-type RR Lyrae stars exhibit a strong bimodal distribution
of light curve shapes in the $gri$ bands, suggesting that this class may include
two distinct subclasses, with each subclass having very small scatter of light
curve shapes. This behavior is further discussed in \S~\ref{P-A distribution}.

\begin{figure}
\epsscale{1.0}
\plotone{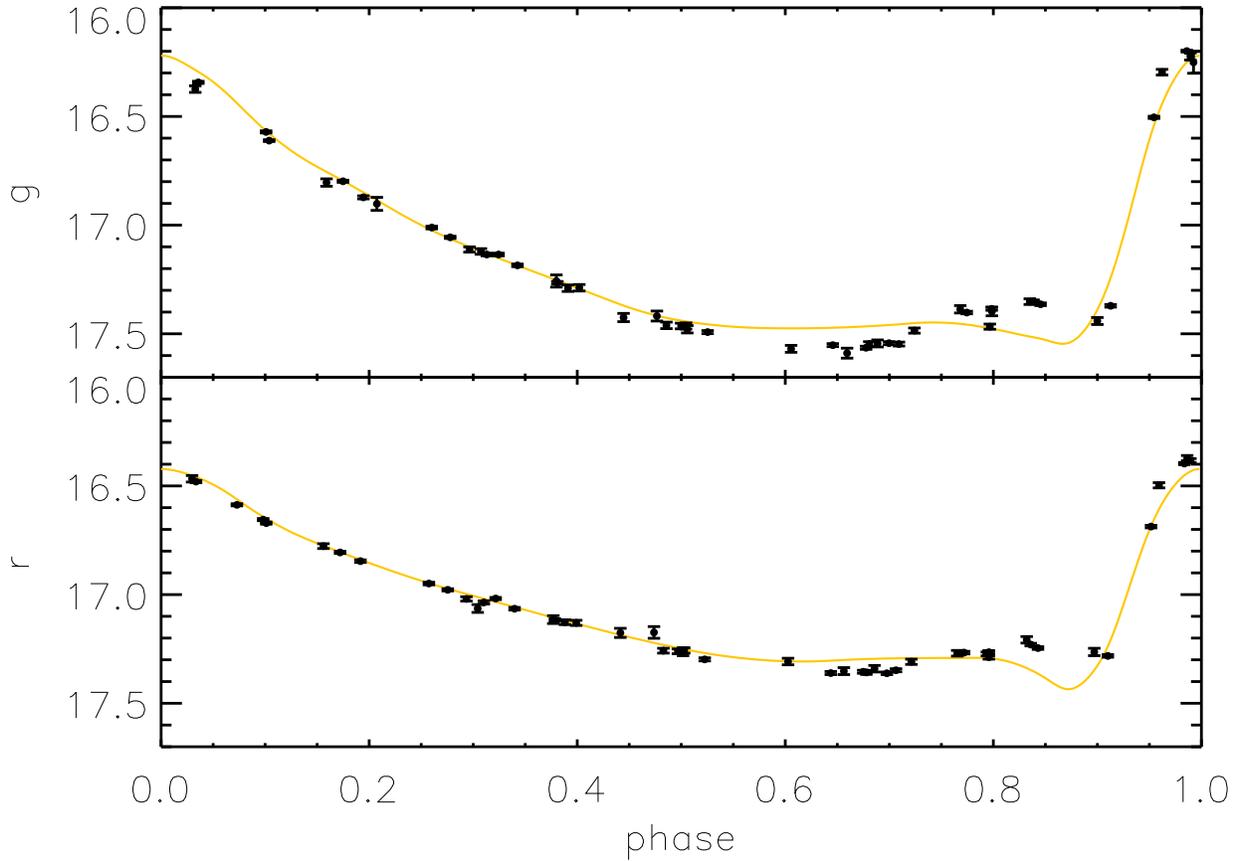}
\caption{
An example of incompleteness in our RR Lyrae light curve template set. The
period-folded $g$- and $r$-band light curves are shown with symbols, and the
best-fit templates are shown with solid lines. The fact that even the best-fit
templates fail to model these light curves at minimum brightness
($0.5<phase<0.9$) suggests that our light curve template set is not entirely
complete.
\label{incomplete_set}}
\end{figure}

We also note that our template set is not entirely complete, as illustrated in
Figure~\ref{incomplete_set}. There are 15 examples of stars (3\%) for which the
current template does not describe their (unique) light curves at the level of
the photometric uncertainty (0.02 mag).

\subsection{A Comparison With Theoretical Template Light Curves\label{comparison_w_theory}}

\begin{figure}
\epsscale{0.9}
\plotone{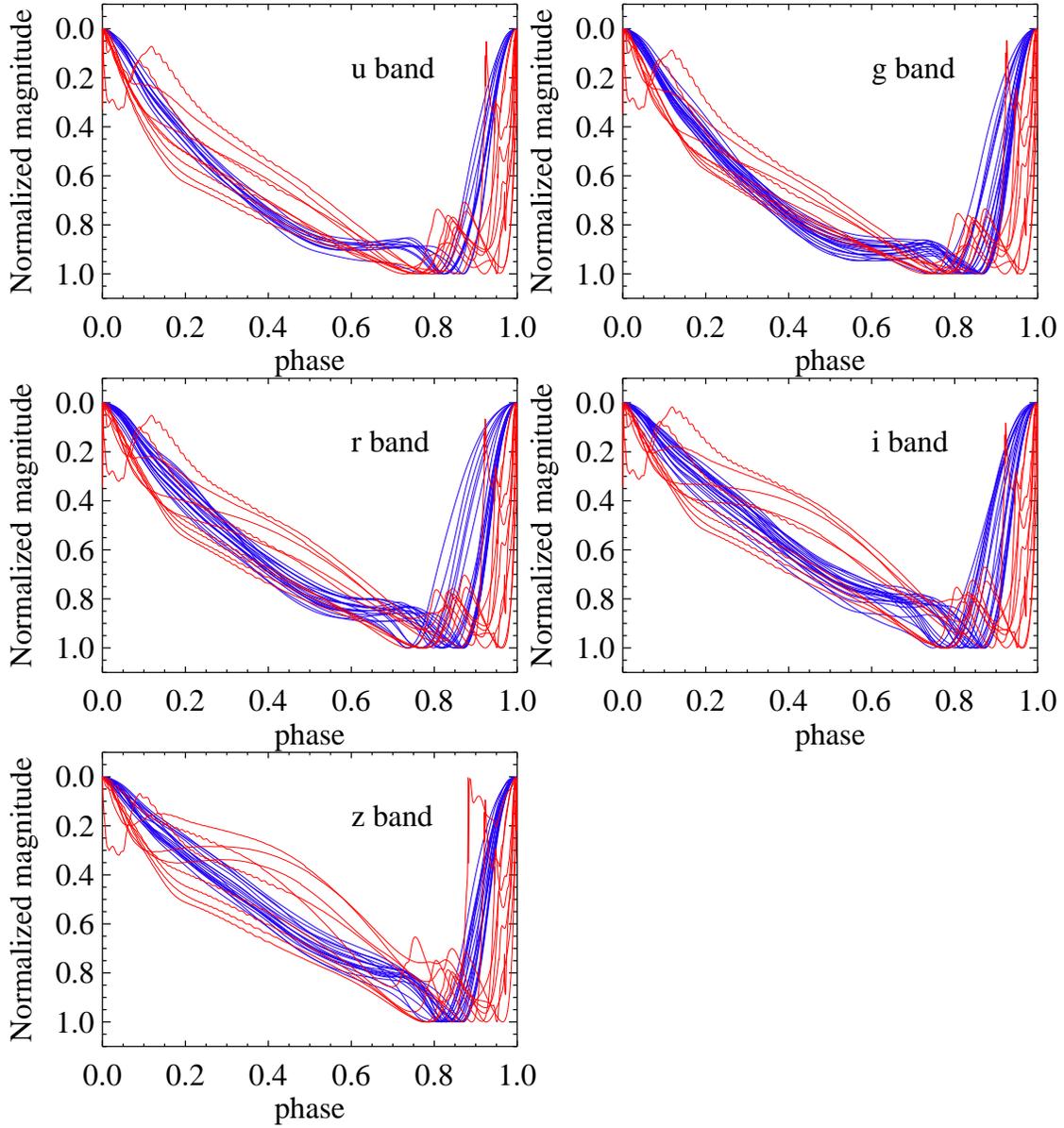}
\caption{
A comparison of empirical ({\em blue lines}) and theoretical ({\em red lines})
RR $ab$-type template light curves in SDSS $ugriz$ band-passes. The theoretical
\citet{mar06} templates were obtained from a RR Lyrae pulsational model with
metallicity $Z=0.001$, luminosity $L=1.61L_\sun$, mass $M = 0.75 M_\sun$, and
effective temperature $T_{eff}=6100-6900$ K. Note how the light curve shape of
empirical and theoretical templates changes significantly between $griz$ bands,
becoming less convex (less curved upward) on the descending brightness branch
(phase$<0.7$) as the band-pass moves to longer wavelengths. The shape does not
change significantly between $u$ and $g$ bands.
\label{griz_template_comparisons}}
\end{figure}

In Figure~\ref{griz_template_comparisons}, we compare empirical RRab templates
to theoretical templates from \citet{mar06}. As the band-pass moves to longer
wavelengths (from the $g$ to $z$ band), the light curve shape of both empirical
and theoretical templates becomes less convex (less curved upward) on the
descending brightness branch (phase$<0.7$). The light curve shape does not
change significantly between $u$ and $g$ bands in neither empirical or
theoretical templates.

\begin{figure}
\epsscale{1.0}
\plotone{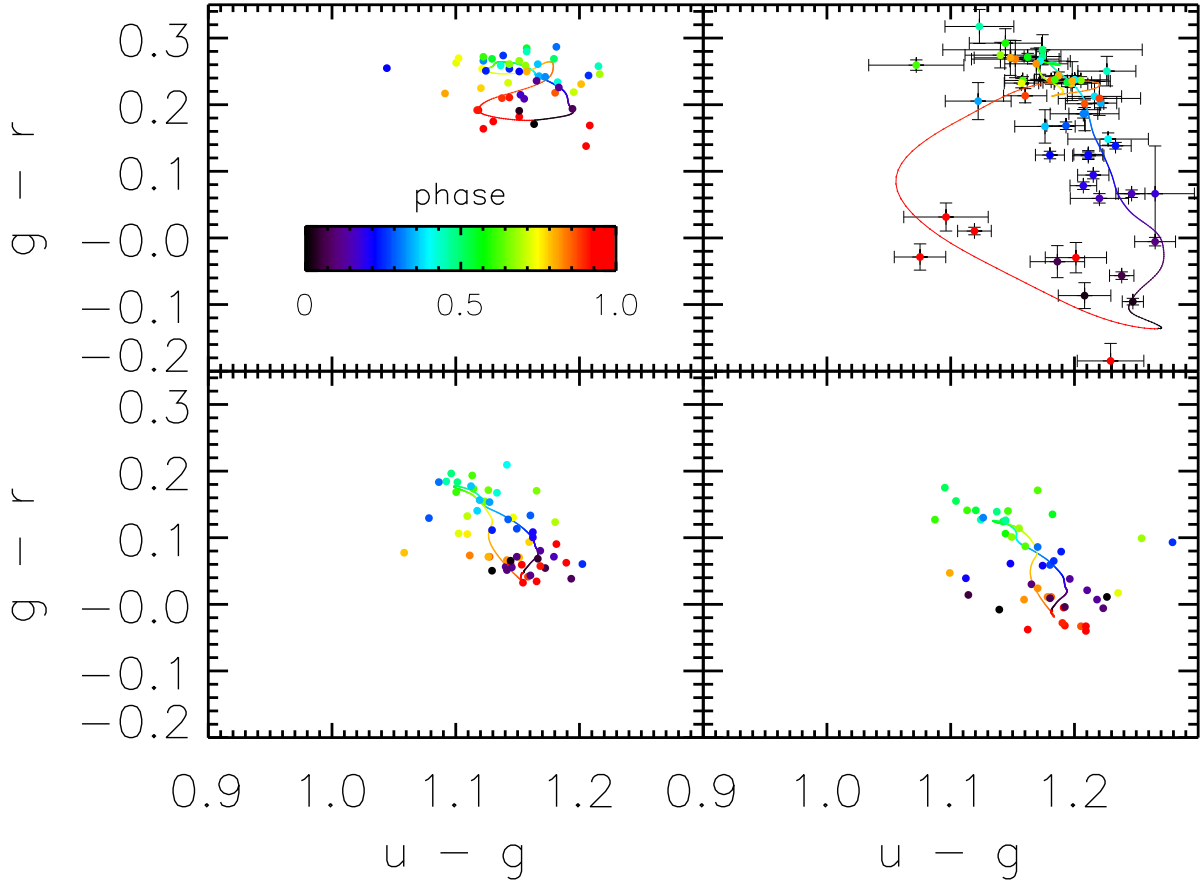}
\caption{
The $u-g$ vs.~$g-r$ color-color loops of four RR Lyrae stars. The top left panel
shows a long-period, low-amplitude type $b$ RR Lyrae stars (ID 276162), the
top right panel shows a short-period, large-amplitude type $a$ RR Lyrae star
(ID 293282), and the bottom panels show two type $c$ RR Lyrae stars (IDs
376465 and 429508), with the one on the left having a longer period and a
lower amplitude than the one on the right. The observed ({\em symbols with
error bars}) and predicted $u-g$ and $g-r$ colors (from the best-fit $ugr$-band
templates, {\em lines}) are color-coded according to the phase of pulsation. The
error bars are not shown in three panels to avoid cluttering. The amplitude in
the $g-r$ color increases with the $g$-band amplitude ($A_{g-r}\sim0.4A_g$,
since $A_g\sim1.4A_r$ and $A_{g-r}\sim A_g-A_r$). The color-color loops for the
rest of the RR Lyrae stars are provided in the electronic edition of the
Journal.
\label{uggr_color_loops}}
\end{figure}

On the other hand, the theoretical templates have a peak after the minimum
brightness that is not observed in our templates. The differences between our
empirical templates and theoretical templates are much larger than the errors in
our templates. The new pulsation models will undoubtedly benefit from
constraints provided by the observational template set derived here. Additional
constraints may also be obtained from comparisons of theoretical (Figure 7 in
\citealt{mar06}) and observed color-color loops (Figure~\ref{uggr_color_loops}
in this work).

\section{Outer Halo Substructure As Traced by Stripe 82 RR Lyrae Stars\label{sec:spatial}}

In this section the spatial distribution of RR Lyrae stars is used to trace the
substructures in the outer halo. First, we estimate the heliocentric distances
of RR Lyrae stars using their best-fit $g$- and $r$-band light curves, then
estimate the sample completeness as a function of magnitude in
Section~\ref{magnitude_completeness}. The observed spatial distribution of RR
Lyrae stars is used to compute their number density distribution, which is then
compared to a model number density distribution. This comparison reveals an
abrupt change in the number density distribution at a distance of $\sim30$ kpc,
suggesting a transition from the inner to outer halo, as well as the presence of
a prominent halo overdensity at 80 kpc.

\subsection{Distance Estimates}

The heliocentric distances, $d$, for RR Lyrae stars are estimated from
measurements of their mean dereddened $V$ magnitudes, $\langle V \rangle$, and
by assuming an average value for their absolute magnitudes, $M_V$
\begin{equation}
d = 10^{(\langle V \rangle-M_V+5)/5}
\end{equation}
For the absolute magnitude in the $V$ band we adopt $M_V = 0.6$, calculated
using the \citet{cc03} $M_V-[Fe/H]$ relation
\begin{equation}
M_V = (0.23\pm0.04)[Fe/H] + (0.93\pm0.12)\label{abs_mag}
\end{equation}
and adopting $[Fe/H]=-1.5$ as the mean halo metallicity \citep{ive08}.

The values of $\langle V \rangle$ are determined by integrating best-fit
template $V$-band light curves after transformation to flux units. The best-fit
template $V$-band light curves are synthesized from the best-fit $g$- and
$r$-band templates ($g_{LC}(\phi)$ and $r_{LC}(\phi)$) using a photometric
transformation given by Equation 10 from
\citet{ive07}:
\begin{eqnarray}
gr_{LC}(\phi) = g_{LC}(\phi)-r_{LC}(\phi) \\
V_{LC}(\phi) = g_{LC}(\phi) + 0.0688gr_{LC}(\phi)^3 -0.2056gr_{LC}(\phi)^2 -0.3838gr_{LC}(\phi) - 0.053
\end{eqnarray}
where $\phi\in[0,1\rangle$ is the phase of pulsation, and where $g_{LC}(\phi)$
and $r_{LC}(\phi)$ are corrected for interstellar extinction. We have also
established that the measured light curves are aligned in phase among the bands
to within the measurement errors (to within 1\% in phase). These
``flux-averaged'' magnitudes are used because pulsation models show that they
better reproduce the behavior of ``static'' magnitudes (the magnitudes the stars
would have if they were not pulsating, \citealt{mar06}). The uncertainty in
$\langle V\rangle$ is mainly due to photometric errors in the $g$ and $r$ bands,
and is estimated to be $\sim0.03$ mag.

Following the discussion by \citet{vz06} (see their Section 4), we adopt
$\sigma_{M_V}=\sqrt{(\sigma_{M_V}^{[Fe/H]})^2+(\Delta M_V^{ev})^2}=0.1$ mag as
the uncertainty in absolute magnitude, where $\sigma_{M_V}^{[Fe/H]}=0.07$ mag is
the uncertainty due to metallicity dispersion ($\sigma_{[Fe/H]}=0.3$ dex,
\citealt{ive08}), and $\Delta M_V^{ev}=0.08$ mag is the uncertainty due to RR
Lyrae evolution off the zero-age horizontal branch \citep{vz06}. The estimated
fractional error in the heliocentric distance ($d/\sigma_d$) of 0.06 is smaller,
by a factor of $\sim2-4$, than the fractional distance error for K giants
\citep{dp01}, M giants \citep{maj03} or main sequence stars \citep{new02,jur08}.
The heliocentric distances and equatorial coordinates of stripe 82 RR Lyrae
stars are listed in Table~\ref{RR_derived_params}.

\subsection{Sample Completeness as a Function of Magnitude\label{magnitude_completeness}}

The fraction of recovered RR Lyrae stars (the completeness) as a function of
apparent magnitude is an important question that needs to be considered before
the spatial distribution of RR Lyrae stars can be discussed. We generate 1000
RRab synthetic $g$-band light curves with 30 data points, amplitudes of 1.0 mag,
periods of 0.6 days, and with flux-averaged magnitudes of
$\langle g \rangle=21.6$ (not corrected for ISM extinction). An RR Lyrae star
with $\langle g \rangle=21.6$ would be located at $\sim120$ kpc, assuming
$M_g=0.6$ and 0.6 mag of ISM extinction in the $g$ band (the highest observed
extinction in stripe 82). Typical random photometric errors were added as a
function of magnitude using Figure 1 from S07. The periods for the synthetic
light curves were estimated using {\em Supersmoother}, and the amplitudes and
mean magnitudes were estimated from best-fit templates.

We find that the periods are recovered with 0.05\% (or better) accuracy, and
flux-averaged magnitudes with 0.03 mag (or better) accuracy. Similar results are
obtained for RRc stars (the amplitude of synthetic light curve is 0.5 mag in
that case). If we define an RR Lyrae as successfully recovered if its
flux-averaged magnitude is estimated with 0.03 mag accuracy, then the stripe 82
RR Lyrae sample is essentially 100\% complete to 120 kpc, even in regions where
the extinction is as high as 0.6 mag ($R.A. > 50\arcdeg$ and $R.A.<308\arcdeg$).

Even though both RRab and RRc types have the same completeness as a function of
apparent magnitude, we only use RRab types when studying the spatial
distribution of RR Lyrae stars, because they represent a homogeneous and robust
sample with a well-studied absolute magnitude vs.~metallicity distribution. A
repeat of the analysis presented below, but with RRc stars included, did not
change our conclusions based exclusively on RRab stars.

\subsection{Observed Spatial Distribution\label{sec:spatDistr}}

Since SDSS stripe 82 is quite narrow in declination ($2.5\arcdeg$ compared to
$118\arcdeg$ in R.A.), we assume $Dec = 0$ for all stars, and describe their
positions in the R.A.$-$heliocentric distance space. The Cartesian coordinates
of stars in this space are defined as
\begin{eqnarray}
x = d\cos(\alpha) \\
y = d\sin(\alpha), \label{coordinates}
\end{eqnarray}
where $d$ is the heliocentric distance in kpc, and $\alpha$ is the right
ascension angle.

\begin{figure}
\epsscale{0.84}
\plotone{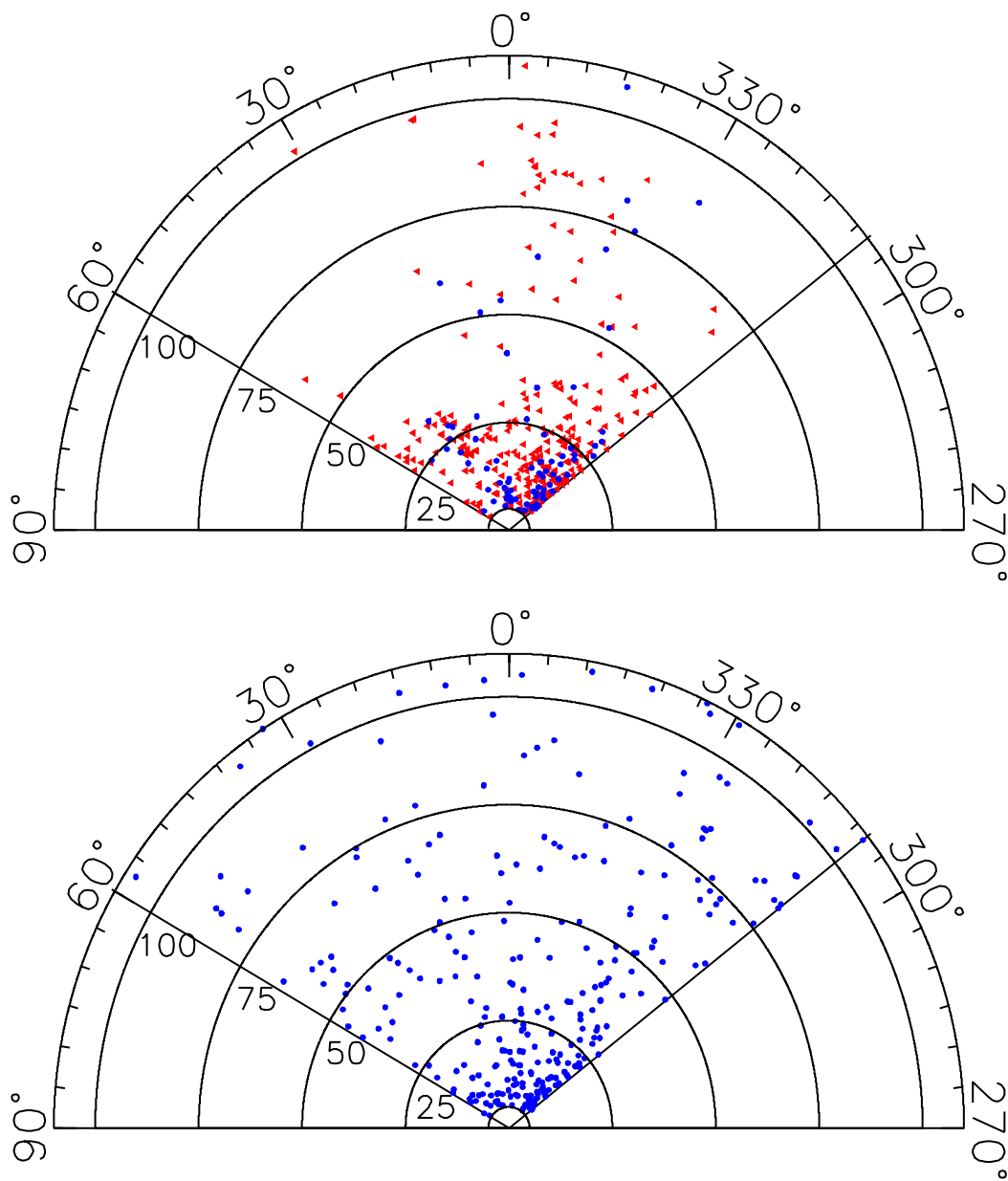}
\caption{
The top panel shows the spatial distribution of 366 RRab stars from stripe 82,
while the bottom panel shows 302 stars randomly drawn from a smooth model
distribution (Eq.~\ref{RR_model}). Both samples of stars have heliocentric
distances between 5 and 120 kpc and $|Dec|<1.25\arcdeg$. The radial axis is the
heliocentric distance in kpc and the angle is the equatorial right ascension.
The circles correspond to $\langle V \rangle$ of 16.3, 17.8, 18.7, and 19.3 mag
(corrected for ISM extinction). The RRab stars shown in the top panel are
divided into Oosterhof I (289 stars, {\em red triangles}) and Oosterhof II
(77 stars, {\em blue dots}) using the selection boundary shown in
Fig.~\ref{gamp_vs_period_colored}.
\label{raw_distribution}}
\end{figure}

The spatial distribution of 366 RRab stars in the R.A.$-$heliocentric distance
space is shown in Figure~\ref{raw_distribution}. For comparison, we also show
the distribution of RR Lyrae stars randomly drawn from a smooth model number
density distribution, based on an analysis of main sequence stars within 20 kpc
by \citet{jur08}
\begin{equation}
\rho^{RR}_{model}(R,Z) = \rho_\sun^{RR}[\frac{R_\sun}{\sqrt{R^2+(Z/q_H)^2}}]^{n_H}\label{RR_model},
\end{equation}
where $\rho_\sun^{RR}=4.2$ kpc$^{-3}$ is the number density of RRab stars at
$R_\sun=8.0$ kpc \citep{vz06}, $q_H=0.64$ indicates the halo is oblate
(``squashed'' in the same sense as the disk), and $n_H=2.77$ is the power-law
index \citep{jur08}. The expected number of RR Lyrae stars (according to
$\rho^{RR}_{model}$) in the 5 to 120 kpc range of stripe 82 is 302. The value of
$\rho_\sun^{RR}$ is within $\pm0.5$ kpc$^{-3}$ from values reported by other
authors \citep{pre91,skk91,wg96}, and the uncertainty in $n_H$ and $q_H$ is
$\la0.2$ \citep{jur08}. We also repeated the following analysis using a
spherical, $1/R^3$ halo model and obtained similar results and conclusions
(i.e.~detection of significant overdensities and the steepening of the halo
density profile).

The two distributions shown in Figure~\ref{raw_distribution} appear remarkably
different. The observed distribution, shown in the top panel, has two overdense
regions within 30 kpc; the one at $R.A.\sim30\arcdeg$ can be identified as the
trailing arm of the Sagittarius dwarf spheroidal (dSph) tidal stream
\citep{ive03}, while the other at $R.A\sim330\arcdeg$ is likely associated with
the Hercules-Aquila Cloud \citep{bel07b}. Beyond 30 kpc, the observed
distribution is quite inhomogeneous in the R.A.~direction, with the majority of
RR Lyrae in the $330\arcdeg$ to $0\arcdeg$ region, and only a few RR Lyrae stars
outside of it. On the other hand, the distribution of stars generated using
Equation~\ref{RR_model} looks more homogeneous, as expected.

\subsection{Bayesian Number Density Maps\label{number_density}}

We compute the local number density of RR Lyrae stars using a Bayesian estimator
developed by \citet[see their Appendix B2 and Equation B6]{ive05}. The Bayesian
metric presented there can be recast in a much simpler form
\begin{equation}
\rho_{Bayes} = C \frac{N_{nn}^k} {N_{stars}} \frac{1}{\sum_{i=1}^{N_{nn}}r_i^k},\label{bayes_simple}
\end{equation}
where $N_{nn}$ is the number of nearest neighbors to which the distance $r$ in a
$k$-dimensional space is calculated, and $N_{stars}$ is the number of stars. The
dimensionless constant $C$ involves complex integrals, but can be easily
determined empirically once the geometry of the space in which densities need to
be calculated is known.

To estimate the constant $C$ we use the following procedure. We populate the
stripe 82 region with $N_{stars}=302$ stars randomly drawn from the smooth model
number density distribution (Equation~\ref{RR_model}), and calculate the
Bayesian metric on a 0.5x0.5 $x-y$ grid assuming $N_{nn}=8$, $k=2$, and $C=1$.
The nearest neighbors are found using an implementation of the nearest neighbor
searching algorithm \citep{ary98}. The number density is then calculated as
\begin{equation}
\rho(x,y) = \rho_{Bayes}(x,y)/d(x,y),\label{rhoBayes}
\end{equation}
where the Bayesian metric is divided by the heliocentric distance to account for
the increase in volume of a grid point at greater heliocentric distances. We
generate 1000 $\rho(x,y)$ density maps, and average them out to obtain
$\rho_{avg}(x,y)=\sum\rho(x,y)/1000$. The constant $C=1030$ is estimated by
requiring the median value of $\rho_{avg}(x,y)$ to equal the median value of
$\rho^{RR}_{model}(x,y)$.

The sparseness of our sample (366 RR Lyrae in a volume of 38,000 kpc$^{3}$)
will result in random (Poisson) number density fluctuations even for a smooth
underlying density distribution. To quantify the scatter in number density maps
and to estimate the fraction of highly overdense and underdense pixels
($|\log(\rho/\rho^{RR}_{model})|>1$) due to Poisson fluctuations, we generate
1000 random RR Lyrae star samples of the same size as the observed sample by
drawing them from the smooth model, and compute $\log(\rho/\rho^{RR}_{model})$
maps for each sample. We find that the median rms scatter in smooth model-based
maps is 0.3 dex (or about a factor of two on the linear scale), and that the
fraction of highly overdense and underdense pixels
($|\log(\rho/\rho^{RR}_{model})|>1$) is only $10^{-6}$. We conclude that the
precision of the Bayesian density estimator, given the sparseness of our sample,
is about 0.3 dex, and that highly overdense and underdense pixels are not likely
to be caused by Poisson fluctuations.

\begin{figure}
\epsscale{0.67}
\plotone{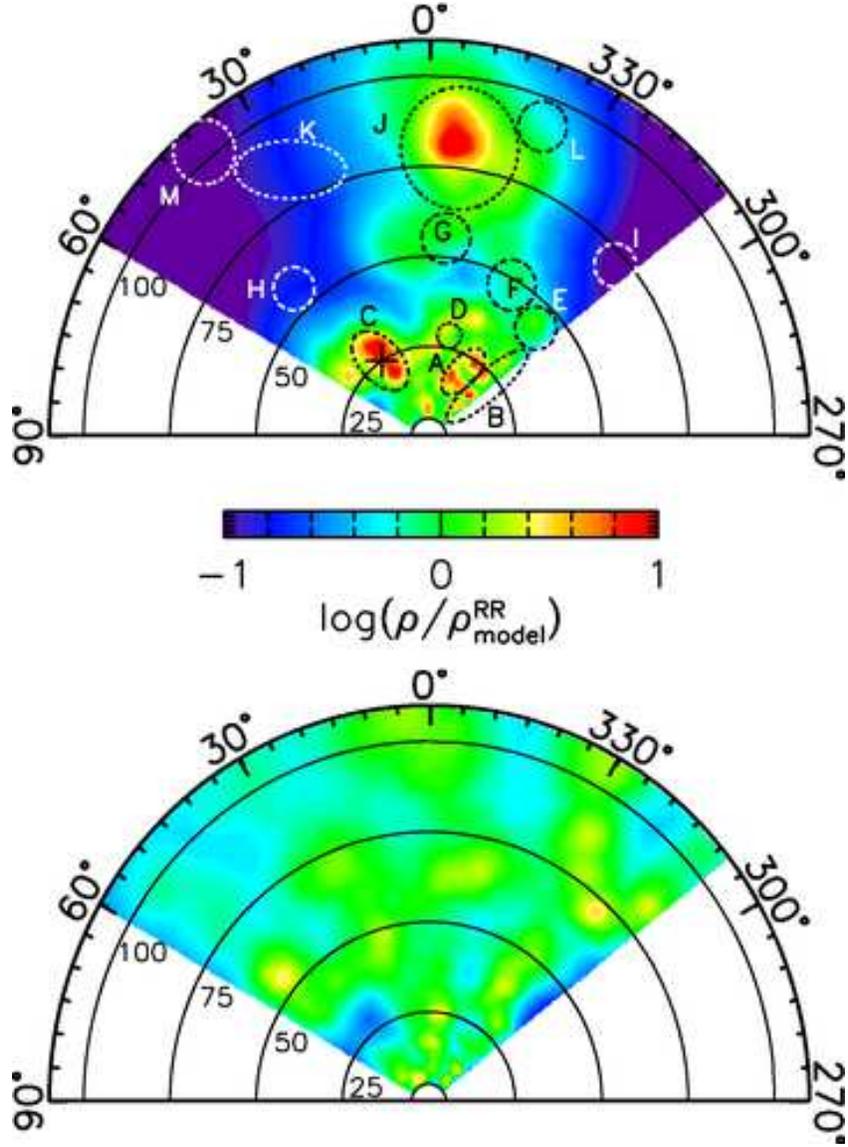}
\caption{
The number density distribution of observed ({\em top}) and smooth model-based
({\em bottom}) RR Lyrae stars from Fig.~\ref{raw_distribution}, computed using
Eq.~\ref{rhoBayes}, and shown relative to the smooth model number density
($\rho^{RR}_{model}$). The $\log(\rho/\rho^{RR}_{model})$ values are color-coded
according to the legend, with values outside the legend's range saturating in
purple and red, respectively. The stripe 82 plane intersects the Sagittarius
dSph tidal stream (trailing arm) at ($\sim25$ kpc, $\sim30\arcdeg$), the
Hercules-Aquila cloud at ($\la25$ kpc, $\sim330\arcdeg$), and the Pisces stream
at ($\sim80$ kpc, $\sim355\arcdeg$). The overdensities detected in S07 are
overplotted and labeled in the top panel. The ``+'' symbol marks the position
where the \citet{ljm05} ``spherical'' model of the Sagittarius dSph tidal stream
(trailing arm) intersects the stripe 82 plane.
\label{density_plots}}
\end{figure}

With the Bayesian density estimator now fully defined, we compute the number
density $\rho(x,y)$ on a $x-y$ grid for the observed distribution of RR Lyrae
stars and for the sample of stars randomly drawn from the smooth model
distribution. The computed number density maps are divided by
$\rho^{RR}_{model}(x,y)$; the results are shown in Figure~\ref{density_plots}.

Figure~\ref{density_plots} ({\em top}) vividly shows the striking difference
between the observed number density distribution of RR Lyrae stars and the
smooth model number density distribution (Equation~\ref{RR_model}). While the
density fluctuations ($\log(\rho/\rho^{RR}_{model})$) in the model-based map
(Figure~\ref{density_plots} [{\em bottom}]) are consistent with Poisson noise
(amplitudes are $<0.3$ dex), the observed ({\em top}) map has several overdense
and underdense regions (by a factor of 10 or more relative to
$\rho^{RR}_{model}$). The Sagittarius dSph tidal stream (trailing arm) is
visible at $\sim25$ kpc and $30\arcdeg$, the Hercules-Aquila cloud is located at
$\la25$ kpc and $\sim330\arcdeg$, and a previously detected, and now confirmed,
structure can be seen at $\sim80$ kpc and $\sim355\arcdeg$ (the ``J''
overdensity in S07 Figure 13). The fraction of pixels with
$|\log(\rho/\rho^{RR}_{model})|>1$ in the observed number density map is about
0.3 (dominated by underdense pixels), or 5 orders of magnitude higher than
expected for Poisson fluctuations. 

Figure~\ref{density_plots} ({\em top}) also shows that some overdensities
detected by S07 are not reproduced in this work. These overdensities were
caused by non-RR Lyrae stars in the S07 candidate RR Lyrae sample
(Table~\ref{S07_clumps} lists the number of RR Lyrae and non-RR Lyrae
stars in each S07 overdensity). About $70\%$ of S07 RR Lyrae candidates are true
RR Lyrae stars, with the re-classified non-RR Lyrae stars dominated by $\delta$
Scuti stars (based on their light curves and periods). These results are in good
agreement with S07 estimates. The overdensities labeled D, F, H, I, K, and L now
have lower numbers of true RR Lyrae stars compared to S07, and are no longer
significant overdensities. These overdensities were also speculated in S07 to be
false detections. The significance of the E, G, and M overdensities decreased as
the number of true RR Lyrae stars decreased compared to S07, and no longer seem
to be real substructures.

The extension of the J overdensity in S07 in the radial direction due to the
uncertainty in distance (the ``finger of God'' effect) is now smaller than in
S07. In S07, the apparent $V$-band magnitudes of RR Lyrae stars were computed
using Equation 3 from \citet{ive05} and using single-epoch measurements:
\begin{equation}
V_{SDSS}^{RRLyr} = r - 2.06(g-r) + 0.355\label{Vzi}.
\end{equation}
While this approximation works fairly well for RRab stars (the rms scatter
around the fit is $\sim0.1$ mag), it produces an offset of $\sim0.3$ mag when
applied to RRc stars, due to a 0.1 mag offset in their $g-r$ colors (RRc stars
have bluer $g-r$ colors than RRab stars). The higher fraction of RRab than RRc
stars in the \citet{ive05} RR Lyrae sample may explain the resulting bias in
Equation~\ref{Vzi} (\citealt{ive05} did not fit Equation~\ref{Vzi} separately
for RRab and RRc stars). In this work we use flux-averaged $V$-band apparent
magnitudes and obtain more precise distance estimates for both RRab and RRc
stars.

The structure at $\sim80$ kpc and $\sim355\arcdeg$ (labeled J in S07) is
most probably a tidal stream, although the possibility that it is a dSph
satellite galaxy cannot be excluded with the data presented here. It is unlikely
that this structure is associated with the Sagittarius dSph stream because it is
$\sim22$ kpc away from the best-fit orbital plane defined by \citet{maj03}. For
comparison, the Sgr dSph trailing tidal stream and the stream detected by
\citet{new03} are both within 2 kpc of the Sgr dSph plane. \citet{wat09} point
out that the location of the J overdensity is close to the Magellanic Plane,
though at a larger galactocentric distance than the Clouds. They also proposed
to name the J overdensity as ``the Pisces overdensity'', following the tradition
of naming stellar streams and structures according to the nearest constellation. We adopt their proposal hereafter.
 
\begin{figure}
\epsscale{1.0}
\plotone{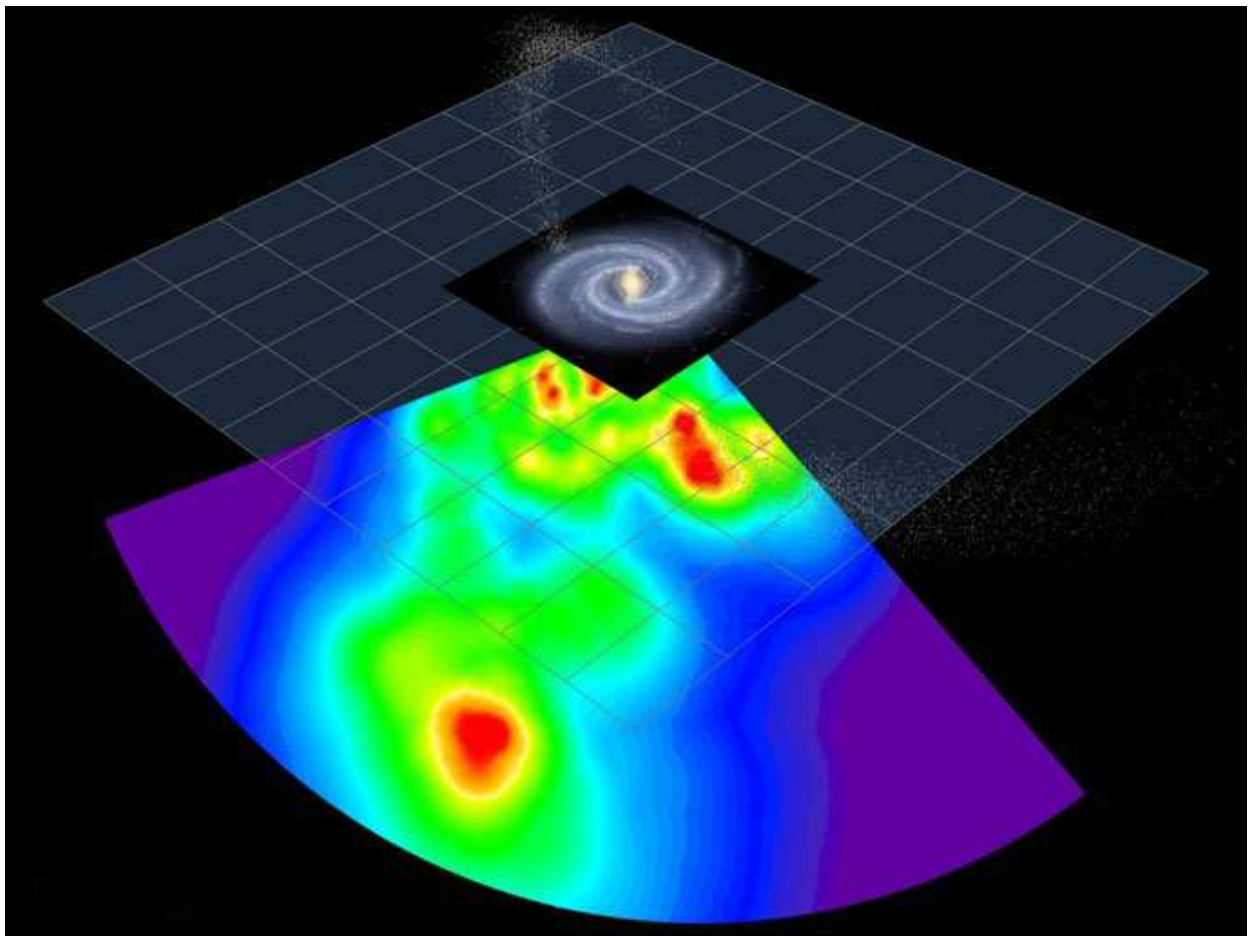}
\caption{
A single frame of an animation showing a fly-by of the stripe 82 plane and of
the scaled Galactic plane (annotated artist's concept by NASA/JPL-Caltech). The
white dots show the Sagittarius dSph and its tidal streams, as modeled by the
\citet{ljm05} ``spherical'' model. The animation is provided in the electronic
edition of the Journal.
\label{snapshot}}
\end{figure}

The width of the Pisces overdensity/stream in the radial direction ($\sim4$ kpc)
is consistent with the scatter due to uncertainty in distance, while the scatter
in the R.A.~direction ($\sim4$ kpc) might be a projection effect (a combination
of the stream's incidence angle and thickness of the stripe 82 plane), or
evidence of the stream's width in the R.A.~direction. To illustrate the scale
and position of this and other detected structures, we provide an animation
showing a ``fly-by'' of the stripe 82 region. The animation is provided in the
electronic edition of the Journal; a single frame is shown as
Figure~\ref{snapshot}.

\subsection{The Steepening of the Halo Density Profile\label{sec:steep}}

\begin{figure}
\epsscale{0.85}
\plotone{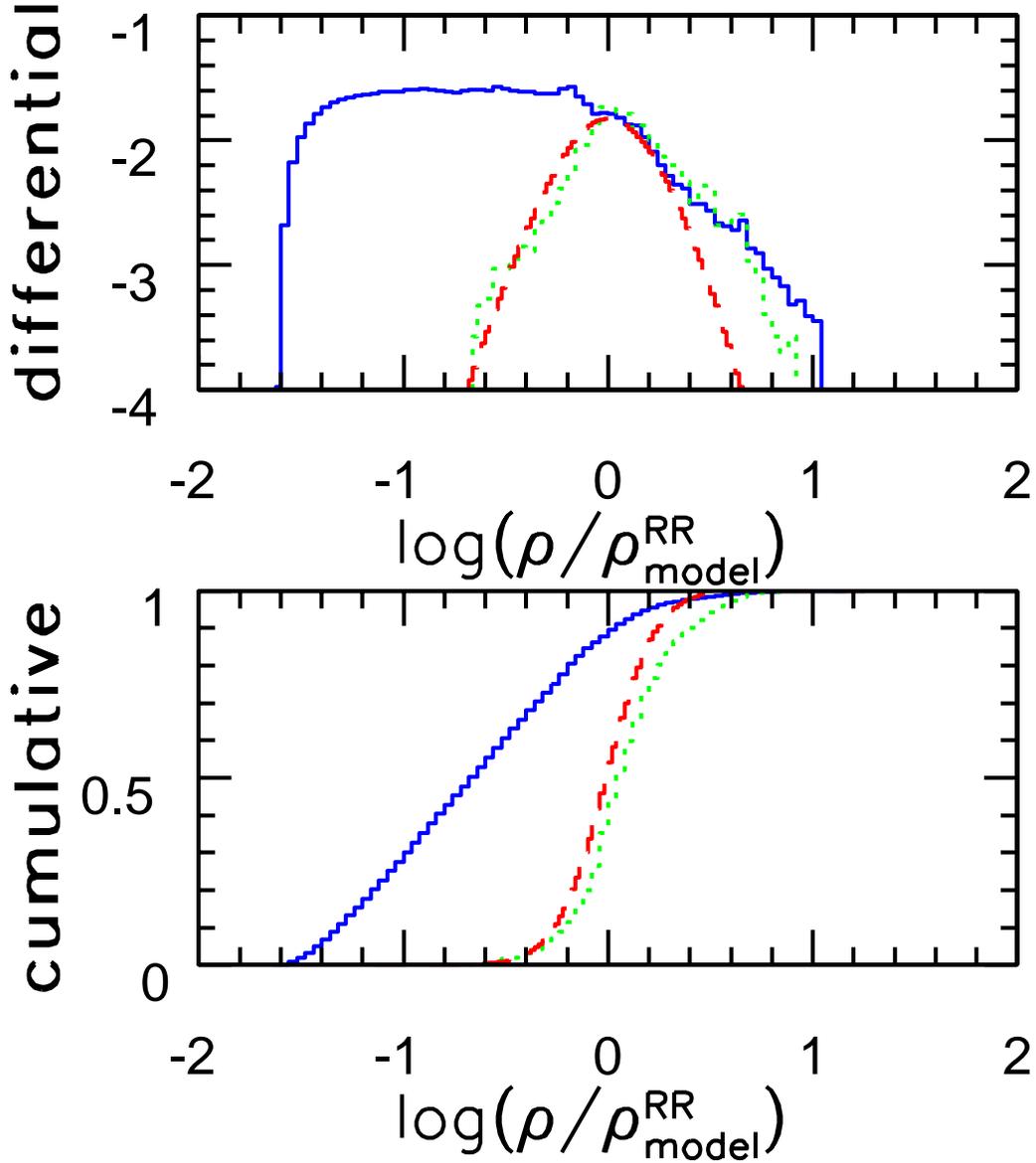}
\caption{
The differential ({\em top}, log y-axis) and cumulative ({\em bottom})
distribution of $\log(\rho/\rho^{RR}_{model})$ in model-based ({\em dashed
line}) and observed number density maps ({\em solid line}, within 110 kpc;
{\em dotted line}, within 30 kpc). The slope of the observed
$\log(\rho/\rho^{RR}_{model})$ distribution in the
$0.3<\log(\rho/\rho^{RR}_{model})<1.0$ range is $-1.65\pm0.05$ within 110 kpc,
and $-1.91\pm0.09$ within 30 kpc. Note significantly higher fractions of
strongly overdense and underdense regions than predicted by a smooth halo model.
\label{ratio_one}}
\end{figure}

The observed number density map (Figure~\ref{density_plots} [{\em top}])
suggests that the smooth model (Equation~\ref{RR_model}) systematically
overestimates the observed number densities at galactocentric distances
$R_{GC}>30$ kpc. We investigate this discrepancy in Figure~\ref{ratio_one}
({\em top}) by comparing the observed $\log(\rho/\rho^{RR}_{model})$
distributions within $R_{GC}<30$ and $R_{GC}<110$ kpc with the distribution
obtained from model-based number density maps. After a 33\% increase in the
model normalization ($\rho^{RR}_\sun$ from Equation~\ref{RR_model}), the
observed and model-based distributions match well for $R_{GC}<30$ kpc. About
80\% of the halo volume within $R_{GC}<30$ kpc is consistent with the smooth
model (within Poisson noise limits, $|\log(\rho/\rho^{RR}_{model})|<0.3$); about
20\% of the halo volume is found in predominantly overdense, substructures. The
overdense tail of the distribution is more pronounced beyond 30 kpc. At these
larger radii, the cumulative distribution (Figure~\ref{ratio_one}
[{\em bottom}]) indicates that most of the probed halo volume is underdense,
relative to the adopted smooth model (which is extrapolated from observations of
main sequence stars within $\sim$20 kpc).

\begin{figure}
\epsscale{1.0}
\plotone{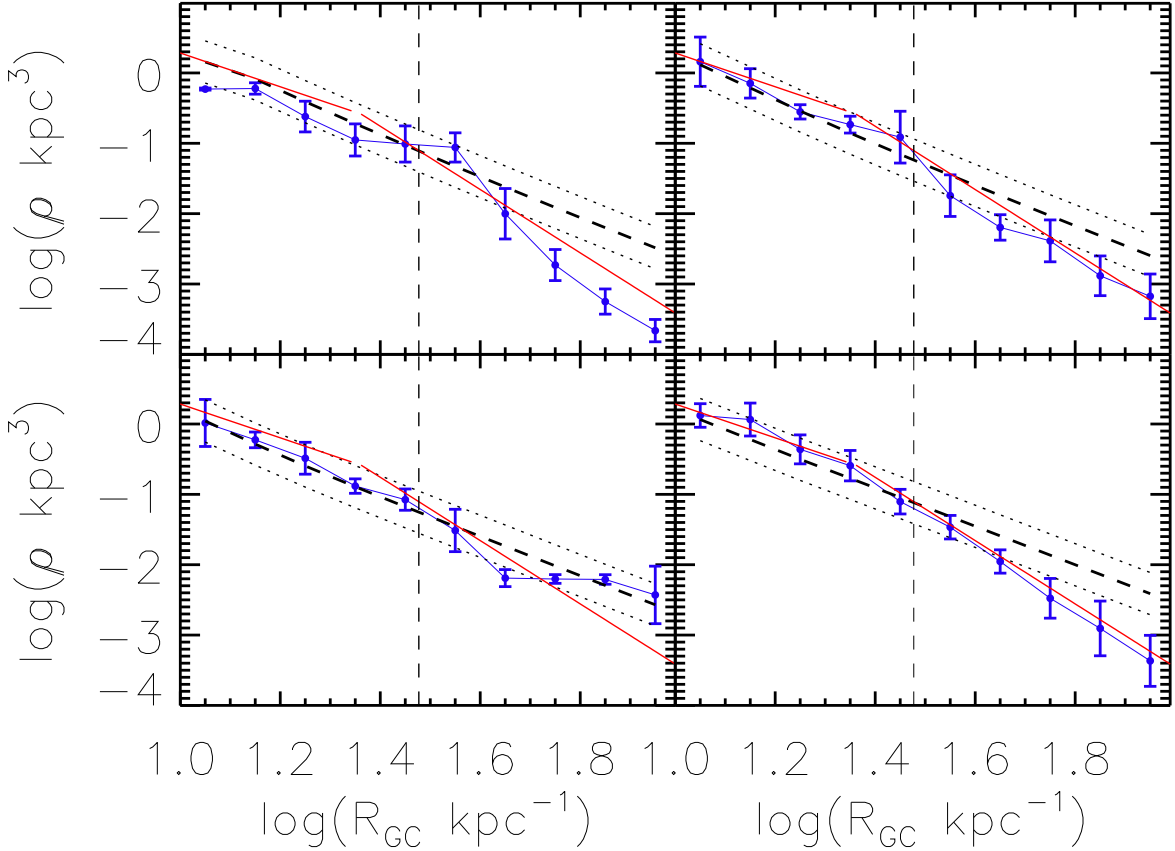}
\caption{
The radial number density profiles for four regions in R.A.~:
$30\arcdeg<R.A.<60\arcdeg$ ({\em top left}), $0\arcdeg<R.A.<30\arcdeg$
({\em top right}), $340\arcdeg<R.A.<0\arcdeg$ ({\em bottom left}), and 
$310\arcdeg<R.A.<340\arcdeg$ ({\em bottom right}). The bottom left panel
contains the Pisces overdensity at $R_{GC}\sim80$ kpc ($\log(R_{GC})\sim1.9$).
The symbols show the observed median number density and the rms scatter around
the median value for pixels shown in Figure~\ref{density_plots} ({\em top}). The
dashed line shows the number density predicted from the smooth model
($\log(\rho^{RR}_{model})$,Eq.~\ref{RR_model}), and the dotted lines show the
$\pm0.3$ dex envelope (a factor of 2) around the predicted values which
corresponds to the sample Poisson noise determined from model-based number
density maps. The solid line shows the prediction from \citet{wat09}. The
$\rho^{RR}_{\sun}$ in Eq.~\ref{RR_model} was increased by 33\% to 5.6 kpc$^{-3}$
to match the observations, and \citet{wat09} predictions (their Eq.~12) had to
be scaled down by a factor of 10 (assuming that their normalization was meant to
be in units of stars per kpc$^3$). The vertical line is added to guide the eye
and shows the $R_{GC}=30$ kpc.
\label{radial_profiles}}
\end{figure}

The apparent need for a steeper density power-law index at $R_{GC}>30$ kpc is
also illustrated in Figure~\ref{radial_profiles}, where we plot the number
density of RR Lyrae stars as a function of position angle and galactocentric
distance. For $R_{GC}>30$ kpc and directions that avoid the Pisces overdensity,
the \citet{wat09} best-fit ``broken'' power-law (based on essentially the same
RR Lyrae sample) provides a good fit with a $n_H=4.5$ power-law slope (apart
from normalization: we had to scale down their normalization, taken from their
arXiv posting, by a factor of 10).

The apparent good agreement between the data and smooth, \citet{jur08} halo
model in the $340\arcdeg<R.A.<0\arcdeg$ direction at $\sim80$ kpc
(Figure~\ref{radial_profiles} [{\em bottom left}]) is due to Pisces overdensity.

To quantify the statistical significance of the underdense regions relative to a
smooth model, we have divided stripe 82 into 6 regions, and for each computed
the expected number of RR Lyrae stars by integrating $\rho^{RR}_{model}$. The
statistical significance of the underdense regions is calculated by finding the
absolute difference between the observed ($N_{obs}$) and expected ($N_{exp}$)
number of RR Lyrae stars (based on the oblate halo
model, Equation~\ref{RR_model}), then dividing it by the Poisson uncertainty
($\sqrt{N_{exp}}$). The results are presented in Table~\ref{stat_signif}; the
underdense regions are statistically significant at the $(1-3)\sigma$ level
compared to the oblate halo model (Equation~\ref{RR_model}). These
underdensities cannot be explained by incompleteness as a function of apparent
magnitude (since there are dozens of RR Lyrae detected up to 100 kpc in the
Pisces stream region) or angle (because SDSS provides very homogeneous coverage
of the sky). The simplest interpretation is that the spatial profile steepens at
$\sim30$ kpc. This hypothesis is also supported by the distribution of halo
main sequence stars, discussed in Section~\ref{sec:MS}.

\begin{figure}
\epsscale{0.8}
\plotone{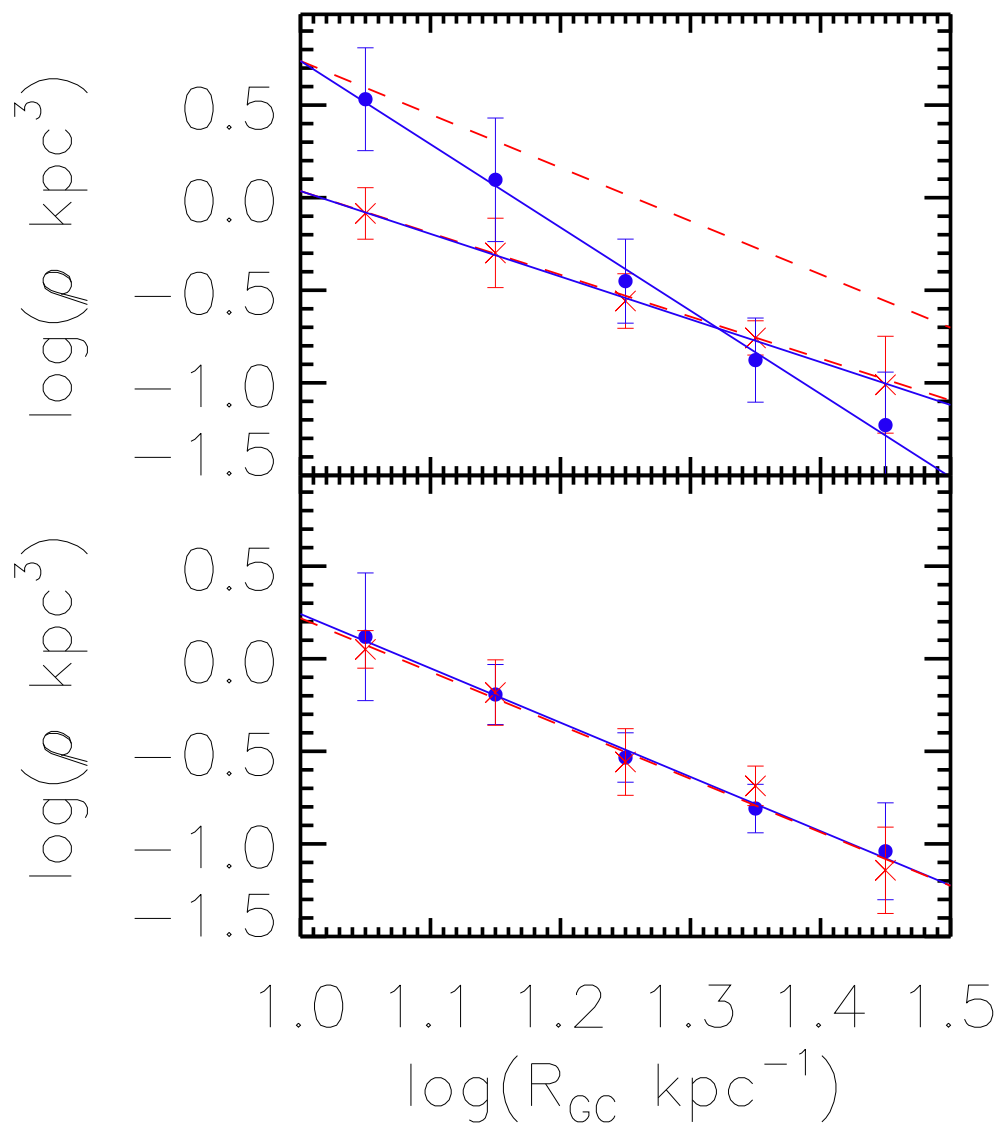}
\caption{
{\em Top}: The radial number density profiles of Oo I ({\em crosses}) and Oo II
RRab stars ({\em dots}) in the $340\arcdeg<R.A.<30\arcdeg$ direction. The
symbols show the observed median number density and the rms scatter around the
median value. The best-fit power laws to the data (power law indices of
$-2.3$ and $-4.5$) are shown as solid lines, with the \citet{mic08} power laws
overplotted as dashed lines. {\em Bottom}: The radial number density profiles of
RRab ({\em dots}) and RRc stars ({\em crosses}) in the
$340\arcdeg<R.A.<30\arcdeg$ direction. The best-fit power laws ({\em solid} and
{\em dashed line}) have consistent power law indices ($-2.9$).
\label{rad_profile_Oo}}
\end{figure}

Using the selection criteria from Section~\ref{P-A distribution}, we have
divided the sample of RRab stars into Oosterhof I (Oo I, 289 stars) and
Oosterhof II (Oo II, 77 stars) subsamples and compared their radial profiles in
Figure~\ref{rad_profile_Oo} ({\em top}). In the direction that does not overlap
with the Sgr tidal stream and the Hercules-Aquila Cloud
($340\arcdeg<R.A.<30\arcdeg$), and for $R_{GC}<30$ kpc, the Oo I subsample shows
a shalower profile than the Oo II subsample. The best-fit power law indices are
$-2.3$ and $-4.5$, respectively. The fit for the OoI subsample is in excellent
agreement with the result by \citet{mic08} obtained over a much larger sky area
(1430 deg$^2$ vs.~$\sim$300 deg$^2$). For the OoII subsample, our best fit is
steeper (Miceli et al. obtained a best-fit index of $-2.9$). The number of stars
with $R_{GC}>30$ kpc is too small (only 23 Oo II stars) to robustly measure the
profile difference beyond the 30 kpc. We also note that the radial number
density profiles of RRab and RRc stars have consistent best-fit power law
indices ($-2.9$), as shown in Figure~\ref{rad_profile_Oo} ({\em bottom}).
Interestingly, the Hercules-Aquila Cloud seems to be dominated by Oo I stars;
the Pisces stream includes only Oo I stars.

\subsection{Period-Amplitude Distribution\label{P-A distribution}}

As suggested by \citet{cat09}, the period-amplitude distribution of RR Lyrae
stars may hold clues about the formation history of the Galactic halo. Catelan
points to a sharp division (a dichotomy first noted by \citealt{oos39}) between
Oosterhof type I (hereafter Oo I, average periods of RRab stars
$\langle P_{ab} \rangle\sim0.55$ days), and Oosterhof type II (hereafter Oo II,
average periods of RRab stars $\langle P_{ab} \rangle\sim0.65$ days) globular
clusters, with only very few clusters in between. On the other hand, the dSph
satellite galaxies and their globular clusters fall preferentially on the
``Oosterhof gap'' ($0.58 < \langle P_{ab} \rangle < 0.62$, see his Figure 5).
In light of these findings, it is interesting to examine the period-amplitude
distribution of stripe 82 RR Lyrae stars in general, and the position of RR
Lyrae stars from the Sgr and Pisces streams in particular, in the
period-amplitude diagram.

\begin{figure}
\epsscale{0.40}
\plotone{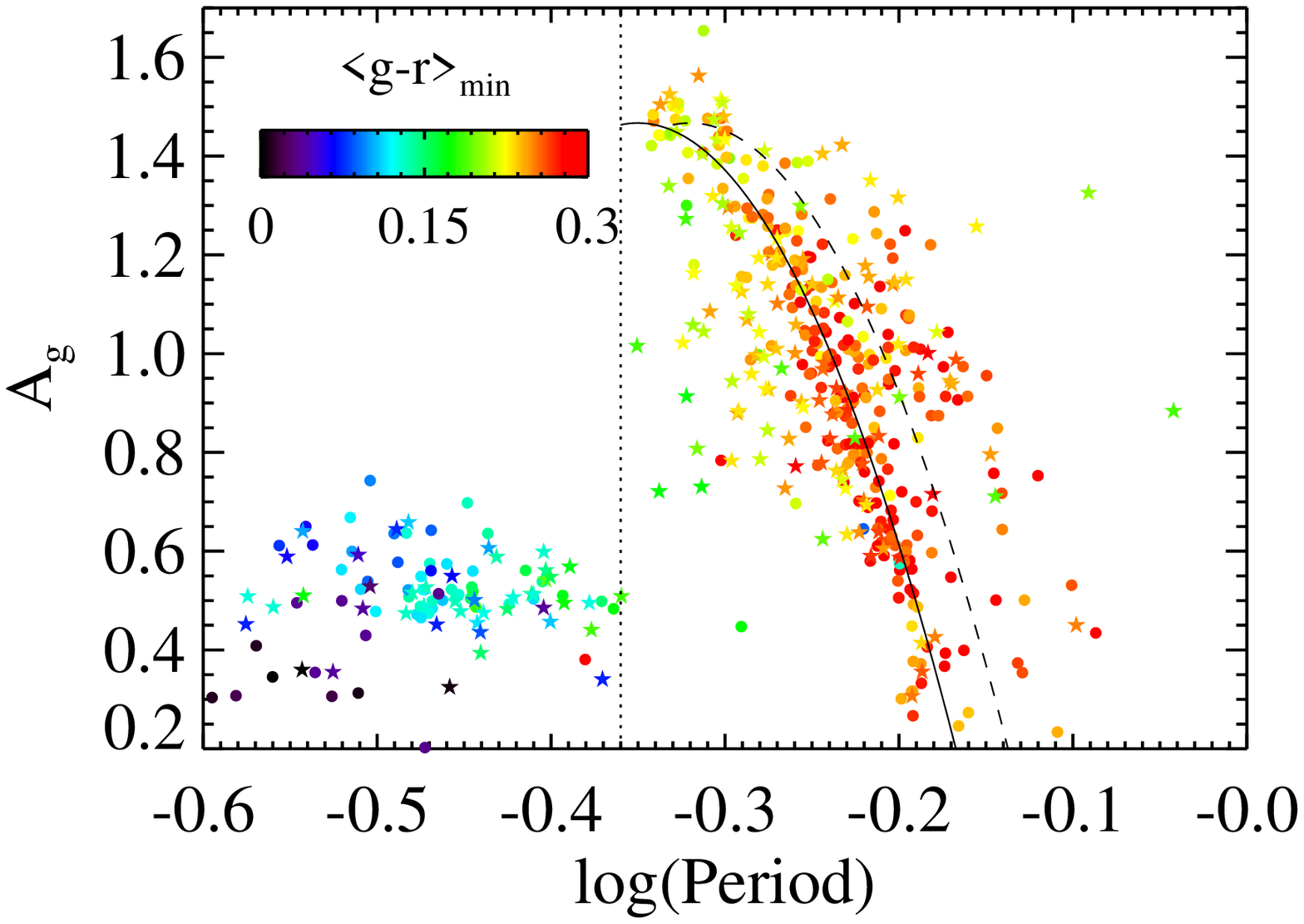}

\plotone{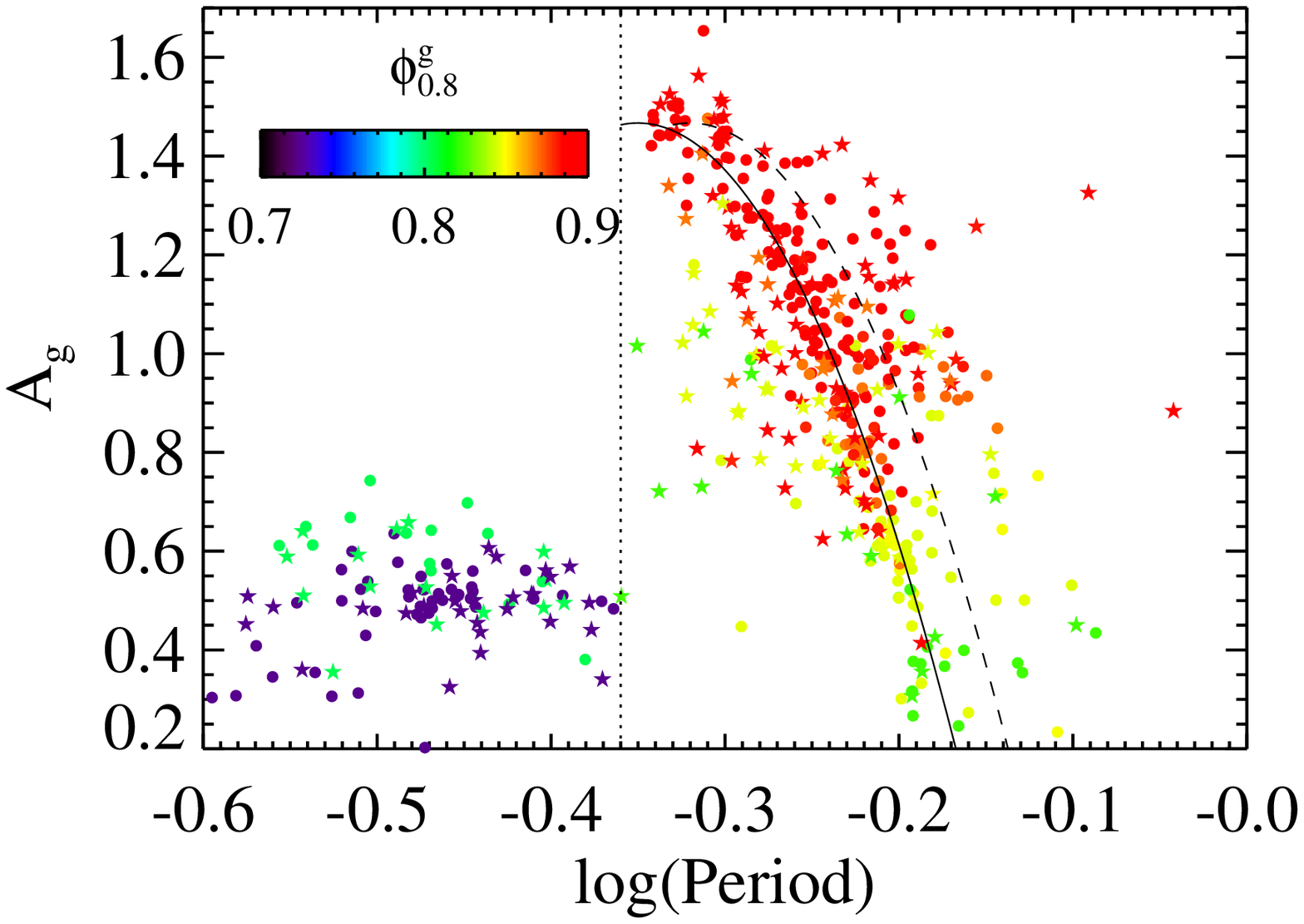}

\plotone{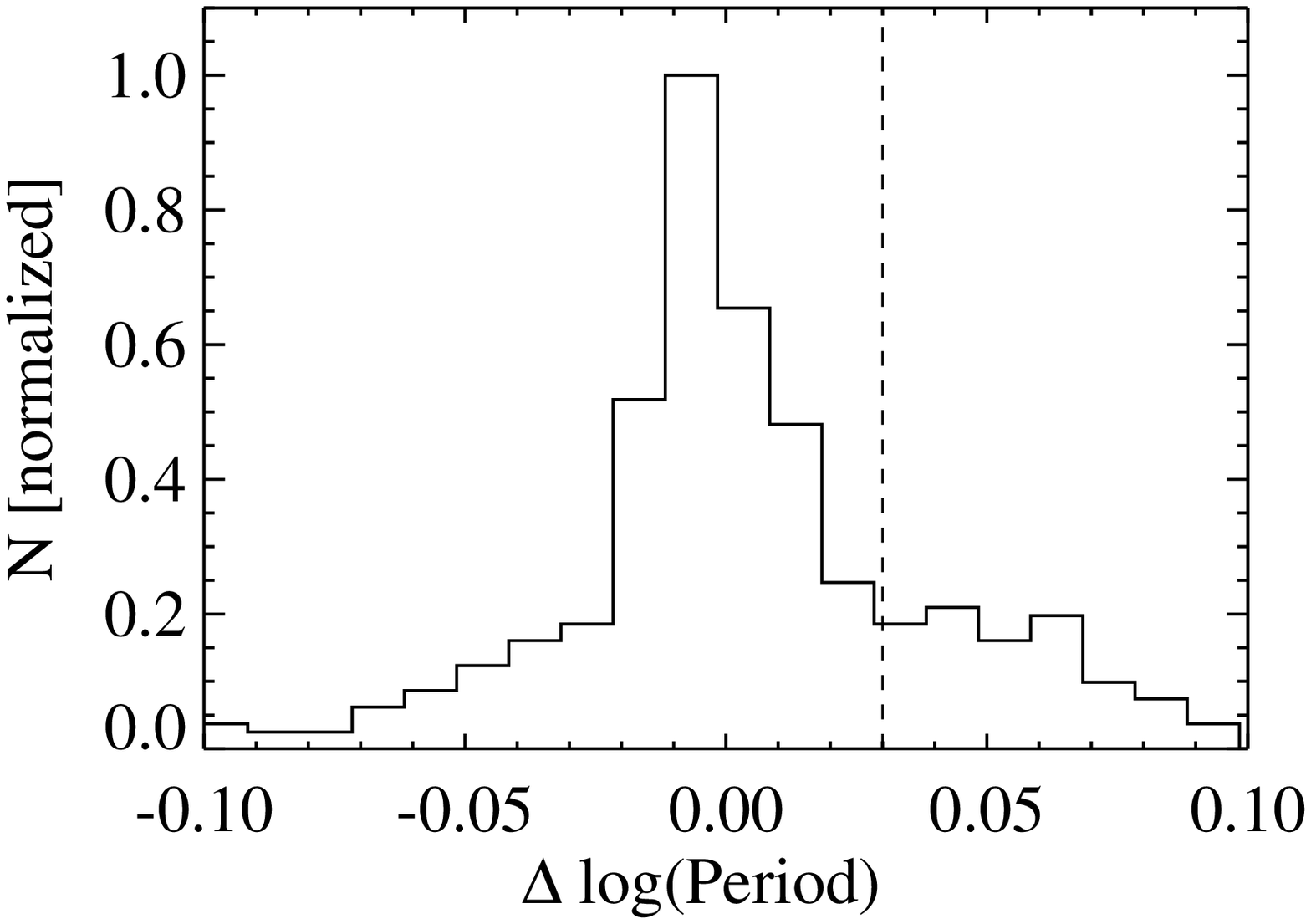}
\caption{
The period-amplitude diagram for RR Lyrae stars with high ($\zeta < 2.3$, {\em
circles}) and low quality $g$-band light curves ($\zeta > 2.3$, {\em stars}),
color-coded by the median $g-r$ color at minimum brightness
($\langle g-r\rangle_{min}$) in the top panel, and by $\phi_{0.8}^g$ in the
middle panel. The vertical line at $\log(Period)=-0.36$ (10.5 hours) divides the
RRc (first overtone, to the left) and RRab stars (fundamental mode, to the
right). The solid line shows the position of the Oo I locus, while the dashed
line (offset by 0.03 in the $\log(Period)$ direction from the Oo I locus line)
separates the Oo I (to the left) and Oo II RRab stars (to the right). The
position of the dashed line was determined from the $\Delta \log(Period)$
histogram ({\em bottom}), where $\Delta \log(Period)$ is the $\log(Period)$
distance (at constant amplitude) from the Oo I locus line. The $g$- and $r$-band
light curves for the star with $A_{g}\sim1.3$ and $\log(Period)\sim-0.09$ are
shown in Fig.~\ref{incomplete_set}.
\label{gamp_vs_period_colored}}
\end{figure}

The period-amplitude (or Bailey) diagram for RR Lyrae stars listed in
Table~\ref{RR_LCparams} is shown in Figure~\ref{gamp_vs_period_colored}. The
stars are separated by the ``quality'' of their $g$-band light curves. A star
with a low quality light curve (a lot of scatter around the best-fit template,
$\zeta>2.3$) may indicate a Blazhko variable RR Lyrae star. The data points are
color-coded by the median $g-r$ color at minimum brightness
($\langle g-r \rangle_{min}$) and by a template parameter $\phi_{0.8}^g$
(which measures the rise time).

We find that $\langle g-r \rangle_{min}$ of RRab and RRc stars becomes redder
(effective temperature at minimum brightness decreases) as the period of
pulsation increases (cooler stars pulsate more slowly). The $g$-band light curve
shape (i.e., the best-fit template) for RRab stars correlates strongly with the
amplitude (and period), becoming on average more asymmetric ($\phi_{0.8}^g$
increases) with increasing amplitude (and decreasing period). Using a linear
fit, we find that the expression
\begin{equation}
    \phi_{0.8}^g = 0.961 + 0.070A_g - 0.249P 
\end{equation}
reproduces the template measured $\phi_{0.8}^g$ with an rms scatter of 0.025
(the measurement error for $\phi_{0.8}^g$ is $\sim$0.01). The distribution of
light curve shapes for RRab stars appears to be continuous.

\begin{figure}
\epsscale{1.0}
\plotone{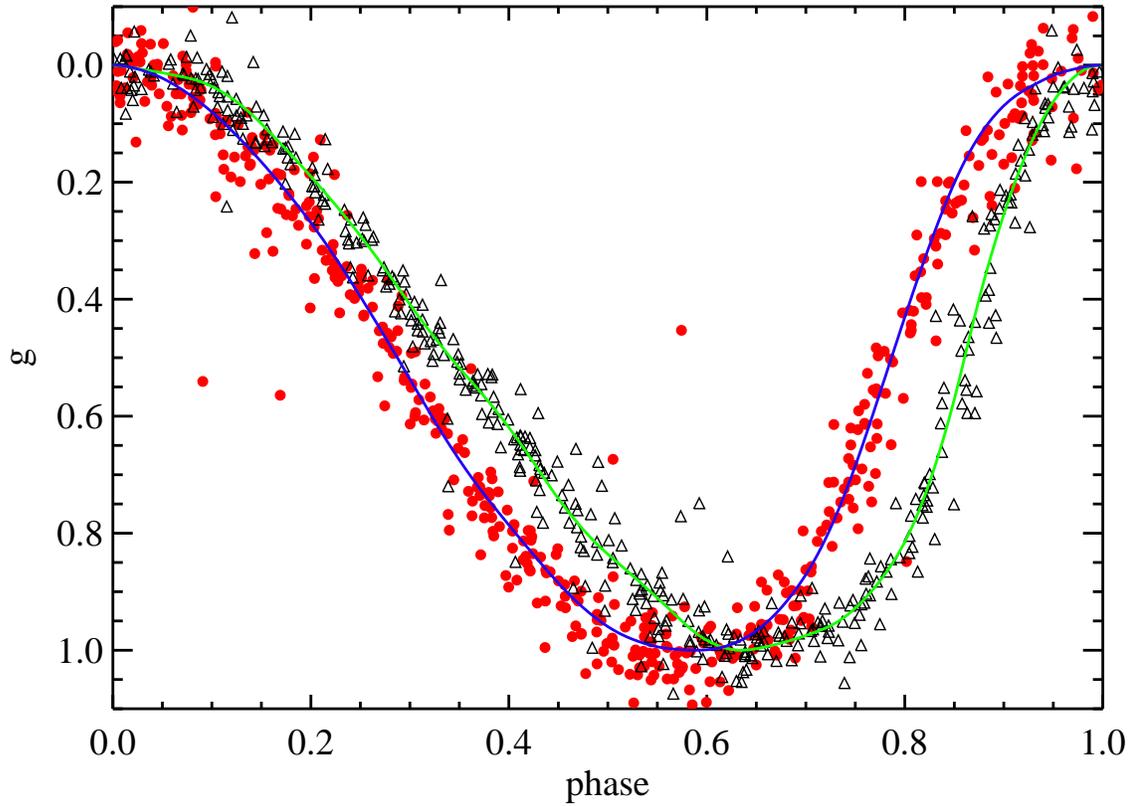}
\caption{
The symbols show high quality $g$-band light curves (normalized and
period-folded) of RRc stars fitted with template 0 (eight stars, {\em dots}) and
1 (nine stars, {\em triangles}), with templates shown as solid lines. The
template 0 and 1 are also shown as the solid and dashed line in
Figure~\ref{template_comparisons} ({\em bottom}). This bimodal distribution of
observed light curves suggests that $c$-type RR Lyrae stars may include two
distinct subtypes.
\label{RRc_bimodal}}
\end{figure}

The shape of RRc stars, on the other hand, appears to be bimodal, a finding
supported by the light curves shown in Figure~\ref{RRc_bimodal}. We emphasize
that this conclusion is independent of template fitting; the only information
needed to fold the light curves in this figure is the best-fit period. The RRc
stars with more asymmetric shapes ($\phi_{0.8}^g\sim0.8$, dashed line template
in Figure~\ref{template_comparisons} [{\em bottom}]) also have on average higher
$g$-band amplitudes than RRc stars with more symmetric curves (by $\sim0.1$
mag). These trends for RRab stars have been known for some time (in the Johnson
$UBVRI$ photometric system; \citealt{sks81}), while the strong bimodality in the
$g$-band light curve shape of RRc stars is apparently a new result.

The period-amplitude distribution of RRab stars in
Figure~\ref{gamp_vs_period_colored} appears to be bimodal, with the two groups
possibly being Oosterhof type I and type II stars. We fit a quadratic line
\begin{equation}
A_g = -3.18 -26.53\log(Period)-37.88\log(Period)^2\label{P-A}
\end{equation}
to high quality light curve stars in the main locus, and note that the fitted
line is very similar to the \citet{ccc05} period-amplitude line of M3 globular
cluster RR Lyrae stars. Since M3 is the prototype Oo I globular cluster, we
label the main locus as the Oo I locus. The stars with low quality light curves
are not used in this fit, as they are probably Blazhko variables that might not
be observed at their maximum amplitude, and therefore tend to scatter towards
lower amplitudes in the Bailey diagram. We calculate the period shift,
$\Delta \log(Period)$ (the $\log(Period)$ distance at a constant amplitude from
the Oo I locus line) as
\begin{equation}
\Delta \log(Period)=\log(Period)-\log(Period\, interpolated\, from\, the\, Oo\, I\, locus).
\end{equation}
The distribution of $\Delta \log(Period)$ values is centered on
$\Delta \log(Period)=0$ by definition (the position of the Oo I locus), and has
a long-period tail. Even though we do not see the clearly displaced secondary
peak that is usually associated with an Oo II component (see Figure 21 in
\citealt{mic08}), hereafter we will refer to stars in the long-period tail as Oo
II stars. Assuming that the distribution around $\Delta \log(Period)=0$ is
symmetric for Oo I stars, we define RRab stars with $\Delta \log(Period)<0.03$
as Oo I, and those with $\Delta \log(Period)>0.03$ as Oo II.

\begin{figure}
\epsscale{1.0}
\plotone{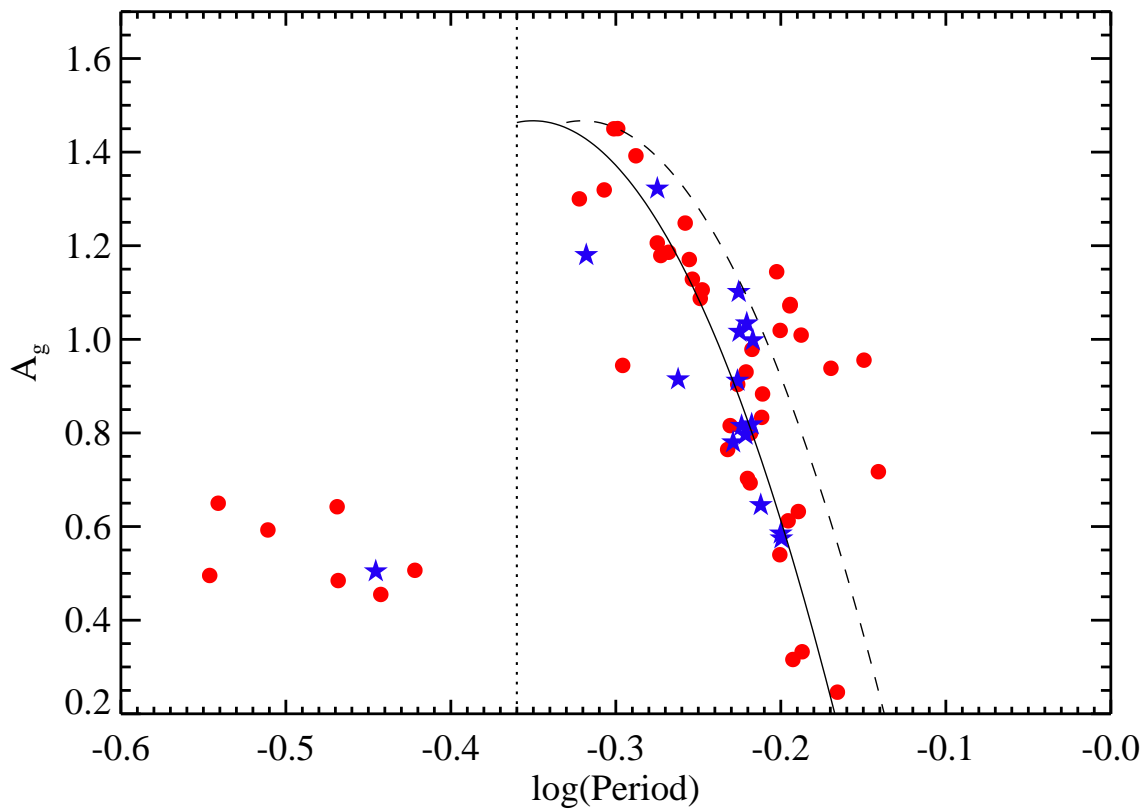}
\caption{
The period-amplitude diagram for RR Lyrae stars selected from the Sagittarius
({\em dots}) and Pisces streams ({\em stars}). The vertical line at
$\log(Period)=-0.36$ divides the RRc (first overtone, to the left) and RRab
stars (fundamental mode, to the right). The solid line shows the position of the
Oo I locus, while the dashed line (offset by 0.03 in the $\log(Period)$
direction from the Oo I locus line) separates the Oo I (to the left) and Oo II
RRab stars (to the right).
\label{P-A_clumps}}
\end{figure}

The period-amplitude distributions of RR Lyrae stars in the Sgr stream
($15\arcdeg<R.A.<44\arcdeg$, $18<d/kpc<35$) and the Pisces stream
($R.A. > 345\arcdeg\, R.A. < 0\arcdeg$, $75<d/kpc<100$) are shown in
Figure~\ref{P-A_clumps}. The Sgr stream has nine RRab Oo II stars while the
Pisces stream has none; if the Pisces stream had the same fraction of Oo II
stars as the Sgr stream, four would be expected. We have already discussed the
differences in the spatial profiles between Oo I and Oo II stars in
Section~\ref{sec:steep}, and provide evidence that Oo II stars have
metallicities about 0.3 dex lower than Oo I stars in the next Section.


\subsection{Metallicity Estimates for RR Lyrae Stars}

In order to obtain a rough estimate of the metallicity of the Pisces and
Sagittarius stream RR Lyrae stars, we consider several methods. The most
accurate metallicity estimates for RR Lyrae stars ($\pm0.2$ dex,
\citealt{viv08, pri09}) can be obtained by employing the $\Delta S$ method
\citep{pre59}, or one of its variants \citep{lay94}, on the observed RR Lyrae
spectra. In the absence of spectra, several photometric metallicity indices can
be used, such as the ultraviolet blanketing index \citep{stu66}, the Str\"omgren
$m_1$ index \citep{ee73}, or the Walraven system (\citealt{bla92} and
references therein). These photometric metallicity indices are less accurate
than the spectroscopic estimates ($\pm0.5$ dex vs.~$\pm0.2$ dex), and are
usually used when the stars of interest are faint, or when samples of RR Lyrae
stars are large. Another method to estimate metallicity for RR Lyrae stars is
based on the light curve shape, as discussed below. 

\subsubsection{Photometric Metallicity\label{Photometallicity}} 

We now attempt to develop a photometric metallicity index similar to Sturch's
ultraviolet blanketing index, but based on the SDSS $ugr$ bandpasses (instead of
the Johnson $UBV$ bandpasses). The basic idea behind this index is that the
correlation between the $U-B$ and $B-V$ color at minimum brightness can be used
to estimate the metallicity of a RRab star. The physics of this estimate is
similar to that of photometric metallicity method used for main sequence stars
(\citealt{ive08} and references therein).

\begin{figure}
\epsscale{0.58}
\plotone{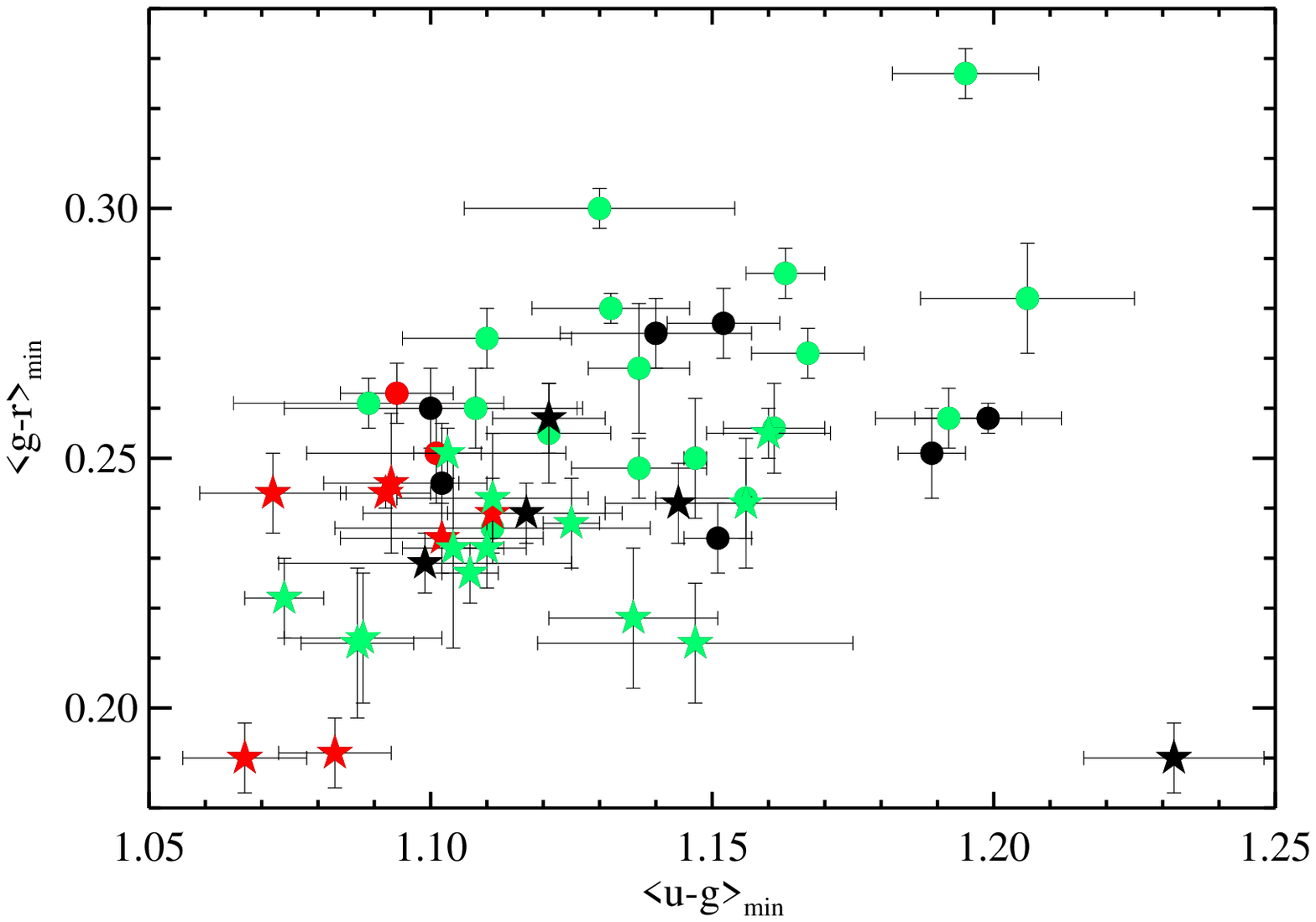}

\plotone{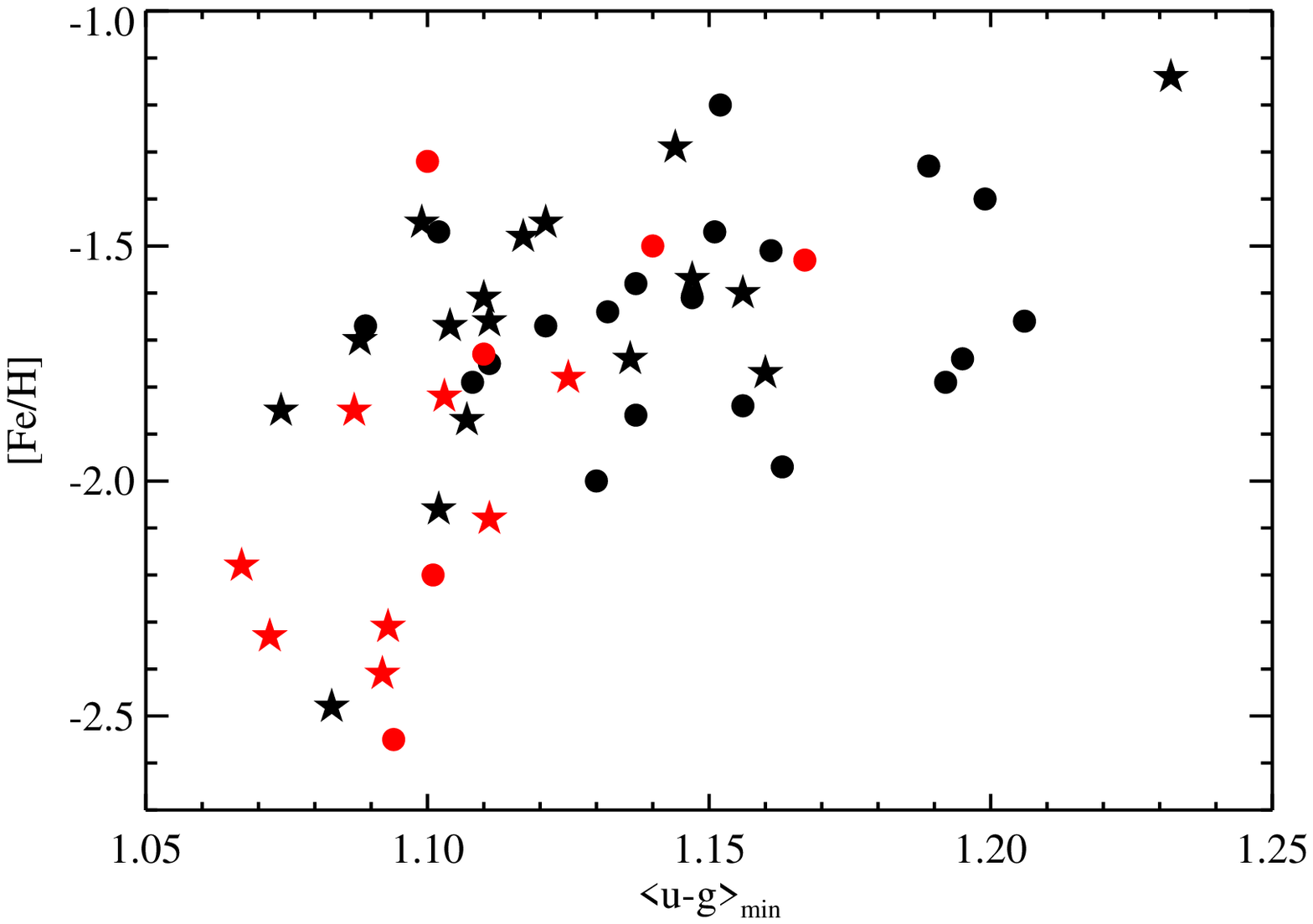}
\caption{
{\em Top}: The $\langle u-g \rangle_{min}$ vs.~$\langle g-r \rangle_{min}$
diagram for 51 RRab stars with high ({\em dots}) and low quality $g$-band light
curves({\em stars}), and with spectroscopic metallicity measurements. The
spectroscopic metallicities were obtained by \citet{dle08} from SDSS spectra
processed with the SEGUE Spectroscopic Parameters Pipeline
\citep[SSPP;][]{lee08}. The $\langle u-g \rangle_{min}$ and
$\langle g-r \rangle_{min}$ ({\em symbols}) show median colors at minimum
brightness ($0.5 < phase < 0.7$), and the error bars show errors in the median.
The symbols are color-coded according to spectroscopically determined
metallicity: $[Fe/H]<-2.0$ ({\em red}), $-1.5<[Fe/H]<-2.0$ ({\em green}), and
$[Fe/H]>-1.5$ ({\em black}). {\em Bottom}: The distribution of Oo I
({\em black symbols}) and Oo II ({\em red symbols}) RRab stars with high
({\em dots}) and low quality $g$-band light curves ({\em stars}) in the $[Fe/H]$
vs.~$\langle u-g\rangle_{min}$ diagram. The Oo II stars are on average more
metal-poor ($[Fe/H]\sim -2.0$) and bluer in $\langle u-g \rangle_{min}$
($\langle u-g \rangle_{min}\sim1.1$) than the Oo I stars ($[Fe/H]\sim -1.7$).
\label{ugmin_vs_grmin}}
\end{figure}

In Figure~\ref{ugmin_vs_grmin} ({\em top}), we investigate if such a correlation
exists between the $u-g$ and $g-r$ colors on a sample of 51 RRab stars with
measured spectroscopic metallicities. The metallicities were obtained from SDSS
spectra with the pulsation phase taken into account, and were processed through
the SEGUE Spectral Parameters Pipeline \citep[SSPP;][]{lee08} by \citet{dle08},
separately from the SDSS Data Release 7 spectra. The random uncertainty in the
De Lee et al.~spectroscopic metallicity is estimated at $\sim0.3$ dex, with
systematic uncertainties of a similar magnitude. We have compared the De Lee et
al.~metallicity values to phase-averaged values listed as part of the SDSS Data
Release 7 and found a median offset of $-0.15$ dex and rms scatter of 0.2 dex.

The conclusion derived from data shown in Figure~\ref{ugmin_vs_grmin} is that
the most metal-poor RR Lyrae ($[Fe/H]<-2.0$) usually have
$\langle u-g \rangle_{min}<1.11$. These metal-poor stars are also predominantly
classified as Oo II stars, as shown in Figure~\ref{ugmin_vs_grmin}
({\em bottom}). The median $\langle u-g \rangle_{min}$ of RRab stars in the
Pisces and Sagittarius streams is $\sim1.13$ mag, resulting in a weak constraint
that their metallicities are in the range $-2.0 < [Fe/H] <-1.5]$ (on the De Lee
metallicity scale).

\begin{figure}
\epsscale{1.0}
\plotone{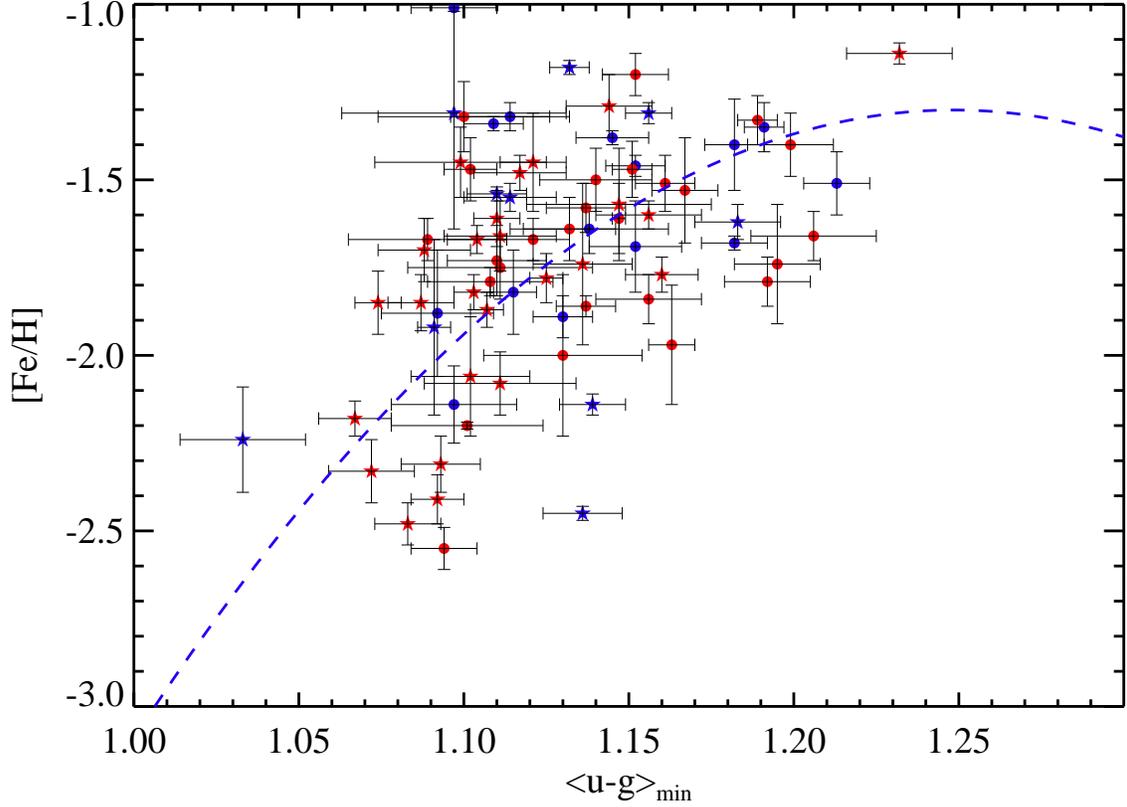}
\caption{
The dependence of spectroscopic metallicity on median $u-g$ color at minimum
brightness ($\langle u-g \rangle_{min}$) for RRab ({\em red symbols}) and RRc
({\em blue symbols}) stars with high ({\em dots}) and low quality $g$-band light
curves ({\em stars}). The best-fit to median $[Fe/H]$ values in 0.02 wide
$\langle u-g \rangle_{min}$ bins is
$[Fe/H]_{photo}=-46.47+72.36\langle u-g\rangle_{min}-28.98\langle u-g\rangle_{min}^2$
({\em dashed line}). The spectroscopic metallicity errors are from
\citet{dle08}, but were not used in the fit as they are most probably
underestimated (they should be $\sim0.3$ dex, De Lee, private communication).
The rms scatter of 76 individual data points around this line is 0.3 dex, and is
consistent with the expected intrinsic scatter in spectroscopic metallicity.
\label{feh_vs_ugmin}}
\end{figure}

Motivated by the $[Fe/H]$ vs.~$u-g$ color distribution shown in the bottom panel
of Figure~\ref{ugmin_vs_grmin}, we tried to estimate the photometric metallicity
using only the $\langle u-g\rangle_{min}$ color. We first compute the median
spectroscopic metallicity in 0.02 wide steps of the $\langle u-g\rangle_{min}$
color, from $\langle u-g \rangle_{min}=1.07$ to $\langle u-g\rangle_{min}=1.25$,
then fit a parabola to these medians (see Figure~\ref{feh_vs_ugmin}). The
best-fit expression is
\begin{equation}
[Fe/H]_{photo} = -46.47 + 72.36\langle u-g\rangle - 28.98\langle u-g\rangle_{min}^2\label{photo_FeH}.
\end{equation}
In the validity range $1.05 < \langle u-g\rangle_{min} < 1.20$, the best-fit
values vary from $-2.4$ to $-1.4$; the largest value of the derivative of the
best-fit is 0.1 dex/0.01 mag for $\langle u-g\rangle_{min}=1.07$. Of course, the
absolute metallicity error (systematic error) in this expression is inherited
from the input data provided by De Lee et al.; it could be as large as several
times 0.1 dex. We emphasize that this estimator should not be applied outside
the color range $1.05 < \langle u-g\rangle_{min} < 1.20$, and that the large
scatter of the calibration sample around the best-fit relation ($\sim0.3$ dex)
should be noted.

Equation~\ref{photo_FeH} predicts $[Fe/H]_{photo}\sim-1.7$ dex for the
metallicity of both the Sgr and Pisces streams. A similar value was obtained by
\citet{vzg05} from spectroscopic observations of RR Lyrae stars in the
{\em leading arm} of the Sagittarius tidal stream. However, the
Equation~\ref{photo_FeH} result for the {\em trailing arm} of the Sgr stream
(discussed here) is lower by 0.5 dex than its metallicity obtained using halo
main sequence stars (in Section~\ref{sec:SgrFeH} below). A possible explanation
for this discrepancy may be a systematic offset in the SSPP metallicity scale
for hotter stars. For 45 RR Lyrae from the Sgr tidal stream
($15\arcdeg<R.A.<44\arcdeg$, $18<d<35$ kpc), we find using the SDSS DR7 SSPP
data a median radial velocity of $-150$ km s$^{-1}$ (with an rms scatter of 35
km s$^{-1}$, indicating a fairly cold stream), and a median metallicity of
$[Fe/H]=-1.5$ (rms scatter of 0.2 dex). If the true median metallicity of the
Sgr tidal stream (trailing arm) is $[Fe/H]=-1.2$, as suggested in
Section~\ref{sec:SgrFeH}, then the SSPP underestimates the metallicities of RR
Lyrae (and possibly blue horizontal-branch) stars by 0.3 dex. The combination of
this $-0.3$ dex offset and a $-0.15$ dex offset between the DR7 and De Lee
metallicities could explain the $0.5$ dex difference between the median
metallicity based on Equation~\ref{photo_FeH} and the main sequence-based value
obtained in the next section. The SSPP results for hotter stars are presently
being refined, based on comparisons with high-resolution spectroscopy, and this
may resolve the apparent discrepancy in the near future.

\subsubsection{Relationship between Metallicity and Light Curve Shape
\label{LCmetallicity}} 

Another potentially useful method for estimating the metallicity of RR Lyrae
stars is based on the Fourier decomposition of light curves \citep{jk96}. Their
result that the Fourier parameters are correlated with the metallicity (see also
\citealt{kc08} for an independent confirmation of this correlation) implies that
our templates should also correlate with metallicity. Furthermore, they have
established that the expansion coefficients are well correlated among
themselves, which is consistent with our finding that light curves are
essentially a one-dimensional family. 

\begin{figure}
\epsscale{1.0}
\plotone{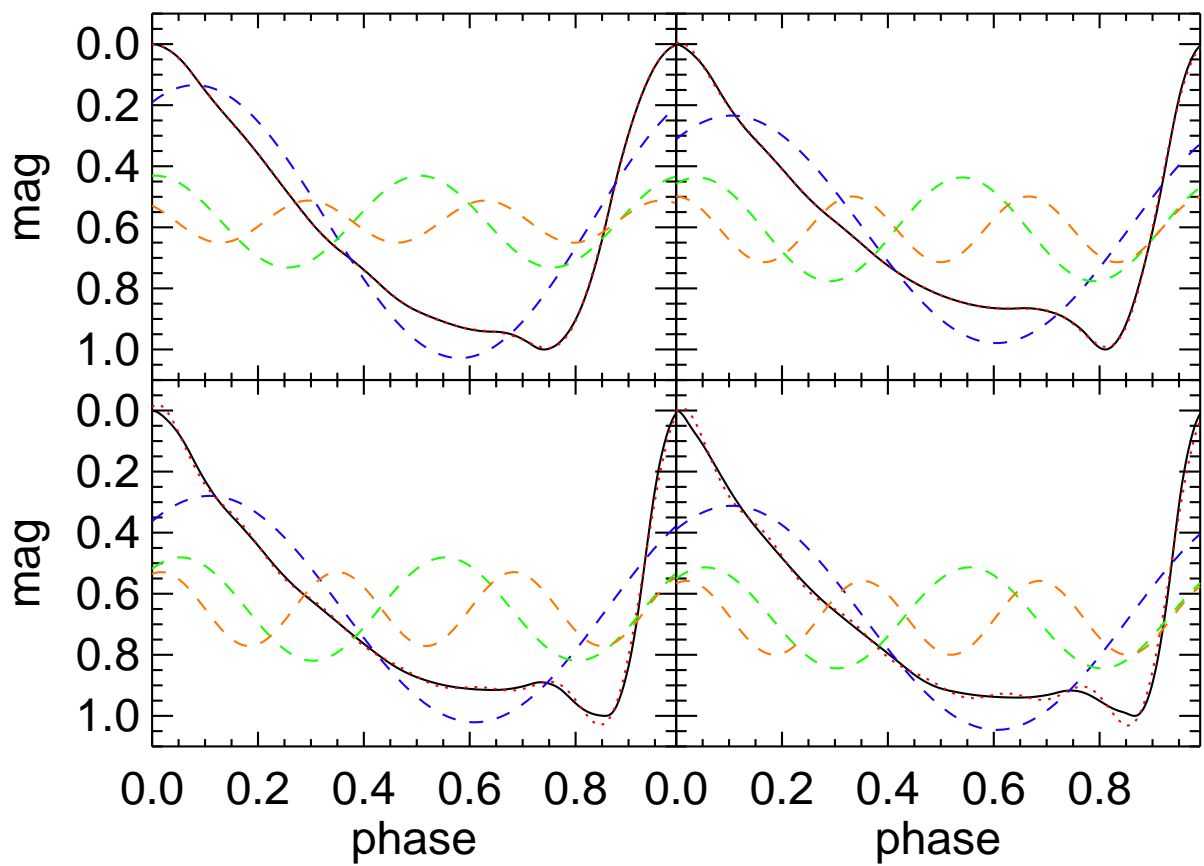}
\caption{Examples of best-fit Fourier expansion with six {\em sine} terms to
light curve templates for $ab$-type RR Lyrae stars. The solid line shows the
template, and the three dashed lines show the first three harmonics (offset for
clarity). The sum of all six harmonics is shown by the dotted line (barely
visible and close to the solid line). These four examples span the full range of
the minimum phase, ranging from 0.75 in the top left panel to 0.87 in the bottom
right panel. For fixed values of period and amplitude, the phase of the minimum
increases as the metallicity decreases.
\label{Fig:Fourier}}
\end{figure}

Following \citet{sc93}, \citet{jk96} parametrized their metallicity-light curve
shape relation using the phase of the third harmonic (for the {\em sine}
expansion), $\phi_{31}$. We have expanded our type-$ab$ templates using six
{\em sine} terms (the fitting residuals for five terms are still comparable to
our template errors, and more than six terms does not result in substantial
best-fit improvement) and found a nearly perfect relation between our $g$-band
template light curve parametrization ($\phi_{0.8}^g$) and the Fourier expansion
($\phi_{31}$): 
\begin{equation}
       \phi_{31} = -6.64 \phi_{0.8}^g + 10.81
\label{Eq:phi31}
\end{equation}
which reproduces directly measured $\phi_{31}$ (from the Fourier expansion) with
a rms of 0.07 (a typical fractional error of 1--2\%), roughly the same as the
error in $\phi_{31}$ due to the template error for $\phi_{0.8}^g$ ($\sim$0.01).
Examples of the best Fourier fits to our templates are shown in
Figure~\ref{Fig:Fourier}. It is evident that the third harmonic controls the
position of the light curve minimum, which in turn determines the value of
$\phi_{0.8}^g$.

With this relation, we can now compute the metallicity estimates using the
\citet{san04} expression (given on the \citealt{zw84} scale; the Jurcsik \&
Kov\'acs scale is about 0.3 dex more metal rich):
\begin{equation}
       [Fe/H]_{San04} = -5.49 - 5.664 P + 1.412 \phi_{31}
\label{Eq:FeHsandage}
\end{equation}
This expression is valid for $P>0.43$ days and $4.7<\phi_{31}<5.7$, and has a
precision of 0.2 dex. The difference between the $[Fe/H]_{San04}$ and the SDSS
spectroscopic metallicity for 51 stars with the latter data has a median offset
of -0.07 dex and rms scatter of 0.33 dex. Given that the errors in SDSS values
are about 0.3 dex, the agreement is satisfactory. Indeed, it is encouraging that
the median offset is so small given that the methods and their calibration are
entirely independent. 

Using the SDSS metallicity for 51 stars, we have also determined a best-fit
linear combination of period, $g$-band amplitude and $\phi_{31}$ 
\begin{equation}
    [Fe/H]_{SDSS} = -0.66 - 3.65 P -0.493 A_g + 0.313\phi_{31}
\label{Eq:FeHsdss}
\end{equation}
which reproduces the spectroscopic metallicity with an rms scatter of 0.26 dex.
The scatter is smaller than for $[Fe/H]_{San04}$ because
Equation~\ref{Eq:FeHsdss} is directly fit to the data. On average,
$[Fe/H]_{SDSS}$ is 0.1 dex higher than $[Fe/H]_{San04}$, with an rms scatter of
their difference of only 0.13 dex; hence, the two expressions are practically
equivalent. Equation~\ref{Eq:FeHsdss}, together with Equation~\ref{Eq:phi31},
implies that for fixed values of period and amplitude, the phase of the light
curve minimum increases as the metallicity decreases; on average,
low-metallicity RR Lyrae have shorter rise time than metal rich RR Lyrae. 

With the light curve based metallicity given by Equation~\ref{Eq:FeHsdss}, we
revisited the $[Fe/H]$ vs.~$u-g$ relationship given by Equation~\ref{photo_FeH}
using a six times larger and much less selection-biased sample. The results show
a much smaller gradient of the $u-g$ color with metallicity, about 0.02 mag/dex,
and call in question the reliability of the photometric metallicity estimator
given by Equation~\ref{photo_FeH}. 

Using metallicity estimates computed by Equations~\ref{photo_FeH} and
\ref{Eq:FeHsandage}, we have compared the metallicity distribution of RR Lyrae
in the Sgr and Pisces streams to the rest of the sample. To within 0.05 dex,
there is no detectable difference. On the other hand, the light curve based
metallicity distributions for Oo I and Oo II stars are different. The median
metallicity for Oo I stars is 0.31 dex higher when using
Equation~\ref{photo_FeH}, and 0.46 dex higher when using
Equation~\ref{Eq:FeHsandage}. However, both the separation of Oo I from Oo II
stars, and the light curve based metallicity use the information about period
and amplitude (e.g., a large period results in Oo II classification and low
metallicity). A somewhat smaller, but still significant ($\sim2\sigma$)
difference in the median metallicity of 0.2 dex is found between 37 Oo I stars
and 14 Oo II stars with available spectroscopic metallicities.

We have also tested light curve-based metallicity estimates using a sample of 11
RR Lyrae stars with accurate light curves, and for which there are reliable
spectroscopic metallicities \citep{viv08}. We fit our $g$-band template set to
their light curves, and apply Equations~\ref{Eq:FeHsandage} and~\ref{Eq:FeHsdss}
to compute $[Fe/H]$. The differences between these light curve-based metallicity
estimates and spectroscopic values show an rms scatter of only 0.14 dex. The
median offset is 0.23 dex, with the spectroscopic values being higher. This
offset may indicate a need to recalibrate the Sandage relation for use with our
templates, because both his relation and the spectroscopic metallicities are
supposed to be tied to the \citet{zw84} scale. 

In summary, given the strong correlations between independent light curve
quantities, period, amplitude, and the shape parameter $\phi_{31}$, and the
results of \citet{jk96} and \citet{kc08}, it is likely that RR Lyrae light
curves contain information about metallicity. However, the currently available
spectroscopic subsample for the SDSS stripe 82 region is insufficient to
convincingly demonstrate this relationship. 

We note that \citet{wat09}, using essentially the same data set as we did, have
claimed that the metallicity for RRab stars can be determined from SDSS light
curves with an rms of 0.25 dex by applying the Jurcsik \& Kov\'acs method, in
apparent conflict with our conclusions. A possible explanation for this
discrepancy is that they fit a large number of free parameters (eight) to small
samples of stars ($\sim$100). We have used the same functional form as proposed
by \citet{wat09} to fit {\em randomly} generated values for $[Fe/H]$ and other
parameters. In 20-30\% of simulated cases, we obtain very similar distributions
in the $[Fe/H]_{fit}$ vs.~$[Fe/H]_{input}$ diagrams as those shown by
\citet{wat09}. 

\section{The Distribution of main sequence Stars in Stripe 82\label{sec:MS}}

The strength of the evidence for a steepening of the halo stellar number density
profile discussed in \S~\ref{sec:steep} is limited by the small size of the RR
Lyrae sample, as well as by difficulties associated with using RR Lyrae stars as
tracers (see Section~\ref{introduction}). A more robust case can be made by
tracing the distribution of the much more numerous main sequence stars, using an
approach such as that discussed by \citet{jur08}. Over most of the SDSS area,
the distribution of blue main sequence stars can be mapped out to a distance
limit of $\sim$20 kpc, which is insufficient to test the transition at $\sim$30
kpc. However, the co-added stripe 82 data (Annis et al., in prep) have a
limiting magnitude about two magnitudes fainter than single-epoch SDSS data, and
can be used to map the number density distribution of blue main sequence stars
out to $\sim$40 kpc.

\begin{figure}
\epsscale{1.0}
\plotone{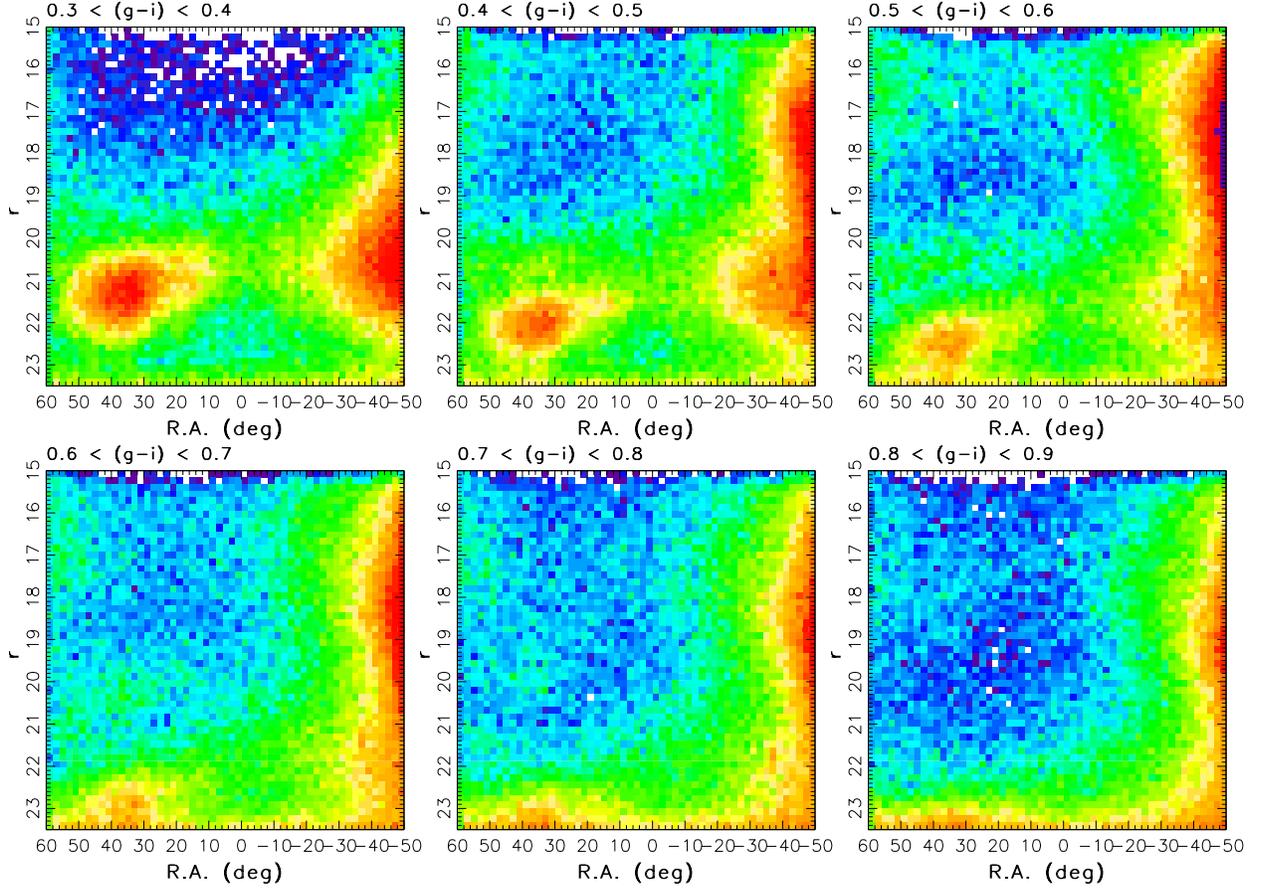}
\caption{
Stellar counts in deep coadded data from stripe 82. Each panel shows stars from
a 0.1 mag wide $g-i$ color bin, as marked on top, with the $g-i$ color
increasing from top left (0.3-0.4) to bottom right (0.8-0.9). Assuming
$[Fe/H]=-1.5$, the median $r$-band absolute magnitude varies from $M_r=4.4$ to
$M_r=6.6$. The counts are shown on a logarithmic scale, increasing from blue to
yellow and red (varying normalization for each panel to emphasize features). The
red vertical features visible at the right edge of each panel are due to disk
stars at $r<20$ and halo stars at fainter magnitudes (including the
Hercules-Aquila Cloud). The stars in the well-defined overdensity at
$R.A.\sim20\arcdeg-50\arcdeg$ belong to the Sgr dSph tidal stream
(trailing arm). The variation of apparent magnitude at which the overdensity is
detected with the $g-i$ color shows that these stars are on the main sequence
(see fig.~\ref{fig:bsFig2}). The distance to the overdensity is in the 25-30 kpc
range.
\label{fig:bsFig1}}
\end{figure}

\begin{figure}
\epsscale{1.0}
\plotone{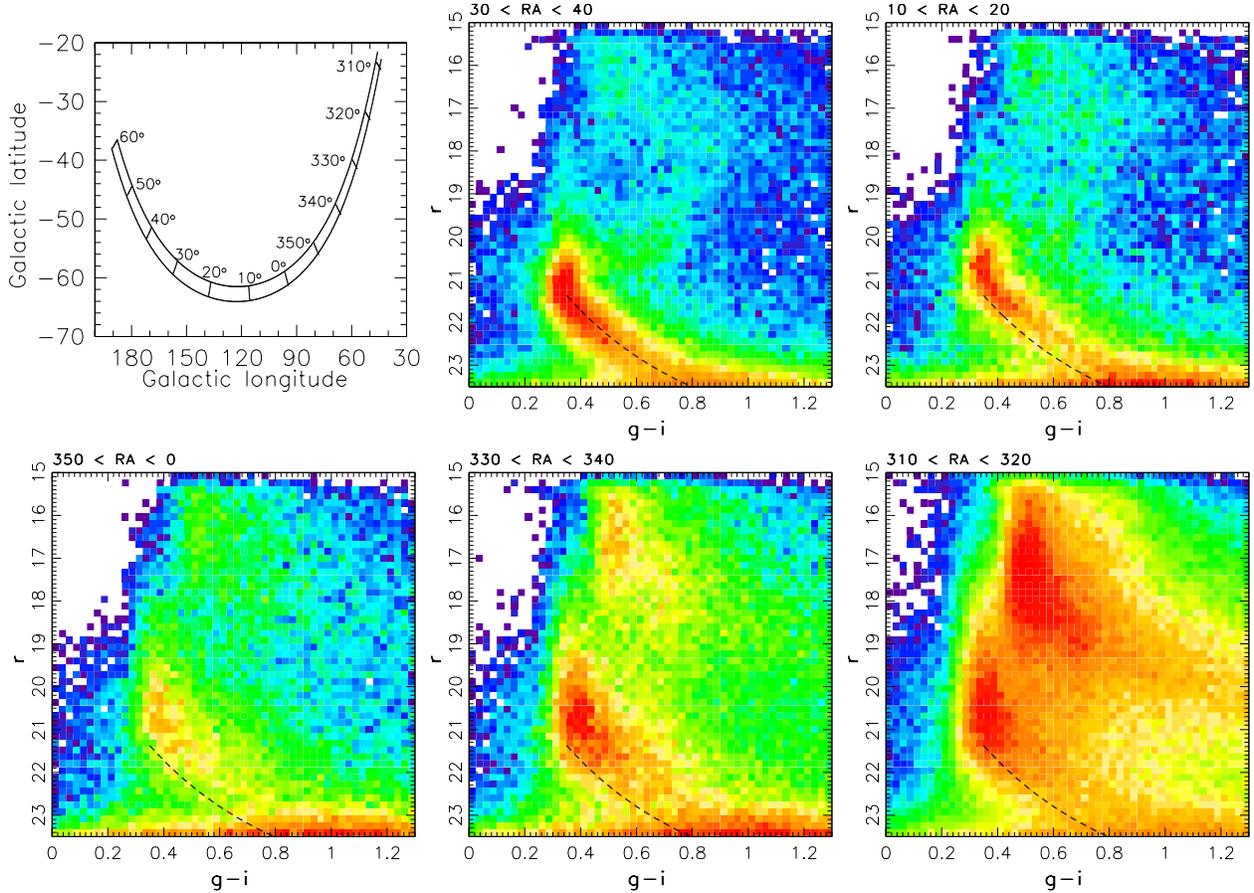}
\caption{
The top left panel shows the stripe 82 footprint in galactic coordinates. The
corresponding R.A. along the stripe is marked (Dec$\sim$0). The other five
panels show the $r$ vs.~$g-i$ color-magnitude (Hess) diagrams for five 25
deg$^2$ large regions from stripe 82 selected by R.A.~(the range is listed on
top of each panel). The top middle panel corresponds to a region intersecting
the Sgr dSph tidal stream; note the main sequence turn-off (MSTO) at
$g-i\sim0.3$ (F type dwarfs) and a well-defined sub-giant branch. The MSTO at
$g-i\sim0.3$ indicates that the main sequence stars in the Sgr trailing arm
are at least 8 Gyr old. The counts are shown on a logarithmic scale, increasing
from blue to yellow and red (with varying normalization to emphasize features).
The dashed lines are added to guide the eye and to show the position of main
sequence stars with $[Fe/H]=-1.2$ at a distance of 30 kpc. Note the lack of
stars with $g-i\sim0.4$ and $r>22$ in most panels. For the median halo
metallicity of $[Fe/H]=-1.5$, $r=22$ corresponds to 28.7 kpc. The feature at
$g-i\sim0.5$ and $r\sim15-18$ is due to thick disk stars. 
\label{fig:bsFig2}}
\end{figure}

Figure~\ref{fig:bsFig1} shows the distribution of stars, in several narrow color
ranges, as a function of apparent magnitude and position along the stripe 82. In
each color bin, stars have approximately the same absolute magnitude, thus more
distant stars are fainter. The data clearly show a well-defined isolated
overdensity of stars at R.A.$\sim20\arcdeg-55\arcdeg$. As the color becomes
redder the overdensity is detected at fainter magnitudes, ranging from
$r\sim21-22$ for the bluest bin and disappearing at the faint end ($r\sim23.5$)
in the reddest bin. This behavior is consistent with the behavior of the
main sequence stars, as illustrated in the top middle panel in
Figure~\ref{fig:bsFig2}. 

\begin{figure}
\epsscale{0.64}
\plotone{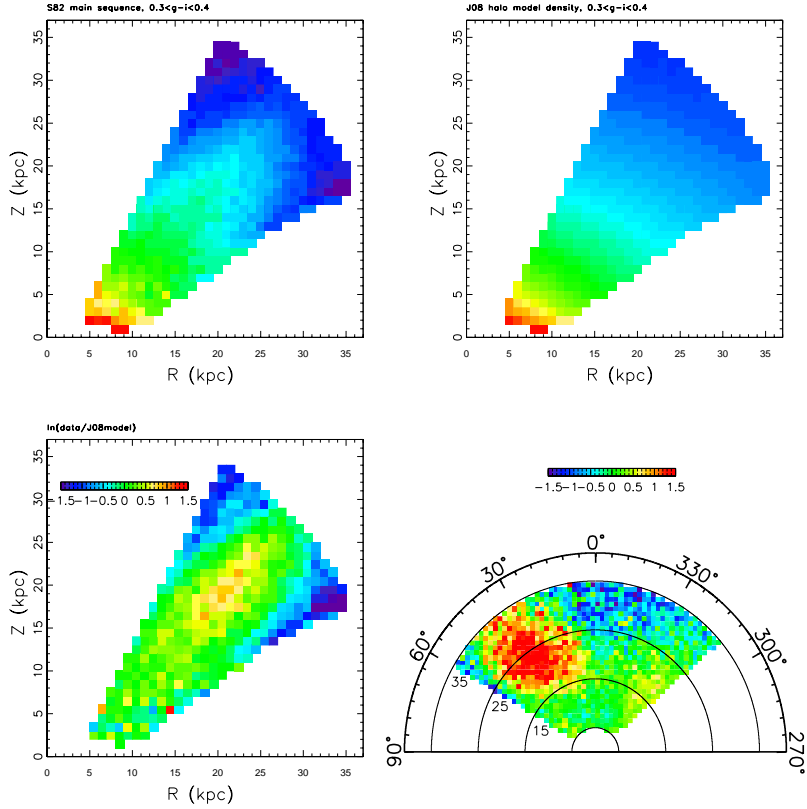}
\caption{A comparison of the observed distribution of main sequence stars in
stripe 82 and predictions based on a smooth oblate halo model from
\citet{jur08}. The top left panel shows the number density of stars with
$0.3<g-i<0.4$ on a logarithmic scale (increasing from blue to yellow and red) in
a galactocentric cylindrical coordinate system (the Sun is at $R=8$ kpc, $Z=0$;
all data are at negative $Z$). The top right panel shows the model prediction.
The bottom left panel shows the ln(data/model) map, color-coded according to the
legend shown in the panel. The overdensity centered on ($R\sim$20 kpc, $Z\sim$20
kpc) is the Sgr dSph tidal stream (trailing arm). Note that at large distances
from the Galactic center, and outside the Sgr dSph tidal stream, the model
overpredicts the observed counts by about a factor of 2 to 3. At galactocentric
distances $\la20$ kpc, the model and data are in excellent agreement. The bottom
right panel is analogous to the bottom left panel, except that the map is shown
in stripe 82 coordinates, with the equatorial position angle (R.A.) marked on
the top. The Sgr dSph tidal stream is seen at $25^\circ<$R.A.$<55^\circ$, with
the maximum value of the overdensity at about five. The Hercules-Aquila cloud is
seen at $310^\circ<$R.A.$<330^\circ$, and distances in the range 10--25 kpc, and
represents a factor of $\sim$1.6 overdensity. The region with
$340^\circ<$R.A.$<5^\circ$ and distances in the range 25--35 kpc represents a
factor of $\sim2$ underdensity, relative to the \citet{jur08} smooth halo model.
\label{fig:bsFig4}}
\end{figure}

The top right and bottom panels in Figure~\ref{fig:bsFig2} exhibit qualitative
evidence for the steepening of the halo density profile; in the color range
$0.3<g-i<0.5$, and away from the Sgr stream, the differential counts of stars
significantly decreases (by about a factor of two) at magnitudes $r>22$. If the
density profile followed the canonical $R^{-3}$ power law, the differential
counts would be approximately constant. In order to quantitatively address this
effect, we compare the observed counts to a simulated sample based on a smooth
oblate halo model from \citet{jur08}. As shown in Figure~\ref{fig:bsFig4}, the
model and data are in excellent agreement at galactocentric distances less than
$\sim$25 kpc. However, at larger distances from the Galactic center, and outside
the Sgr dSph tidal stream, the model {\em overpredicts} the observed counts by
about a factor of two. This mismatch strongly suggests that the halo stellar
number density profile is much steeper beyond 30 kpc than given by the best-fit
model from \citet{jur08} (which was constrained using stars at distances below
20 kpc). Unfortunately, due to the small stripe 82 area, and the effect of the
Sgr dSph tidal stream on measured stellar counts, it is not possible to robustly
derive precise quantitative adjustments to model parameters (e.g., there is a
strong degeneracy between oblateness and density profile power-law index over
such a small area). Nevertheless, this analysis provides {\em the first direct
evidence} that the halo stellar number density is about a factor of two smaller
than predicted by the \citet{jur08} model at $\sim$30 kpc from the Galactic
center. As a rough estimate, in order to induce a factor of two difference in
the count ratio as the galactocentric radius increases from 25 pc to 35 kpc, the
power-law index has to decrease by about two (e.g., from $R^{-3}$ to $R^{-5}$).
This estimate is consistent with the profiles shown in
Figure~\ref{radial_profiles} (beyond 30 kpc).

The bottom right panel in Figure~\ref{fig:bsFig4} shows another overdensity at
$310^\circ<$R.A.$<330^\circ$ and distances in the range 10--25 kpc (about a
factor of 1.6 compared to the smooth halo model; the Sgr dSph tidal stream
produces a peak overdensity of about a factor of five). This overdensity appears
to be the same structure as the Hercules-Aquila Cloud discovered by
\citet{bel07b}, and also seen in the distribution of RR Lyrae stars
(Section~\ref{sec:spatDistr}). 

\subsection{The Metallicity of the Sgr dSph Tidal Stream (Trailing Arm)\label{sec:SgrFeH}}

\citet{ive08} developed a photometric metallicity estimator based on the $u-g$
and $g-r$ colors of main sequence stars. The deep co-added stripe 82 data can be
used to apply this method to a larger distance limit ($\sim$20-30 kpc) than for
nominal SDSS data ($\sim$10 kpc), due to the decreased photometric errors.
Figure~\ref{fig:bsFig3} shows the median metallicity as a function of apparent
magnitude and position along the stripe 82. Throughout most of the volume, the
median metallicity is $[Fe/H]\sim -1.5$, in agreement with results from
\citet{ive08}. However, in the region where the highest concentration of Sgr
dSph tidal stream stars is found, the median metallicity is higher by 0.3 dex.
On the other hand, the median metallicity in the Hercules-Aquila Cloud region is
consistent with the median halo metallicity. 

In regions outside the Sgr tidal stream, and for $r>21$, there is evidence for a
metallicity gradient, and for median values $[Fe/H]< -1.5$. We have analyzed the
metallicity distribution for stars with $320\arcdeg<$R.A.~$<350\arcdeg$
($l\sim60\arcdeg$, $b\sim -40\arcdeg$, $0.3<g-i<0.4$), as a function of
galactocentric radius. In the range 10--20 kpc, we find that the median
metallicity decreases nearly linearly from $[Fe/H]=-1.4$ (with a rms scatter of
0.33 dex) at 10 kpc, to $[Fe/H]=-1.6$ (with a rms scatter of 0.8 dex, dominated
by measurement errors) at 20 kpc. This is a larger gradient (0.02 dex/kpc) than
the upper limit of 0.01 dex/kpc measured by \citet{ive08} at smaller
galactocentric distances. This change of metallicity corresponds to a change
of median $u-g$ color (which provides the signal) from 0.92 mag at 10 kpc (rms
scatter around the median is 0.06 mag) to 0.88 mag at 20 kpc (rms scatter around
the median is 0.17 mag). Clearly, this measurement would greatly benefit
from deeper $u$-band data.

The photometric metallicity measurement for the Sgr dSph tidal stream can be
used to test a novel method for estimating the metallicity of spatially coherent
substructures when the $u$-band data are {\em not} available. The absolute
magnitudes of RR Lyrae and main sequence stars have opposite dependence on
metallicity, with RR Lyrae getting brighter and main sequence stars getting
fainter with decreasing metallicity (see Equation~\ref{abs_mag} in this work and
Equation A2 in \citet{ive08}, respectively). With an assumption that both RR
Lyrae stars and main sequence stars are at the same distance and have the same
metallicity in a well-defined overdensity such as the Sgr dSph tidal stream,
then both distance and metallicity can be simultaneously determined. The
application of this method to the Sgr dSph tidal stream (trailing arm) data is
illustrated in Figure~\ref{fig:bsFig5}. 

A slight difficulty in the application of this method comes from the small RR
Lyrae sample size and the finite width of their apparent magnitude distribution.
We have adopted the mode (with 10 stars in a 0.5 mag wide bin) as a relevant
statistic. The main sequence stars are so numerous that the median and mode are
indistinguishable. A consistent solution is obtained for a distance modulus of
17.2 mag (distance of 27.5 kpc) and $[Fe/H]=-1.2$.

It is encouraging that this metallicity is the same as obtained using the
photometric metallicity method. The photometric metallicity estimate implies
that, with the various error distributions taken into account, the metallicity
difference between RR Lyrae and main sequence stars in the  Sgr trailing arm is
not larger than 0.2 dex. We note that the photometric metallicity scale is tied
to the SDSS spectroscopic metallicity scale for main sequence stars. 

In addition to similarity of their metallicity distributions, RR Lyrae and main
sequence stars in the Sgr trailing arm seem to have similar age distributions.
The top middle panel in Figure~\ref{fig:bsFig2} robustly shows the main sequence
turn-off (MSTO) of the trailing arm at $g-i\sim0.3$ (F dwarfs). Together with
the color-metallicity-age map shown in Figure 28 from \citet{an09}, this MSTO
color and our metallicity estimate ($[Fe/H]=-1.2$) indicate that the main
sequence population in the Sgr trailing arm is at least 8 Gyr old, and thus
consistent with the presumed age of RR Lyrae stars. 

In summary, if both RR Lyrae and main sequence stars are detected in the same
spatially coherent structure, the metallicity can be reliably estimated (to
within $\sim$0.1 dex) by requiring the same distance modulus. We note that the
calibration of both relevant absolute magnitude vs.~metallicity relations is
based on available globular cluster data.

\section{Conclusions and Discussion\label{discussion}}

We have presented the most complete sample of RR Lyrae stars identified so far
in the SDSS stripe 82 data set, consisting of 379 RRab and 104 RRc stars. Our
visual inspection of single-band and color light curves insures that the sample
contamination is essentially negligible. Although the sky area is relatively
small compared to other recent surveys, such as \citet{kel08} and \citet{mic08},
this RR Lyrae sample has the largest distance limit to date ($\sim$100 kpc).

In addition, a large number of well-sampled light curves in the $ugriz$
photometric system have enabled the construction of a new set of empirical
templates. This multi-band empirical template set provides strong constraints
for models of stellar pulsations, through the single-band light curve shape, the
shape variation with wavelength, and through the distributions of the templates.
For example, while $ab$-type RR Lyrae appear to have a continuous distribution
of light curve shapes, $c$-type RR Lyrae exhibit a clear bimodal distribution.
This bimodality suggests that the latter class may include two distinct
subclasses, which would be an unexpected result given that RR Lyrae stars have
been carefully studied for over a century (e.g., \citealt{bai03}). A subclass of
RR Lyrae stars with amplitudes and periods similar to the $c$-type, but with
asymmetric light curves, has been noted in the MACHO data by \citet{alc96}. They
suggested that these stars are pulsating in the second-harmonic mode. 

Our accurate templates will greatly assist the identification of RR Lyrae stars
in data sets with relatively small numbers of epochs. For example, all three
major upcoming wide-area ground-based surveys plan to obtain fewer than 10
epochs (SkyMapper, Dark Energy Survey and Pan-STARRS1 3$\pi$ survey), in the
same photometric system. Without template fitting, the selection based on only
low-order light-curve statistics includes a significant fraction ($\sim$30\%) of
contaminants (dominated by $\delta$ Scuti stars).

The high level of completeness and low contamination of our resulting sample, as
well as the more precise distance estimates, enabled a more robust study of halo
substructure than was possible in our first study. Our main result remains: the
spatial distribution of halo RR Lyrae at galactocentric distances 5--100 kpc is
highly inhomogeneous. The distribution of $\rho/\rho^{RR}_{model}$ suggests that
at least 20\% of the halo within 30 kpc is found in apparently real
substructures. \citet{bel08} find a lower limit of 40\%, but they used a
different metric and different tracers (turn-off stars). Schlaufman et al.~(in
prep) have demonstrated that at least 34\% of inner-halo (in their paper they
explored regions up to 17 kpc from the Galactic center) main sequence turnoff
stars are contained in structures they refer to as ECHOS (elements of cold halo
substructures), based on an examination of radial velocities in the SEGUE
sample. Thus, three separate studies, using very different probes, have come to
essentially identical conclusions.

A comparison of the observed spatial distribution of RR Lyrae stars and
main sequence stars to the \citet{jur08} model, which was constrained by
main sequence stars at distances up to 20 kpc, strongly suggests that the halo
stellar number density profile steepens beyond $\sim$30 kpc. While various
indirect evidence for this behavior, based on kinematics of field stars,
globular clusters, and other tracers has been published (\citealt{car07}; and
references therein), our samples provide a direct measurement of the stellar
halo spatial profile beyond the galactocentric distance limit of $\sim$30 kpc. A
similar steepening of the spatial profile was detected using candidate RR Lyrae
stars by \citet{kel08}. In addition to presenting new evidence for the
steepening of halo stellar number density profile beyond $\sim$30 kpc, we have
confirmed the result of \citet{mic08} that the density profile for Oosterhof II
stars is steeper than for Oosterhof I stars within 30 kpc. 

We have analyzed several methods for estimating the metallicities of RR Lyrae
stars. Using spectroscopic data processed with the SDSS SSPP pipeline as a
training sample, we have established a weak correlation with the $u-g$ color at
minimum light. The rms scatter around the best-fit relation is $\sim0.3$ dex,
but systematic errors could be as large, and the best-fit relation should be
used with care. We have established a relationship between the parametrization
of our template set for $ab$-type RR Lyrae stars and the results based on the
Fourier expansion of light curves. This allowed us to estimate the metallicity
from observed light curves with an estimated rms scatter of $\sim0.3$ dex. This
method failed to uncover a systematic metallicity difference between field RR
Lyrae stars and those associated with the Sgr stream, which is seen in
main sequence stars. Nevertheless, it is likely that this is a promising method
for estimating metallicity, which is currently limited by the lack of a large
and reliable calibration sample. Our templates can be used to bypass noisy
Fourier transformation of sparsely sampled data, as already discussed by
\citet{kin06} and \citet{kk07}.

We introduced a novel method for estimating metallicity that is based on the
absolute magnitude vs.~metallicity relations for RR Lyrae stars and
main sequence stars (calibrated using globular clusters). This method does not
require the $u$-band photometry, and will be useful to estimate metallicity of
spatially coherent structures that may be discovered by the Dark Energy and
Pan-STARRS surveys. While the existing SDSS data are too shallow to apply this
method to the Pisces stream, we used it to obtain a metallicity estimate for the
Sgr tidal stream that is consistent with an independent estimate based on the
photometric $u$-band method for main sequence stars. Our result, $[Fe/H]=-1.2$,
with an uncertainty of $\sim$0.1 dex, strongly rules out the hypothesis that the
trailing arm of the Sgr dSph tidal stream has the same metallicity as halo field
stars ($[Fe/H]=-1.5$), as suggested by \citet{wat09}. \citet{yan09} detected
peaks at $[Fe/H]=-1.3$ and at $[Fe/H]=-1.6$ using SDSS spectroscopic
metallicities for blue horizontal-branch (BHB) stars from the trailing arm.
However, both the Yanny et al.~and Watkins et al.~results could be affected by
systematic errors in the SDSS metallicity scale for BHB stars. Our results
suggest that the SDSS metallicity scale for BHB stars could be biased low by
about 0.3 dex (relative to the SDSS metallicity scale for main sequence stars,
and assuming that RR Lyrae and main sequence stars from the Sgr tidal stream in
stripe 82 area have the same metallicity distributions).

Simulations by \citet{bj05} and \citet{joh08} predict that there should be a 
difference in the chemical composition between stars in the inner halo that was 
built from accretion of massive satellites about 9 Gyr ago, and outer halo
dominated by stars coming from dSph satellites that were accreted in the last 5
Gyr (see Fig.~11 in \citealt{bj05}). These accreted dSph satellites were
presumably more metal-poor than the massive satellites accreted in earlier
epochs (see Fig.~3 in \citealt{rob05}). Other recent simulation studies support
these conclusions. For example, \citet{dh08} find evidence in their simulation
for a strong concentration of (relatively) higher metallicity stars at distances
close to the Galactic center, and the presence of (relatively) lower metallicity
stars at distances beyond 20 kpc from the center. \citet{zol09} find from their
simulations that their inner halos include stars from both in-situ formed stars
and accreted populations, while their outer halos appear to originate through
pure accretion and disruption of satellites. Salvadori et al.~(in prep) have
considered the distribution of ages and metallicities of metal-poor stars in a
Milky-Way like halo, as a function of galactocentric radius, based on a hybrid
N-body and semi-analytic simulation. They find an inner-halo population that is
well-described by a power-law index $n=-3.2$ (for stars with $-2<[Fe/H]<-1$),
and an outer-halo consistent with a much shallower profile, $n=-2.2$. The
relative contributions of stars with $[Fe/H]\le-2$ in their simulation increases
from about 16\% for stars within 7 kpc $< R <$ 20 kpc, to $>40\%$ for stars with
$R >$ 20 kpc.

\begin{figure}
\epsscale{1.0}
\plotone{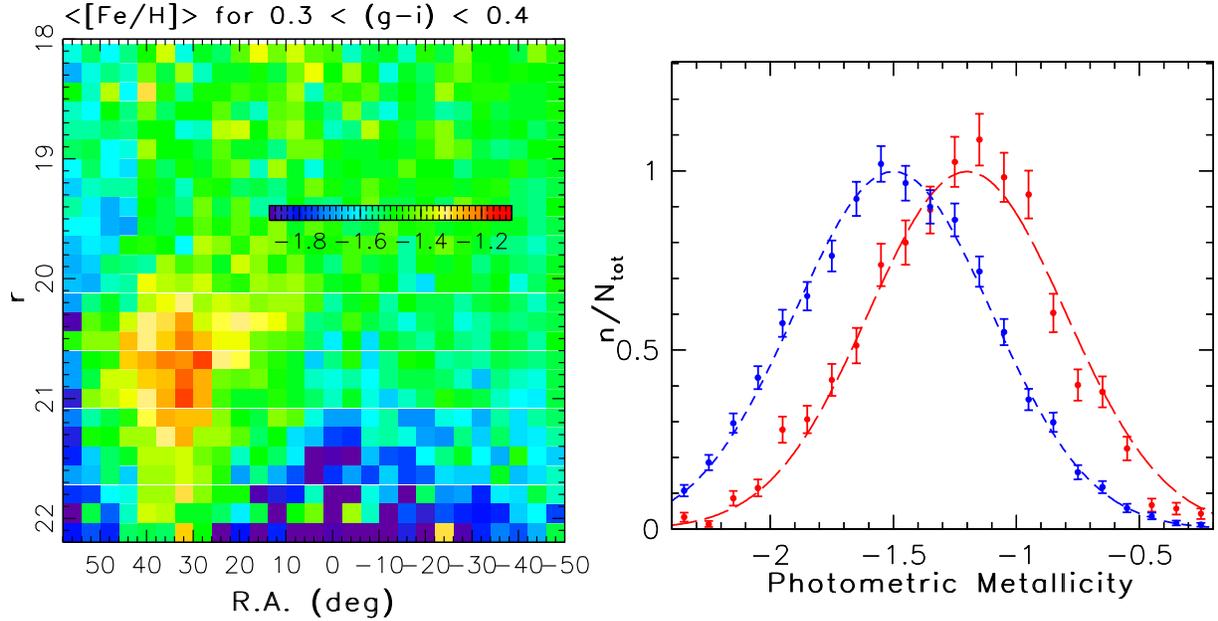}
\caption{
The left panel shows the median photometric metallicity for 88,000 stars with
$0.3<g-i<0.4$ and $18<r<22.2$ in deep coadded data from stripe 82, as a function
of apparent magnitude and position along the stripe. Note the relatively high
metallicity ($[Fe/H]\sim -1.20$) of the Sgr dSph tidal stream region 
(R.A.$\sim30\arcdeg-40\arcdeg$ and $20<r<21$). On the other hand, the median
metallicity in the region with R.A.$<-10\arcdeg$ and $r>20$, where the
Hercules-Aquila overdensity is found, is consistent with the median halo
metallicity ($[Fe/H]=-1.50$). The right panel compares the metallicity
distributions for stars with $20.5<r<21$ selected from two regions: 2,200 stars
with $30\arcdeg < R.A. < 40\arcdeg$ (Sgr dSph tidal stream, right histogram) and
4,500 stars with $-40^\circ < R.A. < -10^\circ$ (control halo sample, left
histogram). The dashed lines are best-fit Gaussians; both have widths
$\sigma=0.4$ dex and are centered on $[Fe/H]=-1.50$ (halo, {\em left}) and
$[Fe/H]=-1.20$ (Sgr dSph tidal stream, {\em right}).
\label{fig:bsFig3}}
\end{figure}

\begin{figure}
\epsscale{0.8}
\plotone{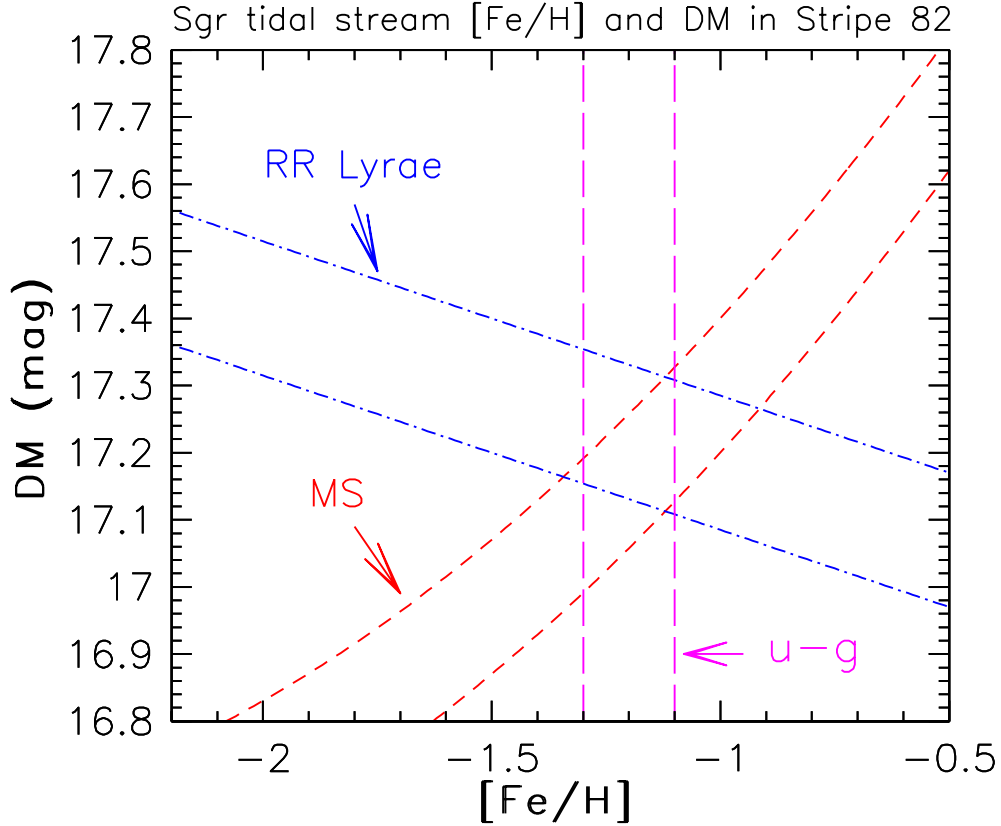}
\caption{A summary of the constraints on the distance and metallicity of Sgr
dSph tidal stream (trailing arm). The dot-dashed lines show a constraint
obtained from the mode of the apparent magnitude distribution of RR Lyrae stars,
with $\pm$0.1 mag adopted as the uncertainty. Diagonal short-dashed lines are
analogous constraints obtained from the median apparent magnitude of
main sequence stars with $0.4<g-i<0.5$. The vertical long-dashed lines mark the
median photometric metallicity for main sequence stars, with $\pm$0.1 dex
adopted as the uncertainty. All three constraints agree if the distance modulus
is $DM=17.2$ mag and $[Fe/H]=-1.2$. A hypothesis that the Sgr dSph tidal stream
(trailing arm) has the same metallicity as halo field stars ($[Fe/H]=-1.5$) is
strongly ruled out.
\label{fig:bsFig5}}
\end{figure}

Our observational results provide some support for these simulation-based
predictions. The faintest main sequence stars ($r\sim21.5$), not including those 
apparently associated with the Sgr stream, exhibit median metallicities at least
0.3 dex lower ($[Fe/H]=-1.8$, see the left panel in Figure~\ref{fig:bsFig3})
than halo stars within 10-15 kpc from the Galactic center \citep{ive08}. Other
studies provide additional support. For example, \citet{car07} and Carollo et
al.~(in prep) have indicated that field stars likely to be associated with the
outer halo exhibit metallicities that are substantially lower ($[Fe/H]=-2.2$)
than those of the inner halo (they place the inner/outer halo boundary at
$\sim$10 kpc). The two metallicity measurements are fully consistent because
photometric $u$-band metallicity estimator is biased high for spectroscopic
values with $[Fe/H]< -2$ (Bond et al., in prep). 

Carollo et al.~(in prep) also found that the inferred density profiles of the
inner- and outer-halo populations differ as well; the inner halo being
consistent with a power-law profile with index $n \sim -3.4$, and the outer halo
having index $n = -2.1$ (the data analysed by Carollo et al. are not suitable
for determination of the relative normalizations of the inner- and outer-halo
populations, due to the selection effects involved). This difference in the
density profiles is very similar to the difference in density profiles for
Oosterhof I and Oosterhof II stars. However, we emphasize that the steeper
profile for the inner halo from Carollo et al.~corresponds to more metal-rich
stars, while we found weak evidence that Oosterhof II stars tend to be more
metal poor than Oosterhof I stars.

Simulations indicate that high surface brightness substructures in the halo
originate from single satellites, typically massive dSph which tend to be
accreted over the last few Gyr \citep{bj05}, and these massive galaxies are
expected to be more more metal-rich than halo field stars \citep{fon08}. The
results from \citet{ive08} and the results presented here seem to support this
prediction. The inner halo has a median metallicity of $[Fe/H]=-1.5$, while at
least two strong overdensities have higher metallicities -- the Monoceros stream
has $[Fe/H]=-1.0$, and for the trailing part of the Sgr tidal stream we find
$[Fe/H]=-1.2$. We emphasize that these three measurements are obtained using the
same method/calibration and the same data set, and thus the measurements of
relative differences are expected to be robust. Simulations by \citet{joh08}
predict that the metallicity of low surface brightness features, such as the
Virgo Overdensity \citep{jur08}, is expected to be lower than the median halo
metallicity. Although the initial estimate by \citet{jur08} (see their Figure
39) claimed a metallicity for this structure of $[Fe/H]\sim-1.5$, a recent study
of the photometric metallicity of stars in this structure by \citet{an09} have
suggested that the mean metallicity is even lower: $[Fe/H]= -2.0$. Both of these
studies are limited by potentially large systematics associated with the
photometric metallicity technique, so more detailed analysis of
spectroscopically determined metallicities for stars in this structure would be
of great value. The sparse information that is available, based on spectroscopic
determinations, supports a mean metallicity between $[Fe/H]= -1.8$ and -2.2
\citep{duf06,viv08,pri09,cas09}. \citet{cas09} report the measurement of a
precise absolute proper motion for the RR Lyrae star RR 167, which appears
highly likely to be associated with the Virgo Overdensity. This proper motion,
in combination with their distance estimate (17 kpc from the Galactic center)
and radial velocity measurement, indicate that the Virgo Overdensity structure
may well be on a highly destructive orbit, with pericenter $\sim 11$ kpc, and
apocenter $\sim 90$ kpc. Thus, the interpretation of the Virgo Overdensity as a
dwarf galaxy, perhaps similar to those that participated in formation of the
outer-halo population, is strengthened.

Our result that (inner-) halo stellar number density profile steepens beyond
$\sim$30 kpc is limited by the relatively small distance limit for main sequence
stars (35 kpc), the sparseness of the RR Lyrae sample ($\sim$500 objects), and
the small survey area ($\sim$300 deg$^2$). Ideally, the halo stellar number
density profile should be studied using numerous main sequence stars detected
over a large fraction of sky. To do so to a distance limit of 100 kpc, imaging
in at least $g$ and $r$ bands (or their equivalent) to a depth several
magnitudes fainter than the co-added SDSS stripe 82 data is required ($r>25$).
Pan-STARRS, the Dark Energy Survey and LSST are planning to obtain such data
over large areas of sky. The LSST, with its deep $u$-band data, will also extend
metallicity mapping of field main sequence stars over half of the sky in the
south; see \citet{ive08} for details. For substructures to be potentially
discovered in the north by Pan-STARRS, the method presented here can be used to
estimate the metallicity of spatially coherent structures even without the
$u$-band data.

\acknowledgments

This research was supported in part by the National Science Foundation under
Grant No. PHY05-51164. B.~S.~and \v{Z}.~I.~acknowledge support by NSF grants AST
61-5991 and AST 07-07901, and by NSF grant AST 05-51161 to LSST for design and
development activity, and acknowledge the hospitality of the KITP at the
University of California, Santa Barbara where part of this work was completed.
\v{Z}.~I. acknowledges the hospitality of the Aspen Center for Physics.
M.~J.~gratefully acknowledges support from the Taplin Fellowship, and NSF grants
PHY-0503584 and AST-0807444. T.C.B. acknowledges partial support from PHY
08-22648: Physics Frontier Center/Joint Institute for Nuclear Astrophysics
(JINA), awarded by the U.S. National Science Foundation. We are grateful to
Kathy Vivas (CIDA) for helpful comments and discussions.

Funding for the SDSS and SDSS-II has been provided by the Alfred P. Sloan
Foundation, the Participating Institutions, the National Science Foundation, the
U.S. Department of Energy, the National Aeronautics and Space Administration,
the Japanese Monbukagakusho, the Max Planck Society, and the Higher Education
Funding Council for England. The SDSS Web Site is http://www.sdss.org/.

The SDSS is managed by the Astrophysical Research Consortium for the
Participating Institutions. The Participating Institutions are the American
Museum of Natural History, Astrophysical Institute Potsdam, University of Basel,
University of Cambridge, Case Western Reserve University, University of Chicago,
Drexel University, Fermilab, the Institute for Advanced Study, the Japan
Participation Group, Johns Hopkins University, the Joint Institute for Nuclear
Astrophysics, the Kavli Institute for Particle Astrophysics and Cosmology, the
Korean Scientist Group, the Chinese Academy of Sciences (LAMOST), Los Alamos
National Laboratory, the Max-Planck-Institute for Astronomy (MPIA), the
Max-Planck-Institute for Astrophysics (MPA), New Mexico State University, Ohio
State University, University of Pittsburgh, University of Portsmouth, Princeton
University, the United States Naval Observatory, and the University of
Washington.

\clearpage


\begin{deluxetable}{rrrrrrrrrrrrrrrrr}
\rotate
\tabletypesize{\scriptsize}
\tablecolumns{17}
\tablewidth{0pc}
\tablecaption{Positions and $ugriz$ photometry for 483 stripe 82 RR Lyrae stars\label{example_photometry}}
\tablehead{
\colhead{$R.A.^a$} & \colhead{$Dec^a$} &
\colhead{$u_{MJD}^b$} & \colhead{$u^c$} & \colhead{$u_{err}^d$} &
\colhead{$g_{MJD}^b$} & \colhead{$g^c$} & \colhead{$g_{err}^d$} &
\colhead{$r_{MJD}^b$} & \colhead{$r^c$} & \colhead{$r_{err}^d$} &
\colhead{$i_{MJD}^b$} & \colhead{$i^c$} & \colhead{$i_{err}^d$} &
\colhead{$z_{MJD}^b$} & \colhead{$z^c$} & \colhead{$z_{err}^d$}
}
\startdata
0.935679 & 1.115859 & 51075.3024 &  18.04 &   0.01 & 51075.3041 & 16.79 & 0.01 & 51075.3007 & 16.65 & 0.01 & 51075.3016 & 16.62 & 0.01 & 51075.3032 & 16.59 & 0.01 \\
0.935679 & 1.115859 & 52196.3030 &  18.54 &   0.02 & 52196.3047 & 17.35 & 0.01 & 52196.3013 & 17.05 & 0.01 & 52196.3022 & 16.94 & 0.01 & 52196.3038 & 16.88 & 0.01 \\
0.935679 & 1.115859 & 52197.3161 & -99.99 & -99.00 & 52197.3178 & 17.21 & 0.01 & 52197.3144 & 16.89 & 0.01 & 52197.3153 & 16.77 & 0.01 & 52197.3169 & 16.75 & 0.01 \\ 
0.935679 & 1.115859 & 52225.1663 &  18.53 &   0.02 & 52225.1679 & 17.34 & 0.01 & 52225.1646 & 17.05 & 0.01 & 52225.1654 & 16.94 & 0.01 & 52225.1671 & 16.90 & 0.01 \\ 
0.935679 & 1.115859 & 52231.2020 & -99.99 & -99.00 & 52231.2037 & 16.98 & 0.01 & 52231.2003 & 16.77 & 0.01 & 52231.2012 & 16.67 & 0.01 & 52231.2029 & 16.65 & 0.01
\enddata
\tablenotetext{a}{Median equatorial J2000.0 right ascension and declination in degrees}
\tablenotetext{b}{Modified Julian date in days}
\tablenotetext{c}{Magnitude is not corrected for ISM extinction, -99.00 if the measurement is unreliable}
\tablenotetext{d}{Magnitude uncertainty, -99.00 if the measurement is unreliable}
\tablecomments{Table~\ref{example_photometry} is published in its entirety in
the electronic edition of the Journal. A portion is shown here for guidance
regarding its form and content.}
\end{deluxetable}

\clearpage

\begin{deluxetable}{ccrrrrrrrrrrrrrrrrrrrrr}
\rotate
\tabletypesize{\tiny}
\tablecolumns{23}
\tablewidth{0pc}
\tablecaption{Light curve parameters for stripe 82 RR Lyrae stars\label{RR_LCparams}}
\tablehead{
\colhead{RR Lyrae$^a$} &
\colhead{Type} &
\colhead{Period$^b$} &
\colhead{$A_u^c$} &
\colhead{$u_0^d$} &
\colhead{$\phi^{u,e}_0$} &
\colhead{$T_u^f$} &
\colhead{$A_g^c$} &
\colhead{$g_0^d$} &
\colhead{$\phi^{g,e}_0$} &
\colhead{$T_g^f$} &
\colhead{$A_r^c$} &
\colhead{$r_0^d$} &
\colhead{$\phi^{r,e}_0$} &
\colhead{$T_r^f$} &
\colhead{$A_i^c$} &
\colhead{$i_0^d$} &
\colhead{$\phi^{i,e}_0$} &
\colhead{$T_i^f$} &
\colhead{$A_z^c$} &
\colhead{$z_0^d$} &
\colhead{$\phi^{z,e}_0$} &
\colhead{$T_z^f$}
}
\startdata
4099  & ab & 0.642 & 0.53 & 17.82 & 51075.27 & 101 & 0.56 & 16.66 & 51075.28 & 101 & 0.40 & 16.56 & 51075.29 & 103 & 0.32 & 16.55 & 51075.29 & 102 & 0.30 & 16.54 & 51075.28 & 100 \\
13350 & ab & 0.548 & 1.03 & 18.24 & 54025.33 & 107 & 1.09 & 16.99 & 54025.32 & 120 & 0.80 & 17.09 & 54025.32 & 112 & 0.64 & 17.15 & 54025.32 & 116 & 0.58 & 17.19 & 54025.32 & 114 \\
15927 & ab & 0.612 & 0.65 & 18.92 & 53680.22 & 102 & 0.70 & 17.65 & 53680.22 & 106 & 0.49 & 17.60 & 53680.22 & 108 & 0.37 & 17.61 & 53680.24 & 104 & 0.35 & 17.62 & 53680.24 & 100 \\
20406 & ab & 0.632 & 0.55 & 16.38 & 54000.28 & 100 & 0.59 & 15.21 & 54000.29 & 101 & 0.43 & 15.12 & 54000.27 & 108 & 0.34 & 15.12 & 54000.29 & 102 & 0.30 & 15.13 & 54000.29 & 100 \\
21992 & ab & 0.626 & 1.11 & 15.53 & 53698.24 & 106 & 1.14 & 14.38 & 53698.24 & 112 & 0.85 & 14.42 & 53698.24 & 114 & 0.66 & 14.52 & 53698.24 & 111 & 0.62 & 14.52 & 53698.24 & 114
\enddata
\tablenotetext{a}{RR Lyrae ID number}
\tablenotetext{b}{Period in days}
\tablenotetext{c}{Amplitude determined from the best-fit template}
\tablenotetext{d}{Mean magnitude corrected for the ISM extinction and determined from the best-fit template}
\tablenotetext{e}{Epoch of maximum brightness determined from the best-fit template}
\tablenotetext{f}{Best-fit template ID number}
\tablecomments{Table~\ref{RR_LCparams} is published in its entirety and with
higher-precision values in the electronic edition of the Journal. A portion is
shown here for guidance regarding its form and content.}
\end{deluxetable}

\clearpage

\begin{deluxetable}{crrrrrrrrrrrrrrr}
\rotate
\tabletypesize{\scriptsize}
\tablecolumns{16}
\tablewidth{0pc}
\tablecaption{Positions and derived parameters of stripe 82 RR Lyrae stars\label{RR_derived_params}}
\tablehead{
\colhead{RR Lyrae$^a$} &
\colhead{R.A.$^b$} &
\colhead{Dec$^b$} &
\colhead{rExt$^c$} &
\colhead{d$^d$} &
\colhead{$R_{GC}^d$} &
\colhead{$\langle u \rangle^e$} &
\colhead{$\langle g \rangle^e$} &
\colhead{$\langle r \rangle^e$} &
\colhead{$\langle i \rangle^e$} &
\colhead{$\langle z \rangle^e$} &
\colhead{$\langle V \rangle^e$} &
\colhead{$\langle u-g \rangle_{min}^f$} &
\colhead{$\sigma_{\langle u-g \rangle_{min}}^f$} &
\colhead{$\langle g-r \rangle_{min}^g$} &
\colhead{$\sigma_{\langle g-r \rangle_{min}}^g$}
}
\startdata
4099  & 0.935679 &  1.115859 & 0.089 & 17.75 & 20.03 & 18.13 & 16.99 & 16.78 & 16.70 & 16.68 & 16.85 & 1.10 & 0.02 & 0.28 & 0.01 \\
13350 & 0.283437 &  1.178522 & 0.080 & 24.77 & 26.55 & 18.84 & 17.68 & 17.54 & 17.50 & 17.50 & 17.57 & 1.16 & 0.01 & 0.26 & 0.01 \\
15927 & 3.254658 & -0.584066 & 0.090 & 29.12 & 30.96 & 19.29 & 18.06 & 17.86 & 17.79 & 17.78 & 17.92 & 1.24 & 0.02 & 0.28 & 0.01 \\
20406 & 3.244369 &  0.218891 & 0.088 & 9.13 & 12.76 & 16.71 & 15.54 & 15.34 & 15.29 & 15.28 & 15.40 & 1.18 & 0.01 & 0.27 & 0.01 \\
21992 & 4.315354 &  1.054582 & 0.077 & 7.35 & 11.54 & 16.19 & 15.04 & 14.91 & 14.86 & 14.85 & 14.93 & 1.13 & 0.01 & 0.25 & 0.01 \\
\enddata
\tablenotetext{a}{RR Lyrae ID number}
\tablenotetext{b}{Median equatorial J2000.0 right ascension and declination in degrees}
\tablenotetext{c}{ISM extinction in the SDSS r band obtained from \citet{SFD98}}
\tablenotetext{d}{Heliocentric and galactocentric distance in kpc}
\tablenotetext{e}{Flux-averaged magnitudes corrected for ISM extinction}
\tablenotetext{f}{Median $u-g$ color (corrected for ISM extinction) at the minimum brightness and the uncertainty in the median}
\tablenotetext{g}{Median $g-r$ color (corrected for ISM extinction) at the minimum brightness and the uncertainty in the median}
\tablecomments{Table~\ref{RR_derived_params} is published in its entirety in
the electronic edition of the Journal. A portion is shown here for guidance
regarding its form and content.}
\end{deluxetable}

\clearpage

\begin{deluxetable}{rrrrrrrrr}
\tabletypesize{\scriptsize}
\tablecolumns{6}
\tablewidth{0pc}
\tablecaption{\citet{ses07} overdensities revisited\label{S07_clumps}}
\tablehead{
\colhead{S07 label$^a$} & \colhead{$N_{tot}^b$} & \colhead{$N_{RRLyr}$} &
\colhead{$N_{non-RR Lyr}^d$} & \colhead{$N_{non-var}^e$} &
\colhead{$N_{RR Lyr}/N_{tot}^f$}
}
\startdata
A & 84  & 76 &  7 & 1 & 0.91 \\
B & 144 & 98 & 44 & 2 & 0.68 \\
C & 54  & 50 &  4 & 0 & 0.93 \\
D & 8   &  5 &  3 & 0 & 0.63 \\
E & 11  &  7 &  4 & 0 & 0.64 \\
F & 11  &  4 &  7 & 0 & 0.36 \\
G & 10  &  5 &  5 & 0 & 0.50 \\
H & 7   &  0 &  7 & 0 & 0.00 \\
I & 4   &  1 &  3 & 0 & 0.25 \\
J & 26  & 12 & 14 & 0 & 0.46 \\
K & 8   &  1 &  7 & 0 & 0.13 \\
L & 3   &  0 &  3 & 0 & 0.00 \\
M & 5   &  1 &  4 & 0 & 0.20 \\
\hline
all & 375 & 260 & 112 & 3 & 0.69 \\  
\enddata
\tablenotetext{a}{Overdensity label from S07}
\tablenotetext{b}{Number of S07 candidate RR Lyrae in the overdensity}
\tablenotetext{c}{Number of true RR Lyrae stars}
\tablenotetext{d}{Number of variable, non-RR Lyrae stars}
\tablenotetext{e}{Number of non-variable stars}
\tablenotetext{f}{Fraction of true RR Lyrae stars}
\end{deluxetable}

\clearpage

\begin{deluxetable}{ccrrr}
\tabletypesize{\scriptsize}
\tablecolumns{5}
\tablewidth{0pc}
\tablecaption{Poisson statistics for RR Lyrae detection in six stripe 82 regions\label{stat_signif}}
\tablehead{
\multicolumn{2}{c}{Region} & \colhead{$N_{exp}^a\pm\sigma^b$} & \colhead{$N_{obs}^c$} & \colhead{Statistical significance$^d$}
}
\startdata
               & $30\arcdeg < R.A. < 60\arcdeg$   & $12\pm4$  & 2 & 2.5 \\
$50 < d/{\rm kpc} < 75$  & $0\arcdeg < R.A. < 30\arcdeg$    & $10\pm3$  & 6 & 1 \\
               & $310\arcdeg < R.A. < 340\arcdeg$ & $19\pm4$  & 8 & 3 \\
\tableline
               & $30\arcdeg < R.A. < 60\arcdeg$   & $9\pm3$  & 0 & 3 \\
$75 < d{\rm kpc} < 100$ & $0\arcdeg < R.A. < 30\arcdeg$    & $8\pm3$   & 3 & 1.5 \\
               & $310\arcdeg < R.A. < 340\arcdeg$ & $14\pm4$  & 4 & 2.5
\enddata
\tablenotetext{a}{Expected number of RR Lyrae calculated by integrating Eq.~\ref{RR_model}}
\tablenotetext{b}{Uncertainty in the number of expected RR Lyrae calculated as $\sqrt{N_{exp}}$}
\tablenotetext{c}{Observed number of RR Lyrae} 
\tablenotetext{d}{Statistical significance in units of $\sigma$, calculated as $|N_{exp}-N_{obs}|/\sigma$}
\end{deluxetable}

\clearpage

\begin{deluxetable}{crrrrrrrrr}
\tabletypesize{\scriptsize}
\tablecolumns{9}
\tablewidth{0pc}
\tablecaption{Positions and derived parameters of RR Lyrae stars in the Sagittarius and Pisces streams\label{Psc_Sgr}}
\tablehead{
\colhead{Stream} &
\colhead{$\langle R.A. \rangle^a$} & \colhead{$\langle d \rangle^b$} &
\colhead{$N_{ab}^c$} & \colhead{$N_{c}^c$} &
\colhead{$N_{OoI}^c$} & \colhead{$N_{OoII}^c$} &
\colhead{$\langle P \rangle_{ab}^d$} & \colhead{$\langle u-g \rangle_{min}^e$}
}
\startdata
Sagittarius & $30\arcdeg$ & 25 & 38 & 7 & 29 & 9 & 0.6 & $1.13\pm0.01$ \\
Pisces & $355\arcdeg$ & 80 & 15 & 1 & 15 & 0 & 0.6 & $1.12\pm0.06$
\enddata
\tablenotetext{a}{Median equatorial J2000.0 right ascension}
\tablenotetext{b}{Median heliocentric distance in kpc}
\tablenotetext{c}{Number of type $ab$, $c$, Oo I, and Oo II RR Lyrae stars}
\tablenotetext{d}{Median period of RRab stars in days}
\tablenotetext{e}{Median $\langle u-g \rangle_{min}$ color of RR Lyrae stars and the uncertainty in the median}
\end{deluxetable}

\clearpage


\clearpage

\clearpage

\clearpage

\clearpage

\clearpage

\clearpage

\clearpage

\clearpage

\clearpage

\clearpage

\clearpage

\clearpage

\clearpage

\clearpage

\clearpage

\clearpage

\clearpage

\clearpage

\clearpage

\clearpage

\clearpage

\clearpage

\clearpage

\clearpage

\clearpage

\end{document}